\documentclass[pre,aps,floats,superscriptaddress,floatfix,twocolumn]{revtex4}
\usepackage{amssymb,amsmath}
\usepackage{amsmath,amssymb}
\usepackage{graphicx}
\usepackage{psfrag}
\usepackage{color}
\usepackage{soul}
\usepackage{dcolumn}
\usepackage{bm}
\usepackage[normalem]{ulem}

\def\beq{\begin{equation}}
\def\eeq{\end{equation}}
\def\bea{\begin{eqnarray}}
\def\eea{\end{eqnarray}}
\begin{document}
\title{Statistical mechanics of phase transitions in elastic media with vanishing thermal expansion}
\author{Sudip Mukherjee}\email{sudip.bat@gmail.com}
\affiliation{Barasat Government College,
10, KNC Road, Gupta Colony, Barasat, Kolkata 700124,
West Bengal, India}
\author{Abhik Basu}\email{abhik.123@gmail.com, abhik.basu@saha.ac.in}
\affiliation{Theory Division, Saha Institute of
Nuclear Physics, 1/AF, Bidhannagar, Calcutta 700064, West Bengal, India}
\date{\today}

\begin{abstract}
 
 We consider a minimal spin model for Ising transitions in an isotropic elastic medium in the zero thermal expansion (ZTE) limit. We set up the elastic theory for this system. We use this theory to identify and study the nature of the fluctuations  in the system near the second order phase transitions at $T_c$  in the zero thermal expansion (ZTE) limit given by $dT_c/dV=0$, where $V$ is the system volume, and explore anomalous elasticity. Allowing for the local strain to couple   {\em asymmetrically} or {\em selectively} with the states of the order parameter, we uncover the dramatic effects of  these couplings on the fluctuations of the local displacements near $T_c$, and also on the nature of the phase transition itself. Near second order phase transitions and with weak asymmetry in the order parameter - strain couplings, the variance of the displacement fluctuations in two dimensions scale with the system size $L$ in a universal fashion as $[\ln (L/a_0)]^{2/3}$; $a_0$ is a small-scale cutoff. Likewise, the correlation functions of the difference of the local displacements at two different points separated by $r$ scale as $[\ln (r/a_0)]^{2/3}$ for large $r$. For stronger selectivity above a finite threshold, this variance diverge as $L$ exceeds beyond a (nonuniversal) size, determined by the model parameters, signaling a transition to a phase with only short range order or the  loss of the positional order of the elastic medium. At dimensions higher than two, for sufficiently weak selectivity, the variance of the displacement fluctuations is $L$-independent corresponding to long range order.  However, if the selectivity parameters rise beyond a dimension-dependent threshold values, again the positional order is lost with a concomitant transition to a phase with short range order. Large values of the order parameter - strain couplings can turn the phase transition into a first order as well.  Our theory establishes a {\em one-to-one correspondence}
between the order of phase transitions and anomalous elasticity near the transitions. Our theory should be a useful guide to possible synthesis of appropriate ZTE materials. 
 
 \end{abstract}
 
 \maketitle
 
 \section{Introduction}\label{intro}
 
 
 
 Phase transitions are ubiquitous in nature and continue to remain central to the subject of equilibrium statistical mechanics for many decades~\cite{stanley,sk-ma}. The Ising model is the simplest model that shows phase transitions between a high temperature ($T$) disordered phase to a low $T$ ordered phase at dimensions $d>1$~\cite{ising}. It and its variants have been used to study phase transitions in a wide class of systems, ranging from magnetic phase transitions between the high $T$ paramagnetic to the low $T$ ferromagnetic phase~\cite{plischke} to phase separation transitions from a high-$T$ well-mixed  phase to a low-$T$ phase separated state~\cite{plischke,safran}. These transitions can be second order through a critical point, or first order with a finite jump in the order parameter~\cite{chaikin}. Phase separation transitions and corresponding nucleation and the growth of  domains are not only significant from statistical mechanics point of views, these are believed to be of paramount importance in host of naturally occurring phenomena, which are of of nonequilibrium origin, e.g., chemical and biological phenomena. For instance, phase separations of proteins are expected to be of vital importance in living biological cells~\cite{cell-bio}. 
 
 Elastic media (e.g., crystals) are broken symmetry phases of systems with continuous translational invariance. These are  characterized by the broken symmetry Goldstone modes or acoustic phonons~\cite{chaikin,landau-elasticity}, which are {\em massless} or long-lived fluctuations. At three dimensions (3D), variance $\langle (\mathbf{u}({\bf x}))^2 \rangle$ of the local displacements $u_i ({\bf x})$ of the position $\bf x$ in the undistorted system is a finite constant proportional to $T$, which means positional long range order (LRO), whereas in two-dimensional (2D) systems, $\langle (\mathbf{u}({\bf x}))^2 \rangle$ grows with the linear system size $L$ as $T\ln\,(L/a_0)$, setting Boltzmann constant $k_B=1$, where $a_0$ is a small-scale cutoff, corresponding to the positional quasi long range order (QLRO); the lack of long range order is a consequence of the Mermin-Wagner-Hohenberg theorem (MWHT)~\cite{mwht}. At higher temperatures, crystals undergo a melting transition into a liquid phase. At 3D, the melting transition is known to be a first order transition~\cite{3d-melting}; at 2D, the transition could be either first or second order~\cite{2d-melting}. 
 
  Statistical mechanics of phase transitions are well-developed and have a long history of study~\cite{stanley}. How elastic degrees of freedom may conspire with the order parameter to affect the macroscopic behavior of a system near a critical point remains a topic of debate. This brings up the question on the nature of phase transitions in elastic media, and in turn the corresponding scaling of the position fluctuations near the phase transition temperature. However, studies on their mutual interplay are relatively few.  For instance, studies in Ref.~\cite{moura} showed that the universal critical scaling of the Larkin-Pikin-Sak model is unaffected by a coupling with an elastic continuum.  In a seminal study, Ref.~\cite{berg-halp} showed that generically an isotropic elastic solid gets unstable in the vicinity of an Ising transition at temperature $T=T_c$, except in the case $dT_c/dV=0$, whence the spin and the elastic degrees of freedom decouple in the long wavelength limit, naturally leading to no mutual effects on each other in their model. Most theoretical studies of phase transitions in elastic media till date either usually concern about the nature and growth of order below the phase transition temperature, or the how the second order transition of the scalar order parameter (of the undistorted system) belonging to the Ising universality class is affected by the displacement fluctuations coming from the background lattice or network.

Recent studies indicate that order parameter-strain couplings could be important in various phenomenologies. For instance, in a cross-linked,
elastic polymer network swollen by a solvent mixture, droplets are found to
grow to a fixed size, controlled by the network stiffness~\cite{style-prx}. More recent experimental studies have revealed that compressive stresses in a polymer network can
suppress phase separation of the solvent that swells it, ultimately stabilizing the mixtures even well beyond the standard liquid-liquid phase-separation boundary~\cite{nat-phys,prl-theory,theory,buddha}. Similarly, the order parameter - elasticity interplay is believed to be of importance to  understand the
ground state and collective excitation of magnetic materials. There has been a growing body of research that considers the various aspects and effects of order parameter - strain coupling in the general context of magnetic 
materials~\cite{mag-results}. In general in any real magnetic crystal, the interplay between the magnetic and the elastic degrees of freedom should exist. Similarly, in a composite  elastic medium (e.g., a binary alloy) or a  of composite system made of a two-component fluid and an embedding elastic network (e.g., a polymer network in a mixed fluid), the order parameter - elasticity interplay should be present. In fact, the question of phase separation in an elastic network is believed to be important in cell biological contexts, e.g., liquid-liquid phase separation is proposed as a candidate mechanism for the formation of membraneless compartments in live biological cells~\cite{membrane-less}.

 Studies on the statistical mechanics of phase transitions in  meta-materials are few and far between. Meta-materials are artificially prepared materials that are designed to have specific properties not found in naturally occurring systems.
 In this paper, we formulate a generic and experimentally testable  theory of phase transitions in a zero thermal expansion (ZTE) medium, coupled with Ising spins. ZTE materials,  a particular type of meta-materials that neither expand nor contract over a range of temperature, can be of diverse origin~\cite{exam}, featuring nearly zero thermal expansion behavior. These materials have enormous potential technical applications in wide-ranging fields,
 e.g., precision engineered parts, microdevices, 
and functional materials, e.g., thermomechanical actuators. {  Our theory should be helpful as a guideline in studies on such ZTE systems. We focus on a ferromagnetic Ising model with nearest neighbor interactions, defined on a deformable lattice. 
We show that the fluctuations near phase transitions in such media can behave very differently from conventional systems with finite thermal expansion. While the scalar order parameter field for the Ising model in the limit of a rigid lattice undergoes a second order universality class belonging to the Ising universality class, we generalize the scope of our study by allowing the order parameter - strain tensor couplings to break the Ising symmetry of the order parameter. In other words, the local strain couples with the order parameter {\em selectively} and {\em asymmetrically}, depending upon the ``two states'' (i.e., two different signs) of the Ising order parameter. This can be generically present, e.g., in a binary fluid mixture, or in biologically relevant systems where the embedding elastic network can chemically interact with the two fluid components in different ways, or in a binary alloy, where the local deformability can explicitly depend upon the excess or deficit of one or the other component.} 

  Specifically, in this work we schematically consider a ferromagnetic Ising model grafted on  Hookean-like spring model for an isotropic elastic medium, e.g., a gel. The spin-lattice interactions are chosen in such a way to ensure vanishing thermal expansions, and correspond to a second order phase transition belonging to the Ising universality class with a critical temperature $T_c$. We construct the Landau-Ginzburg theory of phase transitions in this model. We explore anomalous elasticity near phase transitions in this model, that arises due to the interplay between the local strain and Ising-like order parameter fluctuations in elastic media.  In this work, we focus on systems with {\em zero thermal expansions}, which in our model implies $dT_c/dV=0$~\cite{berg-halp}. In order to generalize the scope of our study, we allow for the local strain to couple selectively or asymmetrically
with the states of the order parameter, which breaks the Ising symmetry of the system through these selectivity-dependent order parameter-strain couplings. This is a situation that can potentially arise in soft matter systems, e.g., in elastic networks immersed in a binary fluid.  Our most surprising result is that unexpected anomalous behavior of the elastic modulii ensues near $T_c$, in contrast to the predictions in Ref.~\cite{berg-halp}. We show that in 2D, the elastic modulii {\em either stiffen} significantly, diverging {\em logarithmically} in the wavevector $q$ in the thermodynamic limit $q\rightarrow 0$ for {\em weak selectivity}, or {\em softens} for {\em strong selectivity}, vanishing at finite length scales indicating selectivity-induced structural phase transitions to a phase with short range order (SRO). At 3D, with sufficiently weak selectivity, the elasticity near $T_c$ is statistically identical to that away from $T_c$. However, with stronger selectivity above a finite threshold, the elastic modulii soften and disappear at finite length scales suggesting structural phase transitions like its 2D counterparts, near $T_c$. Lastly, the order parameter-strain couplings can even turn the second order transition of the rigid lattice system a {\em first order transition} in all dimensions. In this case, the elastic modulii shows {\em finite jumps} across the first order transition temperature. In what follows below, we interchangeably use ``ZTE'' and ``$dT_c/dV=0$''. A brief account of these results are available in the Associated Short Paper~\cite{asc}.
 
 The remainder of this article is organized as follows.  In Section~\ref{summ-tech}, we summarize our principal results. In Section~\ref{free-en-1} we present our microscopic model, and formulate its elastic theory near the phase transition by setting up the corresponding coarse-grained Landau-Ginzburg (LG) free energy. Then in Section~\ref{gauss}, we discuss the properties at the harmonic order of the free energy. Next, in Section~\ref{anharm-th}, we analyze the anharmonic effects in 2D and $d>2$ systems. Then in Section~\ref{first-order}, we discuss how the transition can be turned into a first order one. In Section~\ref{corres}, we set up the correspondence between the order of the transitions and the local displacement fluctuations. We summarize our results and discuss possible future directions in Section~\ref{summ}.  The parameters in this paper, and the equations
defining them (where ever applicable), are summarized in the glossary that constitutes Appendix~\ref{gloss}. Many technical details are available for the interested reader in the subsequent Appendices.

 \section{Summary of the technical results}\label{summ-tech}
 
   We consider Ising spins on a Hookean spring-like model for an isotropic gel. The spin-spring interactions are such that the model has zero thermal expansion (ZTE). We study this system by constructing an LG free energy functional, subject to ZTE, in which we describe an elastic medium with a local displacement field $\bf u(x)$ of a point $\bf x$ in the undistorted system, coupled with an Ising-like continuum order parameter field $\phi ({\bf x})$. We formally define a displacement field ${\bf u}({\bf x})$ that describes the local distortion of the elastic medium, such that ${\bf R}({\bf x})\equiv ({\bf x}+{\bf u}({\bf x}))$ denotes the new, post-fluctuation coordinates in the medium that was originally located at ${\bf x}$. Here, ${\bf x}=(x,y)$ for a 2D system, and ${\bf x}=(x,y,z)$ for a corresponding 3D realization~\cite{chaikin}.
 
  After dropping terms that are irrelevant in the renormalization group (RG) sense, the free energy functional $\cal F$ invariant under a constant shift of $\bf u$, as appropriate for this system is
 \begin{widetext}
\begin{eqnarray}
 {\cal F} &=& \int d^dx \left[\frac{r}{2}\phi^2 + \frac{1}{2}({\boldsymbol\nabla}\phi)^2 + v\phi^4 + \mu (\nabla_i u^T_j)^2 + \frac{\tilde\lambda}{2} (\nabla_i u^L_j)^2 
 + \left(g_1\phi^2 + \overline g_1 \phi\right) (\nabla_i u^T_j)^2 + \left(g_2\phi^2 + \overline g_2 \phi\right) (\nabla_i u^L_j)^2\right],\label{free2-intro}
\end{eqnarray}
\end{widetext}
where, $r=0$ is the mean-field critical point in the rigid lattice limit, $u_i^L({\bf x})$ and $u_i^T({\bf x})$ are the inverse Fourier transforms of $u_i^L({\bf q})$ and $u_i^T({\bf q})$. Here, ${\bf u}^L({\bf q})$ and ${\bf u}^T({\bf q})$ are the projections of ${\bf u (q)}$ along and normal to the wavevector $\bf q$. Couplings $\overline g_1$ and $\overline g_2$ manifestly breaks the Ising symmetry of the problem, and are the {\em selectivity parameters}, since their contributions to $\cal F$ depend upon the sign of $\phi$. Lastly, $\mu$ and $\tilde\lambda$ are, respectively, the shear and bulk modulii of the system.

  We study the fluctuations in the system corresponding to $\cal F$ in (\ref{free2-intro}). Our principal results are given below.
  
  In systems with selectivity, i.e., with nonzero $\overline g_1,\,\overline g_2$, the transition is generically first order similar to the liquid-gas first  order transition. However, again like a liquid-gas transition a second order transition with a critical point at $T_c$ can be accessed.
 
 \subsection{Results on 2D systems}
 
 {  
 We show that at 2D with weak selectivity (i.e., weak $\overline g_1$ and $\overline g_2$) such a thin elastic sheet can significantly stiffen  close to the critical point $T_c$ of the ordering transition, a property not found in a pure (one component) system, or away from critical points. In particular, both $\mu$ and $\tilde \lambda$, the shear and bulk modulii respectively, acquire scale-dependence, diverging as $[\ln (\Lambda/q)]^{1/3}$ in the long wavelength limit, which implies {\em anomalous elasticity}; here $\Lambda$ is an upper wavevector cutoff. This is analogous to anomalous elasticity in 3D equilibrium smectics~\cite{smectics}.  
 As a result, the variance of the local fluctuating displacement field $u_i ({\bf x})$ ($i$ is the Cartesian component), that describes the local deformation or dilation, $\langle [u_i ({\bf x})]^2\rangle$ shows a {\em universal dependence} on  the system size $L$ as $[\ln\,(L/a_0)]^{2/3}$, a significantly weaker $L$-dependence than the well-known $\ln\,(L/a_0)$-dependence found away from $T_c$, or in one-component systems at any $T$; $a_0=2\pi/\Lambda$. Likewise, the two-point correlation function of the difference of the local displacements at two points separated by a distance $r$ scales as $[\ln\,(r/a_0)]^{2/3}$, unlike the well-known $\ln (r/a_0)$-dependence of QLRO. These imply a positional order logarithmically stronger than the usual QLRO. We call this  {\em positional SQLRO}, that forms an altogether {\em new, heretofore unstudied universality class}. The prediction of this SQLRO and the associated universality class is a principal outcome of the present study. The second order transition remains unaffected with the critical exponents belonging to the Ising universality class, as they do in the corresponding rigid system. This is not the only state of the system near $T_c$. With sufficiently strong breaking of the Ising symmetry, the system distabilizes, leading to the loss of any SQLRO positional order (PO); only positional short range order (SRO) is possible.  This instability is driven by the selectivity parameters $\overline g_1$ and $\overline g_2$, which are also the Ising symmetry-breaking couplings. Our detailed results show that as the initial or microscopic value of the dimensionless ratio $\Gamma\equiv \overline g_1^2/(\mu g_1)$ exceeds a finite threshold $\Gamma_{1c}$
 \begin{equation}
  \frac{\overline g_1^2}{\mu g_1}>\Gamma_{1c}\approx 1.5 \label{intro-gamm}
 \end{equation}
the system destabilizes with the attendant loss of SQLRO. An analogous relation exists involving $\overline g_1,\tilde\lambda$ and $g_2$; see later. We further show that in such an
 elastic medium near $T_c$ with (\ref{intro-gamm}) holding good, as soon as the system size $L$ exceeds a threshold value $L_c$, controlled by the microscopic (bare) value of the selectivity parameters, is lost. Our theory gives the following expression of $L_c$:
 \begin{equation}
 L_c=a_0\exp\left[\frac{\mu}{\left(\frac{\overline g_{1}^2}{2\mu} - g_{1}\right)} \frac{2\pi}{T_c}\right].\label{phase-2d-intro1}
\end{equation}

 Equations~(\ref{intro-gamm}) allows us to draw the phase diagrams of the system in 2D, demarcating the phases with PO (i.e., SQLRO) and without PO (i.e., with SRO) which are shown 
  in the schematic phase diagrams in Fig.~\ref{phase-diag1-2d} and Fig.~\ref{phase-diag3-2d} and Fig.~\ref{phase-diag2-2d}. The $\overline g_1$ versus $\mu$ curve for a fixed $g_1$, and the $\overline g_1$ versus $g_1$ curve for a fixed $\mu$, which are the phase boundaries between the ordered (PO) phase and disordered phase (SRO), are obviously parabolas as can be seen by equating $\overline g_1^2/(\mu g_1)$ with $\Gamma_{1c}$; see Eq.~(\ref{intro-gamm}). Furthermore, a phase diagram can be drawn in $\overline g_1^2 - L$ plane showing the regions with PO and SRO by using Eq.~(\ref{phase-2d-intro1}); see Fig.~\ref{phase-diag2-2d}.

 Similar phase diagrams could be drawn in terms of $\overline g_2$ and $g_2,\,\tilde\lambda$ or $L$ (not shown here).
 
 
 \begin{figure}[htb]
  \includegraphics[width=7cm]{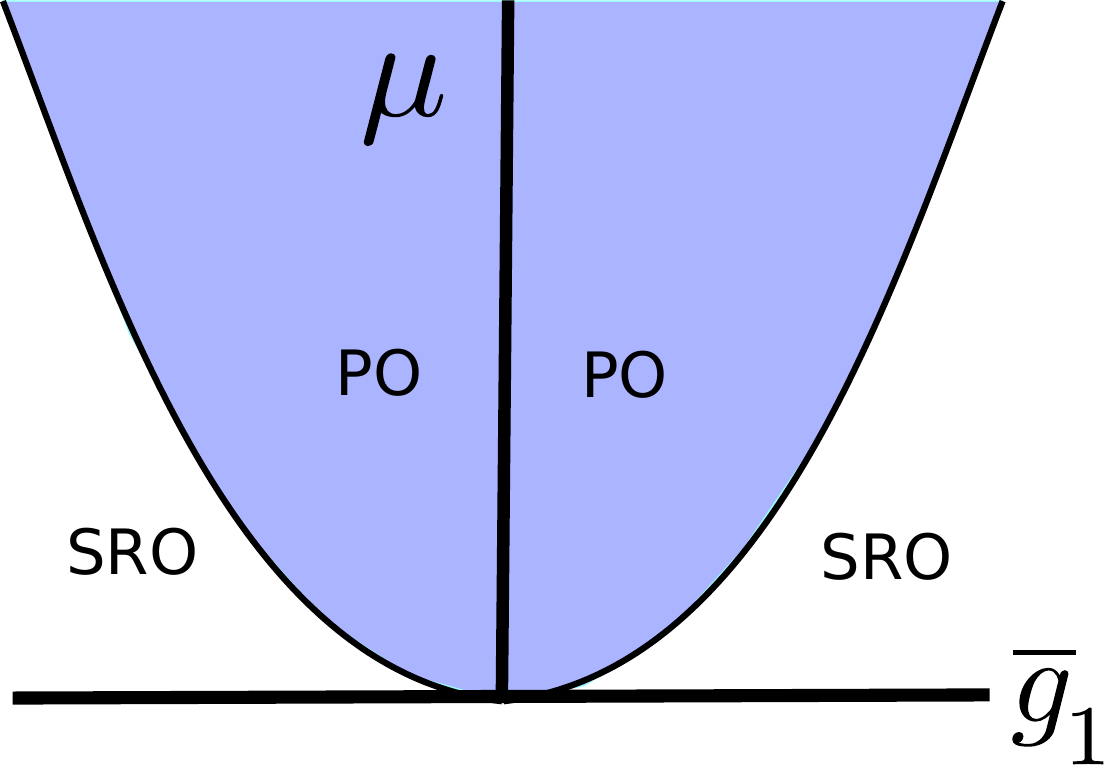}
  \caption{(color online) Schematic phase diagram in the $\overline g_1 -\mu$ plane for $d \geq 2$ near $T_c$. The blue shaded region is where $\mu>0$ corresponding to positional SQLRO with second order transition in 2D. The region outside has $\mu<0$ implying loss of positional order or short range order.  The phase boundary (red) can be obtained by using (\ref{intro-gamm}) above (see text).}\label{phase-diag1-2d}
 \end{figure}
 
 \begin{figure}[htb]
  \includegraphics[width=7cm]{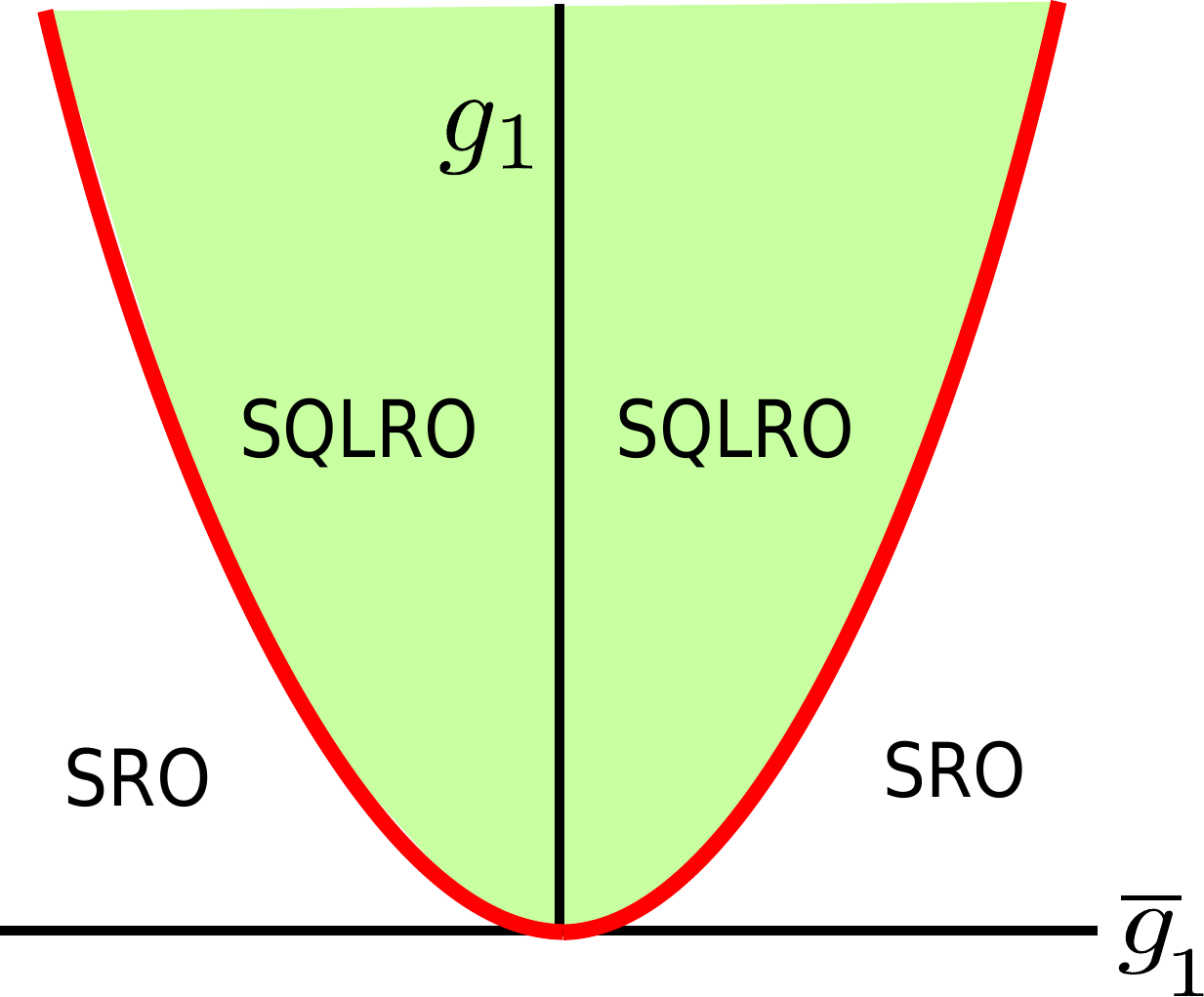}
  \caption{(color online) Schematic phase diagram in the $\overline g_1 - g_1$ plane for $d \geq 2$ near the phase transition. The middle light green region (PO), corresponds to  SQLRO in 2D. The region outside has SRO.  The phase boundary (red) can be obtained by using (\ref{intro-gamm}) above  (see text).}\label{phase-diag3-2d}
 \end{figure}
 
   \begin{figure}[htb]
  \includegraphics[width=7cm]{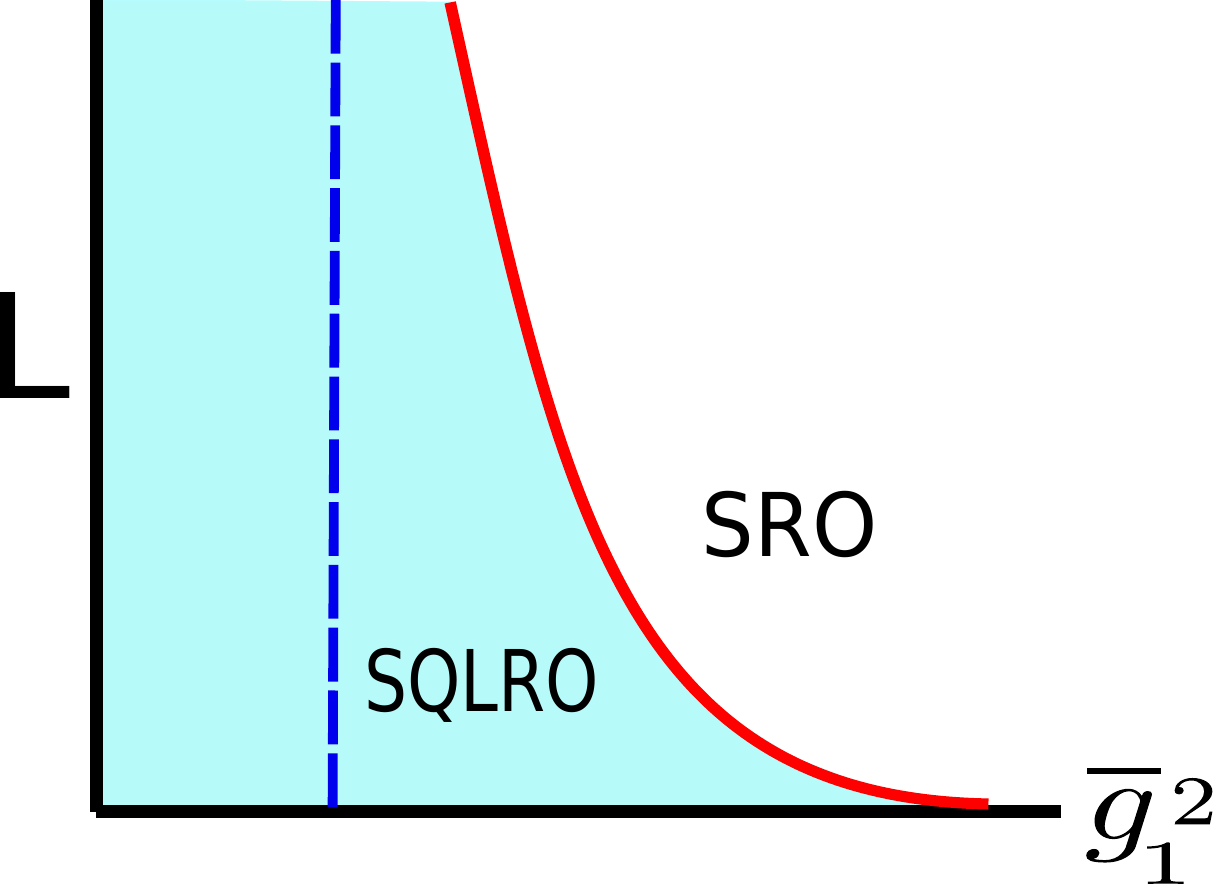}
  \caption{(color online) Schematic phase diagram in the $\overline g_1^2 - L$ plane in 2D near $T_c$. The red curved line corresponding to $L=L_c$, the instability threshold, demarcates regions with PO and  SRO. The region left to the vertical broken blue line corresponds to systems with arbitrarily large $L$ retaining PO. The region between the vertical blue line and the curved red line given by (\ref{phase-2d-intro1}) corresponds to systems having a finite $L<L_c$, a threshold value maintaining PO and is identified with the persistence length or positional correlation length $\xi$; for $L>L_c=\xi$ only SRO is possible (see text). }\label{phase-diag2-2d}
 \end{figure}
 
  The above results assumed a second order transition of the order parameter field. However, the transition of the order parameter field too  can depend very sensitively on the  order parameter - strain couplings.  While this is second order for weak order parameter-strain couplings, it can be turned first order by sufficiently strong order parameter - strain couplings.  
  In the latter case, the system can still show positional order, which in this case would be just conventional QLRO, or can destabilize, and undergoes a 
 transition to SRO for larger couplings. However, the distinctive feature of positional QLRO together with a first order transition is that across the first order transition temperature, there is a {\em finite jump} in the effective elastic modulii, unlike across second order transitions. With a first order transition, the instability in the positional order is, however, independent of the system size $L$, i.e., it sets in independent of the system size.}

 \subsection{ Results on 3D systems:-} 
 
 At 3D, for low selectivity,  there are no infinite renormalizations of $\mu$ and $\tilde\lambda$; therefore, the displacement fluctuations show true positional LRO, indistinguishable from pure 3D elastic media. However, again a sufficiently strong selectivity, exceeding a dimension-dependent threshold, can distabilize the positional order. Unlike its 2D counterpart, this threshold is independent of $L$. In this case, the phase transition of the order parameter is unaffected by the order parameter-displacement couplings, and belongs to the 3D Ising universality class. Stronger selectivity parameters can introduce a first order transition as well, in a way similar to the 2D case, with the elastic modulii displaying finite jumps across the first order transition temperature. These results are shown in the schematic phase diagrams in Fig.~\ref{phase-diag1-2d} and Fig.~\ref{phase-diag2-2d}.

 These results including the phase diagrams could be verified in numerical simulations of the associated Ising spin - lattice discrete model close to the phase transitions. These may also be tested in carefully prepared purpose-built mixed ZTE samples (of magnetic or non-magnetic origins), which undergo phase transitions within the temperature range of ZTE behavior, in future.
 
 

 \section{Landau-Ginzburg free energy}\label{free-en-1}
 
 
 { We start by formulating the coarse-grained Landau-Ginzburg (LG) theory of a spin model for Ising transitions in a deformable ZTE. For simplicity, we schematically consider a conceptual model consisting of ferromagnetic Ising spins with nearest neighbor interactions on a Hookean spring-like model of isotropic gel. We assume the Hamiltonian $\cal H$
 \begin{widetext}
 
 \begin{equation}
  {\cal H}= -\sum_{\alpha\beta}J_{\alpha\beta}\left[({\bf x}_\alpha -{\bf x}_\beta-\tilde {\bf a})^2\right] S_\alpha S_\beta + \sum_{\alpha\beta}S_\alpha J'_{\alpha \beta}\left[({\bf x}_\alpha -{\bf x}_\beta-\tilde {\bf a})^2\right] + \frac{1}{2} \sum_{\alpha\beta}K_{\alpha\beta}({\bf x}_\alpha-{\bf x}_\beta-\tilde {\bf a})^2.\label{ising-hamil}
 \end{equation}
 
 \end{widetext}
Here, $\alpha,\beta$ refer to lattice sites with position vectors ${\bf x}_\alpha,\,{\bf x}_\beta$, and not Cartesian components, and $S_\alpha=\pm 1$ is the Ising spin at lattice site $\alpha$; all the sums over $\alpha,\beta$ are restricted to nearest neighbors. The constant $\tilde a$ is the rest length of the springs. Then in the rigid lattice limit of the lattice, ${\bf y}\equiv {\bf x}_\alpha-{\bf x}_\beta-\tilde {\bf a} =0$, where $y$ is a measure of the strain. Further,  $J_{\alpha\beta}$ is the exchange integral that is assumed to depend quadratically on the strain for small strains: we assume $J_{\alpha\beta}(y) = J_{\alpha\beta}^0 + J_{\alpha\beta}^1 y^2$ for small $y$. Here, $J_{\alpha\beta}^0>0$ is the exchange integral in the limit of an undistorted lattice. We further assume $J^1_{\alpha\beta}>0$ for stability reasons. We have also included an Ising inversion symmetry  breaking term that is linear in $S_\alpha$ but quadratic in the strain, and is formally like a ``local magnetic field''; $J'_{\alpha\beta}(y)={J^\prime_{\alpha\beta}}^0 y^2$ for small $y$. Here, the coupling $J'_{\alpha\beta}$ can be of any sign. As mentioned earlier, such an Ising symmetry breaking coupling, although is not commonly considered in magnetic crystals, routinely appears in the free energies of two-component soft matter systems, where the Ising degree of freedom represents the local difference in the two components of a mixed system; see, e.g. Ref.~\cite{ising-mem}. In that spirit and in order to generalize our theory, we allow for such an anhamornic coupling in our model.  Clearly, in the rigid limit of the lattice  $J_{\alpha\beta}(y=0)=J_{\alpha\beta}^0;\,J'_{\alpha\beta}(y=0)=0$, reducing $\cal H$ in (\ref{ising-hamil}) to the standard ferromagnetic Ising Hamiltonian~\cite{chaikin}. Lastly, $  K_{\alpha\beta}({\bf x}_\alpha-{\bf x}_\beta-\tilde {\bf a})^2$, which is the last term in (4), gives the stretching energy of the Hookean springs. We now construct the LG free energy functional for an isotropic system, assuming that in the rigid limit of the medium the phase transition is of continuous nature, described by a local scalar order parameter $\phi({\bf x})$ representing the local Ising degree of freedom. As discussed above, in an actual physical realization of this model, $\phi$ could be a local magnetic (Ising-type) spin, or the local concentration difference of the two components that make up the elastic medium. The quadratic dependence of the interactions on the strains ensures  ZTE here. We have ignored contributions anhamornic (i.e., non-Hookean) in the strain, since they turn out to be irrelevant in what follows below.}
 
 
 The Landau-Ginzburg free energy functional $\cal F$ for this system is obtained by expanding in terms of the fields and their gradients, assuming small fluctuations~\cite{chaikin}. For a two-component elastic medium, this should have three distinct parts:
 \begin{equation}
  {\cal F}= {\cal F}_\phi + {\cal F}_u + {\cal F}_{u\phi}.\label{free-tot1}
 \end{equation}
 For reasons of analytical manipulations (see later), we write down $\cal F$ in general $d$-dimensions.
Here, ${\cal F}_\phi$ is the free energy functional of an isolated Ising model with all the displacements $u_i({\bf x})=0$ (e.g., on a rigid lattice):
\begin{equation}
 {\cal F}_\phi = \int d^dx \left[\frac{r}{2}\phi^2 + \frac{1}{2}({\boldsymbol\nabla}\phi)^2 + v\phi^4\right].\label{f-phi}
\end{equation}
Here, $v>0$ and $r=T-T_c^0$, where $T$ is the temperature and $T_c^0$ is the mean-field critical temperature of the Ising model.
Further, ${\cal F}_u$ is the elastic free energy of deformation of an isolated elastic medium, which due to the invariance of the system under any translation or rotation can depend on $\bf u(x)$ only through the strain tensor $u_{ij}=\frac{1}{2}(\nabla_i u_j + \nabla_j u_i+\nabla_i u_m \nabla_j u_m)$~\cite{chaikin}. Now, assuming isotropy ${\cal F}_u$ must have the form~\cite{iso}
\begin{equation}
 {\cal F}_u =\frac{1}{2}\int d^dx \left[2\mu u_{ij} u_{ij} + {\lambda} u_{ii}\right].\label{f-u}
\end{equation}
Parameters $\mu$ and $\lambda$ are the well-known Lam\'e coefficients for an elastic medium~\cite{chaikin}. For an incompressible medium, $\lambda$ diverges and  $u_{ii}\rightarrow 0$.

Lastly, ${\cal F}_{u\phi}$ is the free energy of interactions between the local order parameter and the local strain. General symmetry considerations dictate the following form for ${\cal F}_{u\phi}$: 

\begin{eqnarray}
 {\cal F}_{u\phi}&=&\int d^dx \left[g_{1} \phi^2 u_{ij} ^2+ \overline g_{1}\phi u_{ij}^2+g_{20}\phi^2  u_{ii}^2 \nonumber + \overline g_{20} \phi  u_{ii}^2 
    \right],\nonumber \\ \label{f-int}
\end{eqnarray}
to the leading order in gradients and fields. The form of ${\cal F}_{u\phi}$ is chosen in such a way to ensure that it and also the total free energy $\cal F$ are {\em even} (quadratic) in the strain.  The invariance of $\cal F$ under constant shifts of $\bf u$ ensures that  there are no couplings of the form ${\bf u}\cdot {\boldsymbol\nabla}\phi$ in ${\cal F}_{u\phi}$.   The local stress field $\sigma_{ij}$ is the thermodynamic conjugate of $u_{ij}$ and is given by~\cite{chaikin}
\begin{equation}
 \sigma_{ij}=u_{ij}\left[2\mu + g_1 \phi^2 + \overline g_1 \phi\right] + u_{mm}\delta_{ij} \left[\tilde\lambda+g_{20}\phi^2 + \overline g_{20}\phi\right],
\end{equation}
giving strain $u_{ij}$ to vanish identically in the zero stress state with $\sigma_{ij}=0$. Furthermore, $\cal F$ in (\ref{free-tot1}) is constructed in a way such that the thermal average of the strain tensor, $\langle u_{ij}\rangle =0$ identically in the absence of any externally applied stress, as it should be for ZTE materials.

Let us consider the different anharmonic terms included in ${\cal F}_{u\phi}$. The terms on the the right side of (\ref{f-int}) may be interpreted as order parameter-dependent Lam\'e coefficients. In fact, by combining with (\ref{f-u}), we may define effective or local Lam\'e coefficients 
\begin{eqnarray}
\mu(\phi)&=&\mu + g_{1}\phi^2 + \overline g_{1}\phi,\\ \lambda(\phi)&=&\lambda + 2g_{20}\phi^2 + 2\overline g_{20}\phi.
\end{eqnarray}
Thus, the effective Lam\'e coefficients depend on the local order parameter, not only through its magnitude, but also its sign, i.e., by the overall state of the local order parameter.
These anhamornic terms may be alternatively interpreted as follows. The terms $\left[g_1(u_{ij})^2 + g_{20}(u_{ii})^2\right]\phi^2$ can be considered as the local strain-dependent corrections to $T_c^0$ giving $T_c$, the local effective critical temperature:
\begin{equation}
 T_c = T_c^0 - 2  \left[g_1(u_{ij})^2 + g_{20}(u_{ii})^2\right].\label{eff-tc}
\end{equation}
Further, the terms $\left[\overline g_1(u_{ij})^2 + \overline g_{20} (u_{ii})^2\right]\phi$ together effectively act like a local aligning field term $h_\phi \phi$ in $\cal F$, where
\begin{equation}
 h_\phi = - \left[\overline g_1(u_{ij})^2 + \overline g_{20} (u_{ii})^2\right].\label{eff-h}
\end{equation}
that is the analog of an external conjugate field~\cite{conju}.

In order to generalize the scope of our study, we have included Ising ${\cal Z}_2$ symmetry breaking terms $\overline g_{1} \phi  u_{ij}^2$ and $\overline g_{20}\phi u_{ii}^2$ in ${\cal F}_{u\phi}$ to allow for the possibility that the local strain couples with the two states of the Ising degree of freedom differently or {\em selectively}: the magnitudes of the parameters $\overline g_1$ and $\overline g_{20}$ thus give measures of the {\em degree of selectivity} in the model; these are also the parameters that introduce {\em inversion asymmetry} of the Ising order parameter~\cite{cubic}. This could be potentially important, e.g., in a two-component composite ZTE elastic medium, where the local elastic modulii may depend explicitly on the relative concentration of the two components. For instance, in a two-component binary mixture embedded in an elastic medium, the elastic deformations may couple selectively to the two local concentration of the two components~\cite{select,select1}.  These should result into the local elastic modulii depending asymmetrically upon the two states of the order parameter, e.g., relative concentration of the two components in a binary system, or both the sign and amplitude of the  local order parameter. In order to generalize the scope of this study, we allow for such selectivity in composite ZTE elastic medium. These considerations motivate inclusion of the Ising ${\cal Z}_2$ symmetry breaking terms in $\cal F$, which leads to unexpectedly rich behavior~\cite{cubic}. These Ising-symmetry breaking terms in (\ref{f-int}) make it different from its counterparts used in Refs.~\cite{moura,john} (studied in somewhat different contexts though).
Couplings $v,\,g_1,\,g_2>0$ for thermodynamic stability, where as couplings $\overline g_{10},\,\overline g_{2}$ can be of either sign. 



It is convenient to write the fields in the Fourier space as functions of the wavevector $\bf q$, and decompose ${\bf u}({\bf q})$ as the vector sum of ${\bf u}^L({\bf q})$ and ${\bf u}^T({\bf q})$:
\begin{equation}
 {\bf u}({\bf q})= {\bf u}^L({\bf q})+{\bf u}^T({\bf q}),\label{ul-ut}
\end{equation}
where ${\bf u}^L({\bf q})$ and ${\bf u}^T({\bf q})$ are projections of ${\bf u}({\bf q})$ along and perpendicular to the wavevector $\bf q$. Thus
\begin{equation}
 u_i^L({\bf q}) = Q_{ij}({\bf q}) u_j({\bf q}),\;u^T_i({\bf q})= P_{ij}({\bf q}) u_j({\bf q}), 
\end{equation}
where $Q_{ij}({\bf q})= q_iq_j/q^2$ is the longitudinal projection operator, and $P_{ij}({\bf q})=\delta_{ij}-q_iq_j/q^2$ is the transverse projection operator, which project, respectively, any vector onto the space parallel and perpendicular to $\bf q$. Free energy $\cal F$ can then take the form, after dropping cubic or higher order terms in $\nabla_i u_j$ that are irrelevant in the renormalization group (RG) sense
\begin{widetext}
\begin{eqnarray}
 {\cal F} &=& \int d^dx \left[\frac{r}{2}\phi^2 + \frac{1}{2}({\boldsymbol\nabla}\phi)^2 + v\phi^4 + \mu (\nabla_i u^T_j)^2 + \frac{\tilde\lambda}{2} (\nabla_i u^L_j)^2 
 + \left(g_1\phi^2 + \overline g_1 \phi\right) (\nabla_i u^T_j)^2 + \left(g_2\phi^2 + \overline g_2 \phi\right) (\nabla_i u^L_j)^2\right],\label{free2}
\end{eqnarray}
\end{widetext}
where $u_i^L({\bf x})$ and $u_i^T({\bf x})$ are the inverse Fourier transforms of $u_i^L({\bf q})$ and $u_i^T({\bf q})$; $\tilde\lambda=\lambda+2\mu$, $g_2=g_{20}+g_1,\;\overline g_2=\overline g_{1}+ \overline g_{20}$. In (\ref{free2}), $\overline g_1$ and $\overline g_2$ are the selectivity parameters. The corresponding partition function is given by
\begin{equation}
 {\cal Z}=\int {\cal D}\phi {\cal D}u_i^T{\cal D}u_i^L \exp \left(-\beta {\cal F}\right), \label{partition}
\end{equation}
where $\beta\equiv 1/T$ with the Boltzmann constant $k_B=1$. Dimensions and estimates for these parameters are available in Appendix~\ref{param-esti}.  It can be shown that with the form of ${\cal F}$ as given in (\ref{free2}), $dT_c/dV=0$~\cite{dTc}. Our theory differs from the one studied in Ref.~\cite{berg-halp} due to the presence of the spin-lattice anhamornic terms, which are irrelevant in the presence of a finite thermal expansion, and are not considered in Ref.~\cite{berg-halp}. We note that (\ref{free2}) does not contain any term that is odd in strain, for such a term would lead to $\langle u_{ij}\rangle \neq 0$, violating the condition of ZTE.

\section{Gaussian theory}\label{gauss}

Ignoring the anharmonic terms, the free energy (\ref{free2}) reduces to
\begin{eqnarray}
 {\cal F}_g&=&\frac{1}{2}\int \frac{d^dq}{(2\pi)^d} [(r+q^2)|\phi({\bf q})|^2 + 2\mu |{\bf u}^T({\bf q})|^2 \nonumber \\&+& \tilde \lambda |{\bf u}^L({\bf q})|^2]; \label{free-gauss}
\end{eqnarray}
see Appendix~\ref{free-en} for more details.
This gives for the displacement correlation functions at the harmonic order
\begin{eqnarray}
 \langle u^L_i({\bf q}) u^L_j({\bf -q})\rangle &=& \frac{T\delta_{ij}}{\tilde\lambda q^2},\label{uL-corr},\\
 \langle u^T_i({\bf q}) u^T_j({\bf -q})\rangle &=& \frac{T\delta_{ij}}{2\mu q^2}\label{uT-corr}
\end{eqnarray}
at all temperatures $T$.
Equations~(\ref{uL-corr}) and (\ref{uT-corr}) give
\begin{eqnarray}
 \langle (u^T_i)^2\rangle &=& \frac{T}{2\pi\tilde\lambda} \ln ( L/a_0),\label{uT-sys}\\
 \langle (u^L_i)^2\rangle &=& \frac{T}{4\pi\mu}\ln ( L/a_0) \label{uL-sys}
\end{eqnarray}
in 2D.
 Similarly, the correlation functions of the elastic distortions are given by
\begin{eqnarray}
C_{uu0}^T\equiv \langle \left[{\bf u}^T({\bf x}) - {\bf u}^T({\bf x'})\right]^2\rangle &\approx&\frac{T}{4\pi\mu}\ln ( r/a_0),\label{corr-T}\\
C_{uu0}^L\equiv  \langle \left[ {\bf u}^L({\bf x}) - {\bf u}^L({\bf x'})\right]^2\rangle &\approx&\frac{T}{2\pi\tilde\lambda}\ln (r/a_0),\label{corr-L}
\end{eqnarray}
in the limit of large separation $r\equiv |{\bf x-x'}|$ in 2D,
Equations~(\ref{uT-sys}-\ref{corr-L}) correspond to positional quasi-long range order (QLRO). At $d>2$, 
\begin{eqnarray}
 \langle (u^T_i)^2\rangle &=& \frac{T}{2\pi\tilde\lambda} \Lambda,\label{uT-sys3}\\
 \langle (u^L_i)^2\rangle &=& \frac{T}{4\pi\mu}\Lambda, \label{uL-sys3}
\end{eqnarray}
which imply positional long range order (LRO); where $\Lambda$ is an upper wavevector cutoff,  $\Lambda=2\pi/a_0$. 
From the free energy (\ref{free2}) it is clear that the local order parameter $\phi({\bf x})$ introduces corrections to the elastic modulii $\mu$ and $\lambda$; alternatively, the elastic distortions at different points in the system interact via the order parameter fluctuations. At any temperature away from $T_c$, the fluctuations of $\phi({\bf x})$ are {\em short-ranged}, and hence, only short-ranged interactions between local displacement fields are generated.  In contrast, near $T_c$, fluctuations of $\phi({\bf x})$ are scale-invariant and {\em long-ranged}, leading to the local displacement fields interacting via effective {\em long-range} interactions. Interestingly, this takes the system potentially out of the jurisdiction of MWHT. Whether close to $T_c$ these interactions lead to a phase that is {\em more ordered} or {\em less ordered} cannot however be inferred without detailed calculations. Below we calculate the precise quantitative system size dependence of $ \langle (u^L_i)^2\rangle$ and $ \langle (u^T_i)^2\rangle$, and the analogs of the correlation functions defined above in (\ref{corr-T}) and (\ref{corr-L}), which should reveal the nature of order near $T_c$. Due to the large critical point fluctuations, na\"ive perturbation theory fails. To circumvent this problem, we resort to the perturbative RG framework that we discuss below.

\section{Anharmonic theory} \label{anharm-th}

Anhamornic effects are likely to substantially alter the Gaussian theory results (\ref{uT-sys}), (\ref{uL-sys}), (\ref{corr-T}), (\ref{corr-L}), (\ref{uT-sys3}) and (\ref{uL-sys3}) near the critical point. 
We discuss here in details the RG analysis, that we execute at the one-loop order, of the free energy (\ref{free2}).  By construction ${\bf u}^T\cdot {\bf u}^L=0$. This allows us to mutually completely decouple the RG calculations for $\mu,\,g_1,\,\overline g_1$ and $\tilde \lambda,\,g_2,\,\overline g_2$ at the one-loop order.

 Before we embark on the RG calculations, we note that the anhamornic coupling constant $v$ has $d_c=4$ as the critical dimension, where as $d_c=2$ for $g_1,\,\overline g_1,\,g_2$ and $\overline g_2$; see Appendix~\ref{upp-dim}. Since we will perform this RG in an expansion around the critical dimension $d_c=2$, it is useful to consider $\cal F$ to arbitrary $d$ dimensions (see above), instead of considering specific physical dimensions of $d=2,3$. 

 Since the couplings $g_1,\,\overline g_1,\,g_2$ and $\overline g_2$ are irrelevant (in the RG sense) near $d_c=4$, we can  conclude that a second order transition for the order parameter field $\phi$ at dimensions $4\geq d\geq 2$, belongs to the Ising universality class~\cite{chaikin}, and is unaffected by the elastic deformations; see also Ref.~\cite{john} for similar discussions. That  a second order transition is possible even in the presence of the inversion symmetry breaking terms in $\cal F$ [see Eq.~(16)] can be argued in a manner that is exactly analogous to the existence of a second order liquid-gas transition; see Sec.~VI for more details. Here, we assume the existence of a critical point and proceed to examine its consequence. It now remains to calculate how the order parameter fluctuations affect the lattice deformation fluctuations. As already explained, away from the critical point, effects of the $\phi$-fluctuations are {\em small}, which leave the scaling of the variances $\langle ({u}^L_i({\bf x}))^2\rangle$ and $\langle ({u}^T_i({\bf x}))^2\rangle$ and the corresponding correlation functions unchanged from their forms in the harmonic theory, whereas close to the critical point, the observed scaling may change. Hence, below we focus only on the critical region and set $T=T_c$. Being closed to the critical point, we can then set up the renormalized perturbation theory   solely for the  displacement fields ${\bf u}({\bf x})$ and expand in the powers of the coupling constants $g_1,\,\overline g_1,\,g_2,\,\overline g_2$ for a given configuration of $\phi$, and subsequently averaging over the Boltzmann distribution of $\phi$, controlled by ${\cal F}_\phi$ given above~\cite{john}.

\begin{widetext}

\begin{figure}[htb]
 \includegraphics[width=7cm]{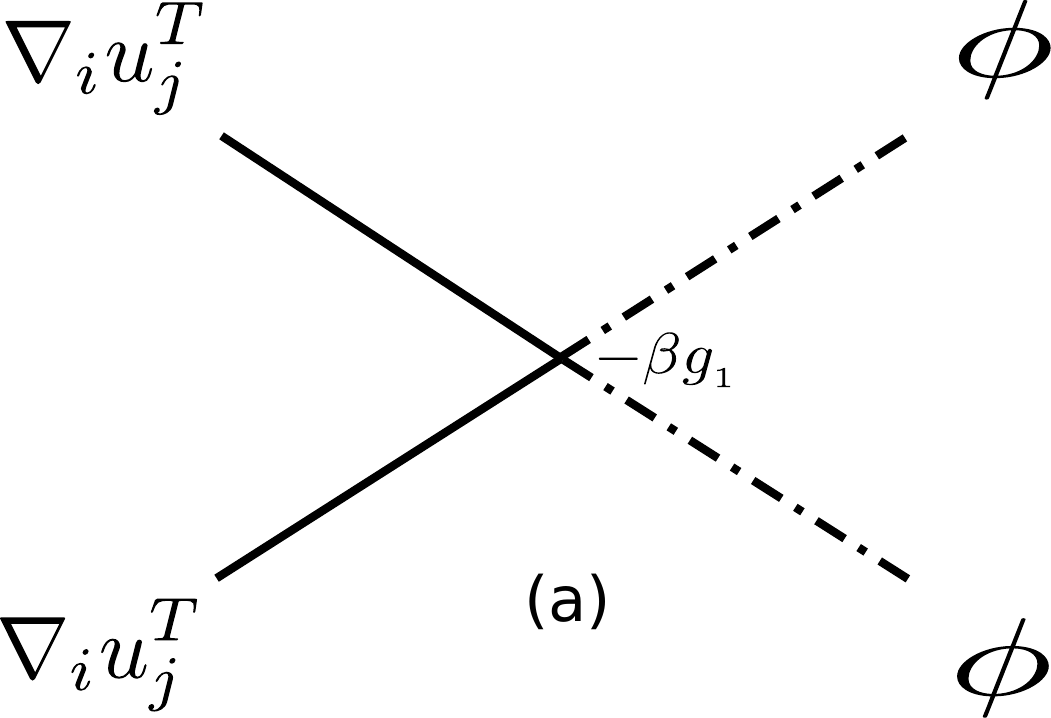}\hfill\includegraphics[width=7cm]{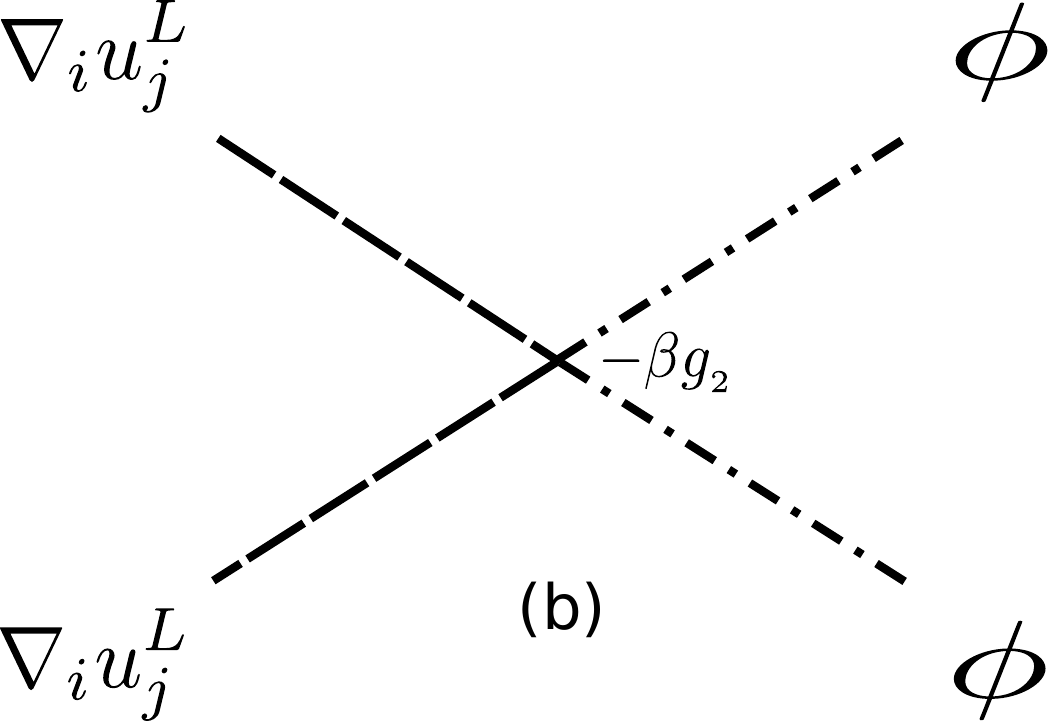}\vspace{.3cm}
  \includegraphics[width=7cm]{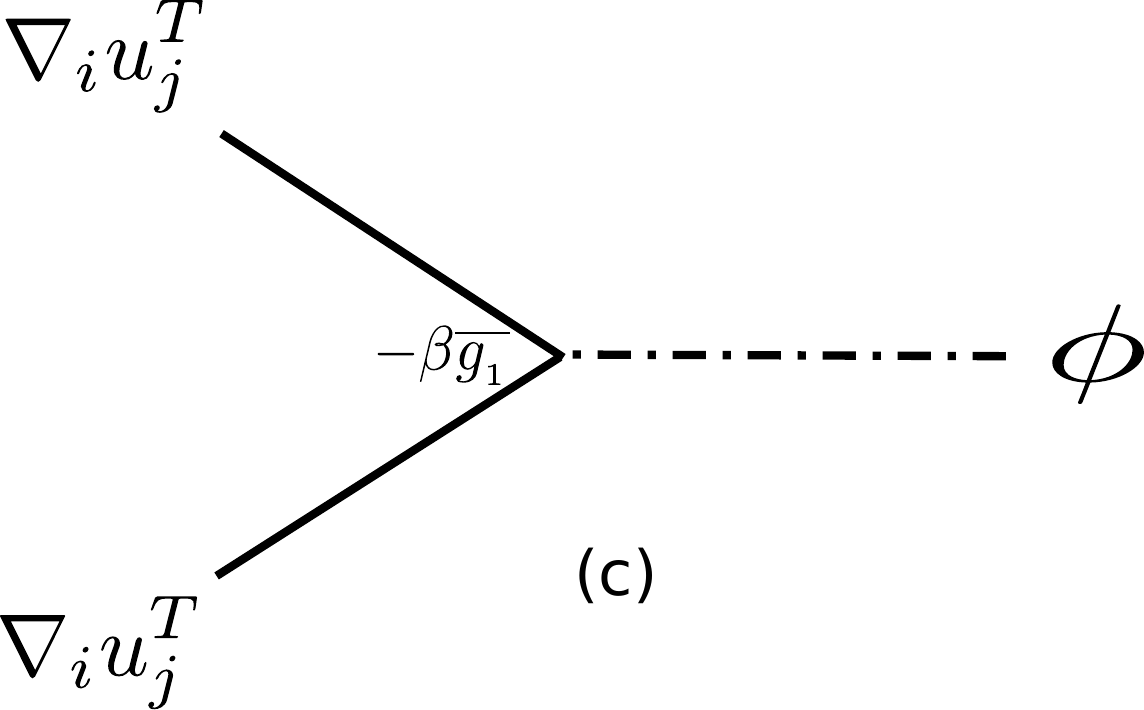}\hfill\includegraphics[width=7cm]{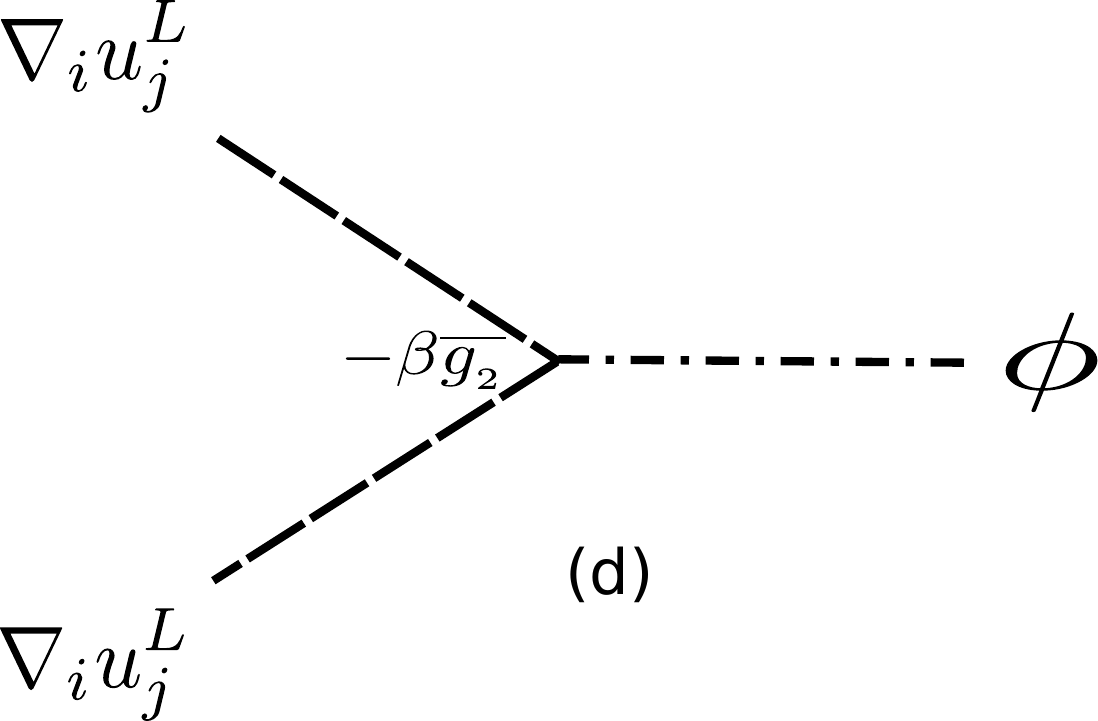}
 \caption{Vertices for the Feynman graphs (a) $g_1\phi^2 (\mathbf{u}^T({\bf x}))^2$, (b) $g_2\phi^2 (\mathbf{u}^L({\bf x}))^2$, (c) $\overline g_1\phi (\mathbf{u}^T({\bf x}))^2$, 
 (d) $\overline g_2\phi (\mathbf{u}^L({\bf x}))^2$. }\label{vertex}
\end{figure}

\end{widetext}

We employ the Wilson momentum shell procedure~\cite{sk-ma,chaikin}. This method consists of tracing over the short wavelength 
Fourier modes
of $\phi(\bf {x})$ and $u_i({\bf x})$, followed by a rescaling of lengths.
In particular, we  follow the standard approach of initially restricting wavevectors  
to lie in a bounded spherical Brillouin zone: $|{\bf q}|<\Lambda$. The fields ${\phi}(\bf{x})$ 
and $u_i({\bf x})$ are separated into high and low wave vector parts
$\phi({\bf x})=\phi^<({\bf x})+\phi^>(\bf{x})$ and $u_i({\bf x})=u_i^<({\bf x})+u_i^>({\bf x})$
where $\phi^>({\bf x})$ and $u_i^>({\bf x})$ are non-zero only in the large wave vector  (short wavelength) 
range $\Lambda
e^{-dl}<|{\bf q}|<\Lambda$, while $\phi^<({\bf x})$ and $u_i^< ({\bf x})$ have support in the small 
wave vector (long wavelength) range $|{\bf q}|<e^{-d l}\Lambda$.
We then integrate out $\phi^>({\bf x})$ and $u_i^>({\bf x})$. This integration is done perturbatively 
in the anharmonic couplings $g_1,\,\overline g_1,\,g_2$ and $\overline g_2$ in (\ref{free2}); as usual, this perturbation theory 
can be represented by Feynman graphs, with the order of perturbation theory 
reflected by the number of loops in the graphs we consider.  We confine our study to the one-loop renormalized theory here. The Feynman graphs or the vertices representing the anhamornic couplings $g_1\phi^2 (\mathbf{u}^T({\bf x}))^2,\,\overline g_1\phi (\mathbf{u}^T({\bf x}))^2,\,g_2\phi^2 (\mathbf{u}^L({\bf x}))^2$ and $\overline g_2\phi (\mathbf{u}^L({\bf x}))^2$ are illustrated in Fig.~\ref{vertex}.

\begin{figure}[htb]
 \includegraphics[width=6cm]{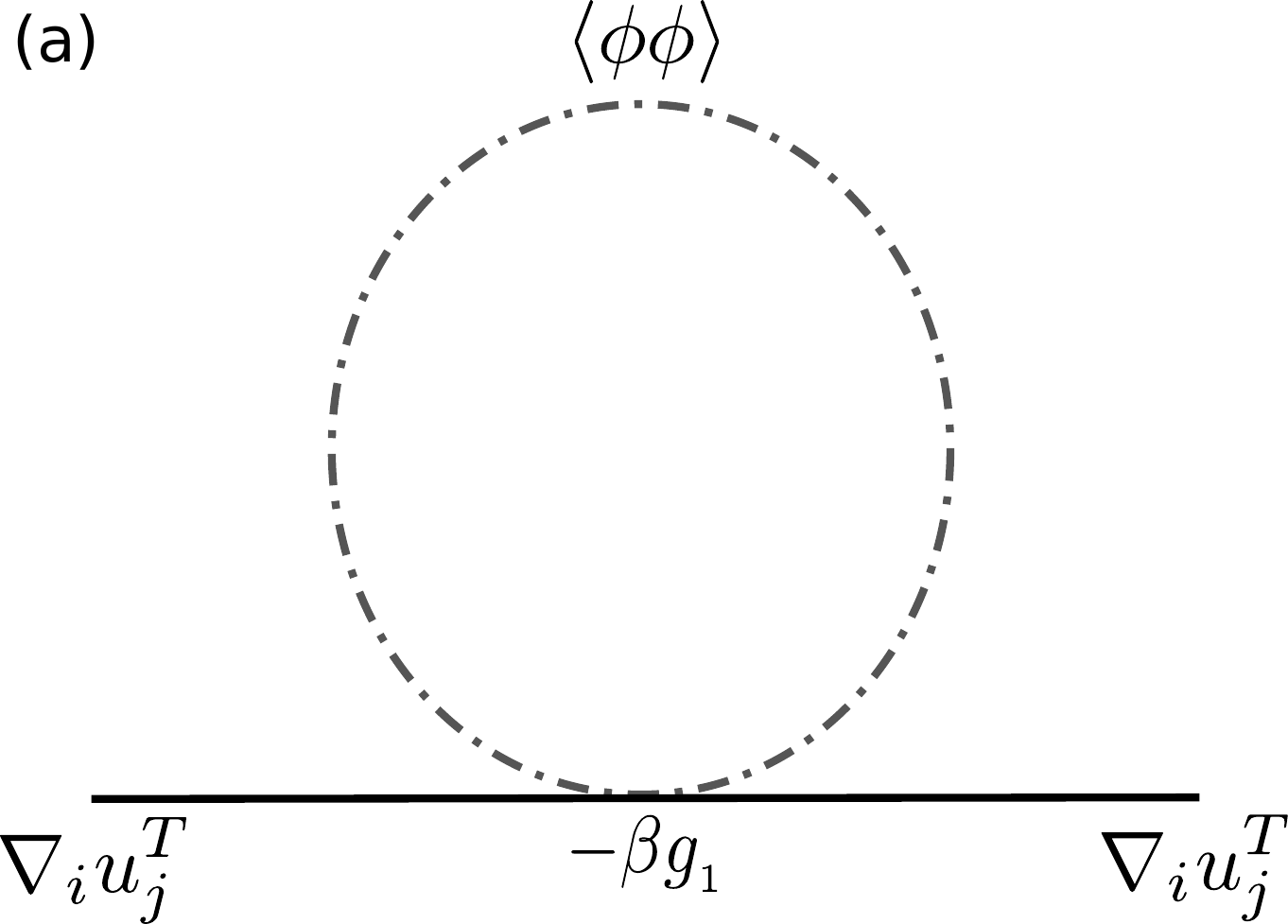}\vspace{.5cm}
  \includegraphics[width=7cm]{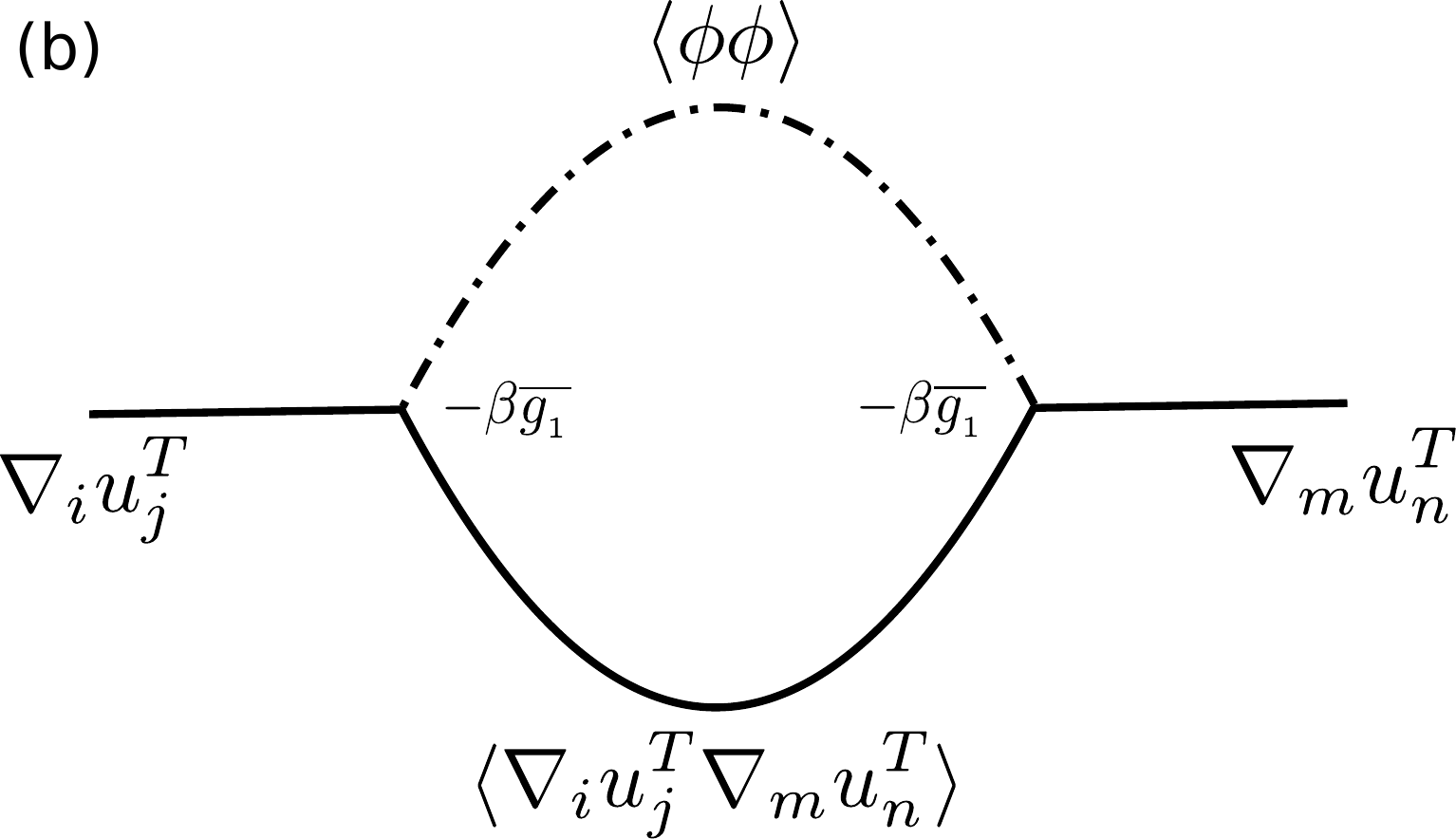}
 \caption{One-loop diagrams that contribute to the fluctuation corrections of $\mu$. Diagram (a) comes from the nonlinear coupling $g_1$, where as diagram (b) comes from $\overline g_1$ (see text).}\label{mu-diag}
\end{figure}

Next to the above perturbative step, we rescale lengths in order to restore the upper cut off back to $\Lambda$: ${\bf x}={\bf x}'b,\,b=\exp(dl)$. We then rescale the long wavelength parts of the fields according to $u_i({\bf x}) = \zeta_u u_i ({\bf x}')$ and $\phi({\bf x}) = \zeta_\phi \phi({\bf x'})$. We determine $\zeta_u$ by demanding that under the rescaling $\mu,\,\tilde\lambda$ do not scale. This gives $\zeta_u=b^{1-d/2}$. We further determine $\zeta_\phi $ by demanding that the coefficient of $\int d^dx ({\boldsymbol\nabla}\phi)^2$ remains unity under rescaling. This gives $\zeta_\phi=b^{1-d/2}$; see Appendix~\ref{upp-dim}. We restrict ourselves to a one-loop calculation. At this order, both $\mu$ and $\tilde\lambda$ receive two fluctuation corrections each,  originating from non-zero $g_1,\,\overline g_1,\,g_2$ and $\overline g_2$. The relevant Feynman diagrams for $\mu$ are given in Fig.~\ref{mu-diag}.


 There are two similar Feynman diagrams that renormalize $\tilde\lambda$; they are discussed in Appendix~\ref{feyn} (see Fig.~\ref{lam-diag}).

\begin{figure}[htb]
 \includegraphics[width=6cm]{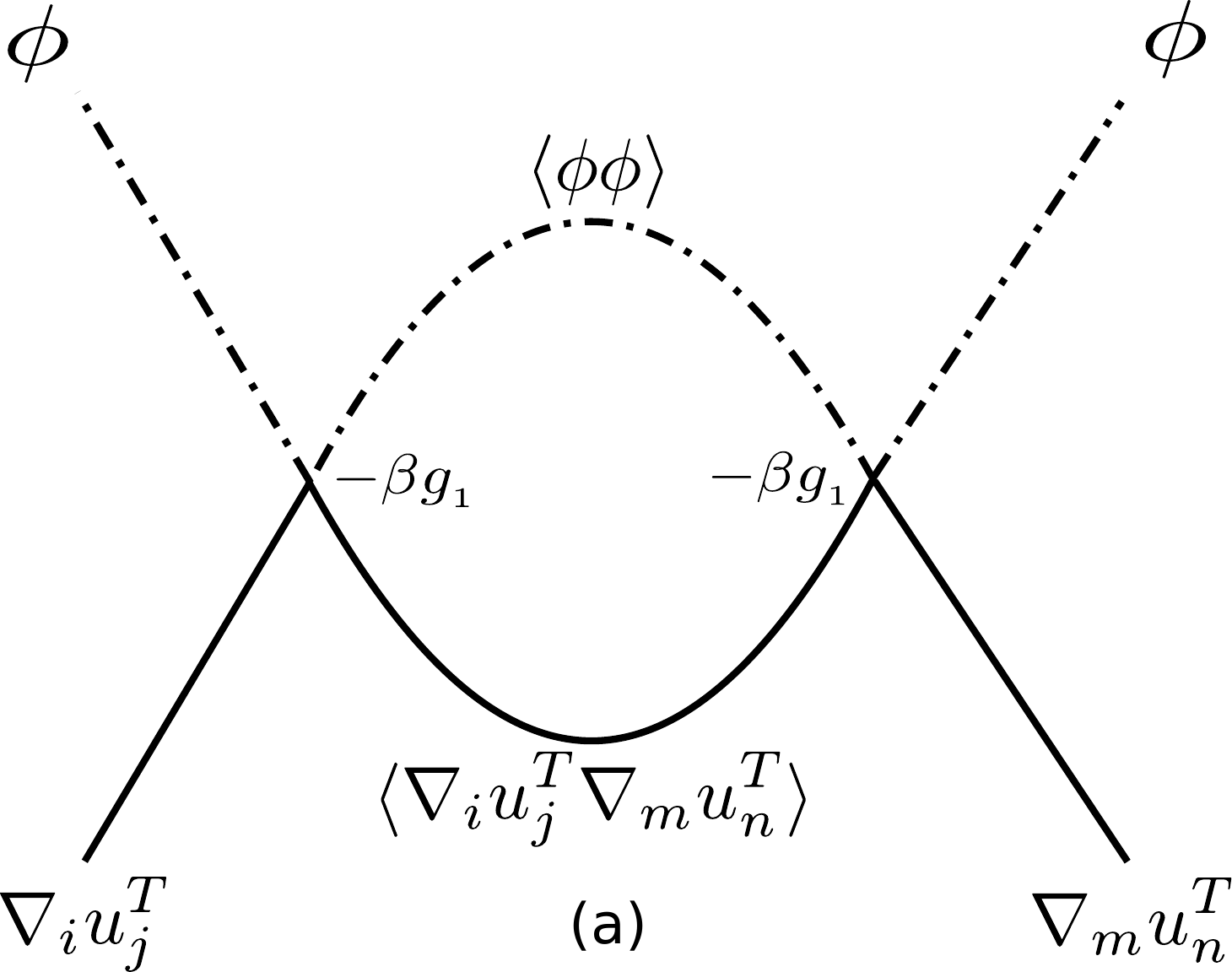}\\
 \includegraphics[width=6cm]{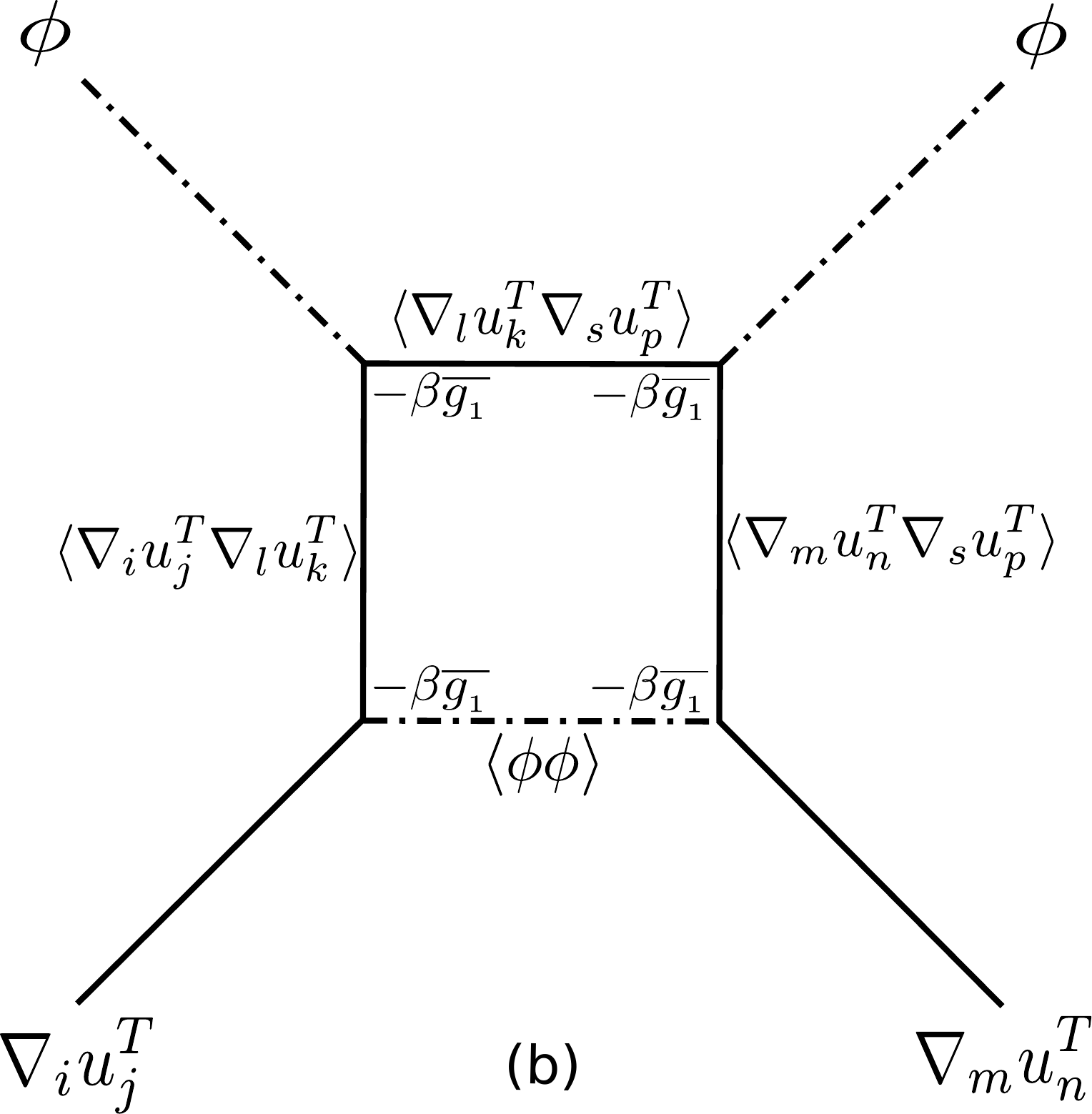}
 \caption{One-loop diagrams that contribute to the fluctuation corrections of $g_1$. Diagram (a) depends only on $g_1$, whereas diagram (b) depends only on $\overline g_1$.}\label{vertex-diag}
\end{figure}

\begin{figure}[htb]
 \includegraphics[width=7cm]{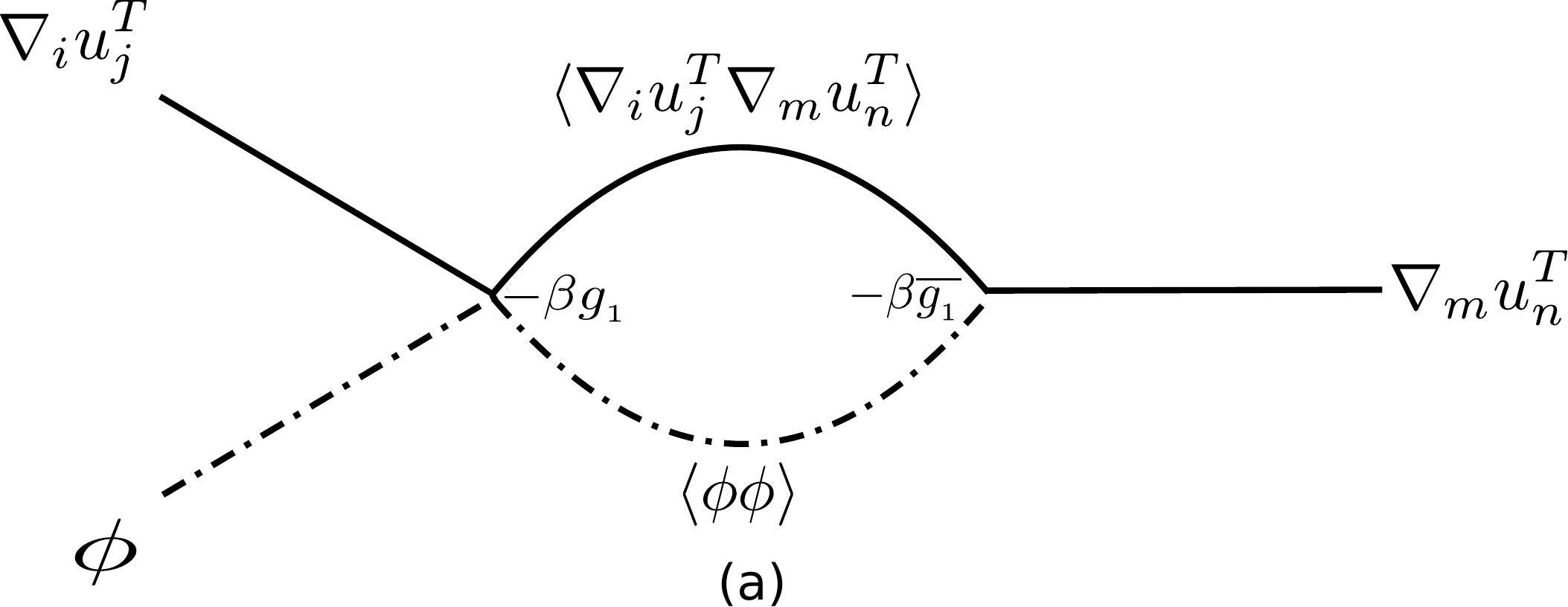}\\
 \includegraphics[width=7cm]{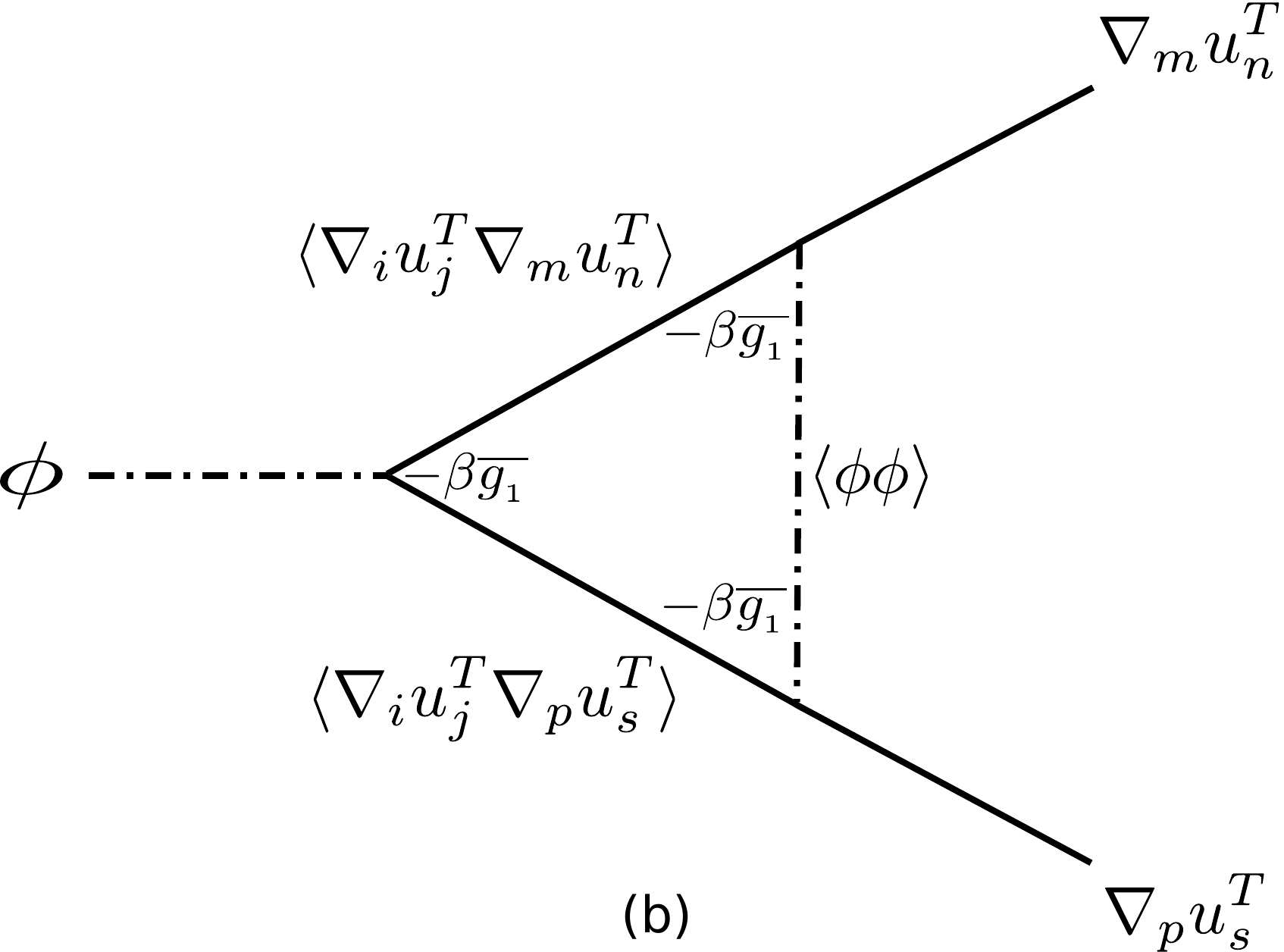}
 \caption{One-loop diagrams that contribute to the fluctuation corrections of $\overline g_1$. Diagram (a) depends both $g_1$ and $\overline g_1$, whereas diagram (b) depends only on $\overline g_1$.}\label{vertex-diag1}
\end{figure}

Likewise coupling constants $g_1,\,\overline g_1$ are each renormalized at the one-loop order by the Feynman graphs illustrated in Fig.~\ref{vertex-diag} and Fig.~\ref{vertex-diag1}, respectively.
Evaluation of these Feynman diagrams are discussed in more details in Appendix~\ref{feyn}. The corresponding Feynman graphs for $g_2,\,\overline g_2$ are given in Appendix~\ref{feyn}. As shown there, {\em all} the one-loop diagrams are proportional to $\langle (\phi^>({\bf x})^2 (x)\rangle = \int_{\Lambda/b}^\Lambda \frac{d^dq}{(2\pi)^d}\langle |\phi ({\bf q})|^2\rangle$. We now use the well-known relation
\begin{equation}
 \frac{\partial \langle \phi^2({\bf x})\rangle }{\partial T} \sim -C_v,\label{sp-heat}
\end{equation}
where $C_v$ is the specific heat at constant volume~\cite{john}. As $T\rightarrow T_c$, $C_v\sim |T - T_c|^{-\alpha}$, where $\alpha$ is the specific heat exponent. At 2D, $\alpha =0$ exactly~\cite{2d-ising}, which corresponds to a logarithmic divergence in $C_v$ as $T\rightarrow T_c$: $C_v\sim \ln (|T-T_c|/T_c)$. For $4>d>2$, $\alpha$ is non-zero, and not known exactly, but known perturbatively (or numerically, e.g., $\alpha=0.11$ in 3D)~\cite{3d-ising-expo}, giving $C_v\sim |T-T_c|^{-\alpha}$. This gives at 2D
\begin{equation}
 \langle \phi^2({\bf x})\rangle \sim \ln (|T-T_c|/T_c),
\end{equation}
where as at higher dimensions,
\begin{equation}
 \langle \phi^2 ({\bf x})\rangle\sim |T-T_c|^{-\alpha+1}.
\end{equation}
Now as $T\rightarrow T_c$, correlation length $\xi\sim |T-T_c|^{-\nu}$. The correlation length exponent $\nu=1$ is again known exactly at 2D, or at higher dimension, $\nu$ is known perturbatively or numerically: $\nu\approx 0.63$ at 3D. This gives $\langle \phi^2({\bf x})\rangle\approx T_c\ln \xi \times {\cal O}(1)$ at 2D, and $\langle \phi^2({\bf x})\rangle\sim T_c\xi^{(-\alpha +1)/\nu}$. These are discussed in Appendix~\ref{feyn}. As already mentioned above, one-loop renormalization of $\mu,\,g_1,\,\overline g_1$ are fully decoupled from those of $\tilde\lambda,\,g_2,\,\overline g_2$.  The result is the following recursion relations
\begin{eqnarray}
 \frac{d\mu}{dl}&=&T_c{g_1}-\frac{T_c\overline g_1^2}{2\mu},\label{mu-flow}\\
 \frac{d\tilde \lambda}{dl} &=&2T_cg_2 -\frac{2T_c\overline g_2^2}{\tilde \lambda},\label{lam-flow}\\
 \frac{dg_1}{dl}&=&-\epsilon g_1-\frac{2T_cg_1^2}{\mu} - \frac{T_c\overline g_1^4}{8\mu^3},\label{g1-flow}\\
 \frac{d\overline g_1}{dl}&=&-\frac{\epsilon}{2}\overline g_1+\frac{T_c\overline g_1^3}{2\mu^2}-\frac{2T_cg_1\overline g_1}{\mu}, \label{g1-overline-flow},\\
 \frac{dg_2}{dl}&=&-\epsilon g_2-\frac{4T_cg_2^2}{\tilde\lambda} - \frac{T_c\overline g_2^4}{\tilde\lambda^3},\label{g2-flow}\\
 \frac{d\overline g_2}{dl}&=&-\frac{\epsilon}{2}\overline g_2+\frac{2T_c\overline g_2^3}{\tilde\lambda^2}-\frac{4T_cg_2\overline g_2}{\tilde\lambda}; \label{g2-overline-flow}
\end{eqnarray}
here, $\epsilon\equiv d-2$. 
To proceed further, we define two dimensionless effective coupling constants
\begin{eqnarray}
 &&\alpha_{1}\equiv \frac{T_cg_{1}S_d}{(2\pi)^d\mu}\Lambda^\epsilon,\;\beta_1\equiv \frac{T_c\overline g_1^2S_d}{(2\pi)^d\mu^2}\Lambda^\epsilon,\label{eff-coup1}\\
 &&\alpha_{2}\equiv \frac{T_cg_{2}S_d}{(2\pi)^d\tilde\lambda}\Lambda^\epsilon,\;\beta_2\equiv \frac{T_c\overline g_2^2S_d}{(2\pi)^d\tilde\lambda^2}\Lambda^\epsilon. \label{eff-coup2}
\end{eqnarray}
Here, $S_d$ is the surface area of a $d$-dimensional sphere of unit radius.
As we will see below, the one-loop perturbation theory that we set up here is actually an expansion in $\alpha_1,\,\alpha_2,\,\beta_1$ and $\beta_2$ up to the linear order.

Next we obtain the RG recursion relations for $\alpha_1,\,\beta_1,\,\alpha_2,\,\beta_2$ by using the flow equations (\ref{mu-flow}-\ref{g2-overline-flow}):
\begin{eqnarray}
 \frac{d\alpha_{1}}{dl}&=&-\epsilon\alpha_1 - 3\alpha_1^2-\frac{\beta_1^2}{8} + \frac{\alpha_1\beta_1}{2},\label{flowa1}\\
 \frac{d\beta_1}{dl}&=&-\epsilon\beta_1 + 2\beta_1^2 - 6\alpha_1\beta_1,\label{flowb1}\\
 \frac{d\alpha_{2}}{dl}&=&-\epsilon\alpha_2 - 6\alpha_2^2-\beta_2^2 + 2\alpha_2\beta_2,\label{flowa2}\\
 \frac{d\beta_2}{dl}&=&-\epsilon\beta_2 + 8\beta_2^2 - 12\alpha_2\beta_2.\label{flowb2}
\end{eqnarray}
At 2D, we must set $\epsilon=0$ in (\ref{mu-flow}-\ref{g2-overline-flow}) and (\ref{flowa1}-\ref{flowb2}). 

\subsection{Two-dimensional system: thin elastic sheet}

To study the system at 2D, we use the flow equations (\ref{flowa1}-\ref{flowb2}) and set $\epsilon=0$. Clearly, the only fixed points are $\alpha_1=0,\,\beta_1=0$ and $\alpha_2=0,\,\beta_2=0$. By exploiting the decoupling between ${\bf u}^T$ and ${\bf u}^L$, we separately focus on the phases and their stability in the $(\alpha_1,\,\beta_1)$ plane, which we work out in details below, and  which suffices for incompressible systems.  An identical analysis holds in the $(\alpha_2,\,\beta_2)$ plane.  

\begin{figure}[htb]
 \includegraphics[width=6cm]{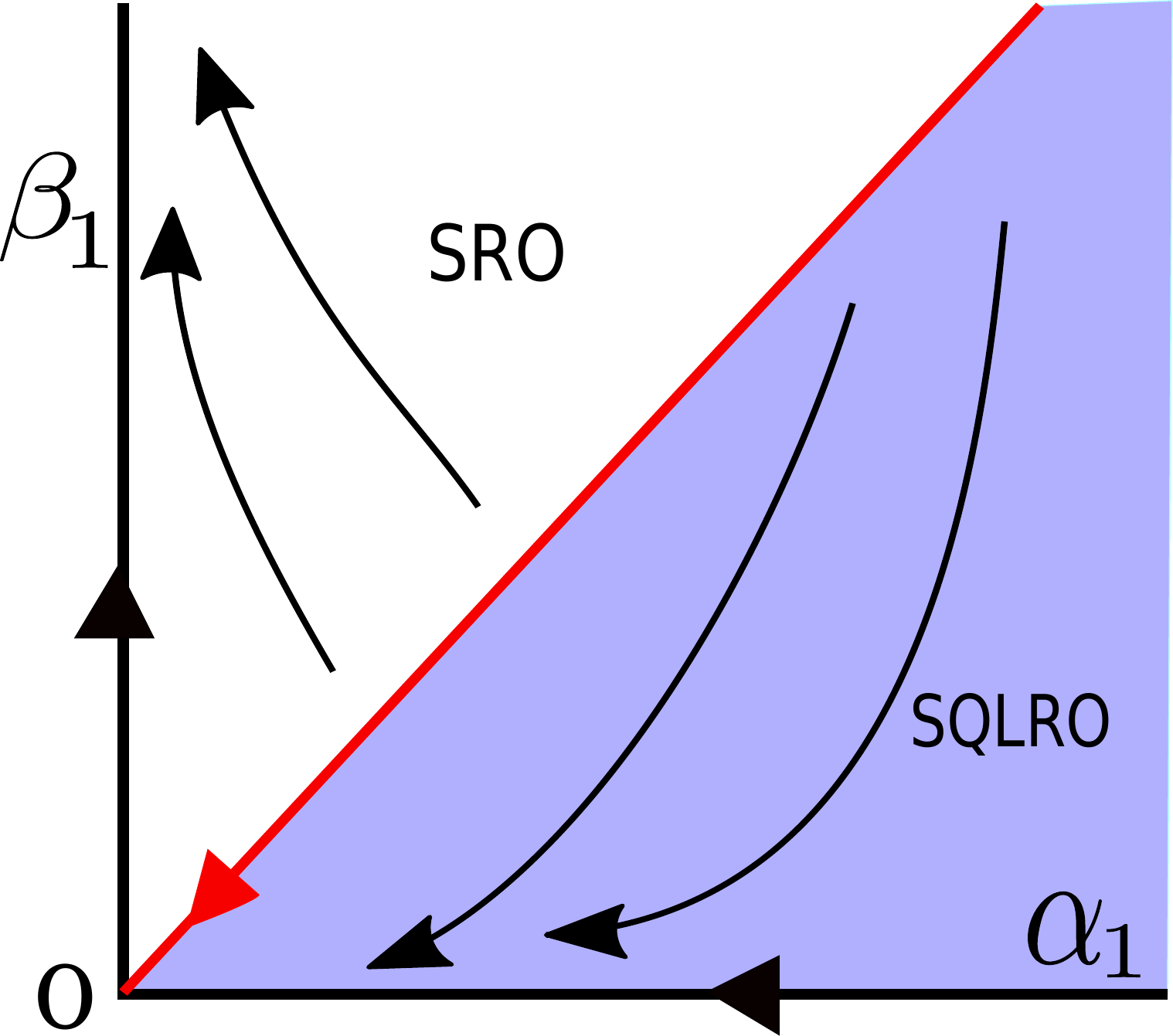}
 \caption{RG flow diagram in the $\alpha_1-\beta_1$ plane in 2D. The origin O is the only fixed point. The red line is the separatrix given by Eq.~(\ref{sep-eq}). Arrows indicate the flow directions (see text).}\label{sep-2d}
\end{figure}

Interestingly, the fixed point (0,0) is attractive (i.e., stable) along the $\alpha_1$-direction, but repulsive (i.e., unstable) along the $\beta_1$-direction. Qualitatively thus, with a sufficiently large initial $\beta_1 (l=0)\equiv \beta_{10}= T_c[\overline g_1(l=0)/\mu(l=0)]^2$ (i.e., $\beta_{10}$ is the ``bare'' or unrenormalized value of $\beta_1$) much larger than the initial $\alpha_1(l=0)\equiv\alpha_{10}=T_c g_1(l=0)/\mu(l=0)$ (again, $\alpha_{10}$ is the ``bare'' or unrenormalized value of $\alpha_1$), the system can become unstable, whereas a sufficiently small $\beta_{10}\ll \alpha_{10}$ may not be able to destroy the stable ordered phase observed for $\beta_1=0$. The question is, then where is the separatrix located in the $\alpha_1-\beta_1$ plane, that separates the stable phase from the unstable phase? Since this putative separatrix must pass through the fixed point (0,\,0), its general equation should be of the form $\beta_1=\alpha_1 f(\alpha_1)$, with $f(\alpha_1)=\Gamma_1 + $ higher order terms in $\alpha_1$. Consistent with our lowest order perturbation theory, we set $\beta_1=\Gamma_1\alpha_1$ as the separatrix that passes through the origin and set out to calculate $\Gamma_1$. Flow equations (\ref{flowa1}) and (\ref{flowb1}) may be written as
\begin{eqnarray}
 \frac{d\ln\alpha_1}{dl}&=&\alpha_1 (-3 -\frac{\Gamma_1^2}{8} +\frac{\Gamma_1}{2}),\\
 \frac{d\ln\beta_1}{dl}&=&\alpha_1 (2\Gamma_1 - 6).
 \end{eqnarray}
Thus,
\begin{eqnarray}
 \frac{d}{dl}\ln \Gamma_1 &=& \frac{d\beta_1}{dl} - \frac{d\alpha_1}{dl}\nonumber \\
 &=&\alpha_1 (\frac{3\Gamma_1}{2} - 3 + \frac{\Gamma_1^2}{8})=0,
\end{eqnarray}
giving the threshold $\Gamma_{1c}$ as
\begin{equation}
 \Gamma_{1c}=\frac{1}{2}\left[-12+\sqrt{240}\right].\label{gamma-val}
\end{equation} 
Thus, if 
\begin{equation}
\beta_1=\Gamma_{1c}\alpha_1=\frac{1}{2}\left[-12+\sqrt{240}\right]\alpha_1, \label{sep-eq}
\end{equation}
equivalently,
\begin{equation}
 \frac{\overline{g_1}^2}{\mu g_1}=\Gamma_{1c}\approx 1.5,\label{border}
\end{equation}
initially, this will continue to hold under renormalization. Points below this locus necessarily flow to the origin, where as points above this locus do not. They flow {\em away} from the origin, until they leave the regime of validity of our perturbation theory. See Fig.~\ref{sep-2d} for a  schematic flow diagram.

The plot of (\ref{border}) in the $\overline g_1-\mu$ for a fixed $g_1$ gives the phase diagram (\ref{phase-diag1-2d}). Similarly, the plot of (\ref{border}) in the $\overline g_1-g_1$ plane gives the phase diagram (\ref{phase-diag3-2d}).

Let us first focus on the scaling properties when $\alpha_{10},\,\beta_{10}$, the initial values of the $\alpha_1,\,\beta_1$, lie {\em below} the separatrix. While all points below the line (\ref{sep-eq}) eventually flow to the origin, we still need  to find out the manner in which these points may approach the origin. That is to say, an arbitrary point $(\alpha_1(l),\,\beta_1(l))$ below (\ref{sep-eq}) may, during its approach to the origin, either {\em go away} from (\ref{sep-eq}), or {\em move towards} it, depending upon whether the locus (\ref{sep-eq}) is {\em attractive} or {\em repulsive}. To know that let us assume $\beta_{10}=\Gamma (l=0)\alpha_{10}$, with $\Gamma_1(l=0)=\Gamma_{1c}-\delta_1(l=0)$, where small $\delta_1(l=0)>0$ corresponding to an initial point below the separatrix (\ref{sep-eq}). Therefore,
\begin{eqnarray}
 &&\frac{d}{dl}\ln \Gamma_1(l)=\frac{1}{\Gamma_1}\frac{d\Gamma_1}{dl}=-\Gamma_{1c}\frac{d\delta_1}{dl}\nonumber \\ &=& -\alpha_1 (\frac{\Gamma_{1c}^2}{8} -\frac{5\delta_1}{4} +\Gamma_{1c} - 3) \nonumber \\&=& -\alpha_1\frac{\delta_1}{4} \left(\Gamma_{1c} + 6\right) <0,\label{delta-eq}
\end{eqnarray}
to the leading order in small $\delta_1$. Equation~(\ref{delta-eq}) implies that $\delta_1(l)$ rises, or equivalently, $\Gamma_1(l)$ decreases as the renormalization group ``time'' $l$ increases, corresponding to a {\em repulsive} separatrix. This further means
\begin{equation}
 \frac{d}{dl}\ln \Gamma_1 = \frac{1}{\Gamma_{1c}}\frac{d}{dl}\left(\frac{\beta_1}{\alpha_1}\right) <0. 
\end{equation}
Therefore, a point $(\alpha_1,\,\beta_1)$ that is slightly away from the separatrix (\ref{sep-2d}) and lies on the stable side of it, not only flow towards the origin, it does so by {\em moving away} from the separatrix as the RG time $l$ increases. 
We thus conclude that in the long wavelength limit, $\beta_1(l)\ll \alpha_1(l),\,l\rightarrow\infty$. In this limit, then by using (\ref{flowa1}), we find
\begin{equation}
 \frac{d\alpha_1}{dl}= - 3\alpha_1^2,
\end{equation}
giving
\begin{equation}
 \frac{-1}{\alpha_1}=-3l -\frac{1}{\alpha_{10}}\approx -3l,\label{alpha-sol1}
\end{equation}
for large $l$, where $\alpha_{10}$, a constant of integration, is the ``initial value'' of $\alpha_1(l)$. Therefore,
\begin{equation}
 \alpha_1(l)=\frac{1}{3l}=\frac{T_cg_1(l)}{\mu(l)}\label{alpha-sol}
\end{equation}
for large $l$.
Likewise, we can find out $\beta_1(l)$ from (\ref{flowb1}). Using $\beta_1(l)\ll\alpha_1(l)$ in the long wavelength limit, (\ref{flowb1}) gives
\begin{equation}
 \frac{d\beta_1}{dl}=-6\alpha_1\beta_1=-\frac{2}{l}\beta_1.
\end{equation}
This gives
\begin{equation}
 \beta_1(l)\sim \frac{1}{l^2} \ll \alpha_1(l)\sim \frac{1}{l},
\end{equation}
which provides {\em aposteriori} justification of $\beta_1(l)\ll \alpha_1(l)$ that we have claimed above. We now calculate the renormalized shear modulus in the long wavelength limit. Using (\ref{mu-flow}) together with (\ref{alpha-sol}) above and using $\beta_1(l)\ll \alpha_1(l)$,
\begin{equation}
 \frac{d\mu}{dl}=\frac{\mu}{3l},
\end{equation}
in the limit of large $l$, which upon integration gives for the scale-dependent, renormalized, scale-dependent shear modulus $\mu (q)$,
\begin{equation}
 \mu(q)\approx \mu_R\left[\ln (\Lambda/q)\right]^{1/3},\label{lambda-q}
\end{equation}
for small enough $q$, where we have used $\ell = \ln b= \ln (\Lambda/q)$. Thus, $\mu(q)$ clearly diverges in the long wavelength limit $q\rightarrow 0$.  Here, $\mu_R$, a constant of integration, is the amplitude of the scale-dependent, renormalized shear modulus. Equation~(\ref{lambda-q}) is the analog of anomalous elasticity in 3D equilibrium smectics~\cite{smectics}. Thus,
\begin{equation}
 \langle |{\bf u}^T({\bf q})|^2\rangle\approx \frac{T_c}{2\mu_R |\ln (\Lambda/q)|^{1/3} q^2},\label{ren-uT}
\end{equation}
for sufficiently small $q$. Equation~(\ref{lambda-q}) shows anomalous elasticity that arises due to the coupling with the 
critical order parameter, and illustrates the new universality class.
What is the range of wavevectors over which (\ref{ren-uT}) is valid? While the Landau-Ginzburg free energy (\ref{free2}) is valid for wavevector $ q\leq \Lambda$, the upper wavevector limit of the validity of the corresponding renormalized free energy (and hence the correlation function (\ref{ren-uT})) { must be} {\em smaller} than $\Lambda$, for high wavevectors close to $\Lambda$, effects of renormalization would be small, and the harmonic theory should suffice there. Enough RG ``time'' $l$ must be spent in order for the fluctuation effects to become dominant, so as to acquire substantial renormalization of the harmonic theory scaling. Equivalently, one might ask how big the system must be for it to display (\ref{ren-uT})? As estimate of that crossover scale $\xi_{NL}$ may be obtained 
from (\ref{alpha-sol1}) by noting that the ``crossover RG time'' $l_{NL}$, at which the anharmonic effects become substantial, is given by the condition $3l_{NL}\sim 1/\alpha_{10}$. This gives
\begin{equation}
 \Lambda\xi_{NL}\approx\exp[\mu/(3g_1 T_c)]\times {\cal O}(1).\label{nonlin-length}
\end{equation}
Thus,  $\xi_{NL}$ that sets the scale at which anhamornic effects become important, depends sensitively on the model parameters and also $T_c$. For instance, if we consider two systems having same values for the  model parameters, but their respective $T_c$'s differing by a factor of 2, $\xi_{NL}$ will differ by a factor of $e^2\approx 7.4$. In fact, if the system size $L<\xi_{NL}$, the system does not get enough ``renormalization group'' time $l$ to have substantial renormalization of the model parameters; as a result conventional QLRO ensues. On the other hand for $L>\xi_{NL}$, the system gets enough RG time for substantial renormalization of the model parameters, and hence SQLRO follows.

We now calculate the variance $\langle (\mathbf{u}^T({\bf x}))^2 \rangle$ near $T=T_c$ that involves inverse Fourier transform of $\langle |{\bf u}^T({\bf q})|^2\rangle$. 
Inverse Fourier transform of (\ref{ren-uT}) gives
\begin{eqnarray}
 &&\langle (\mathbf{u}^T({\bf x}))^2 \rangle = \int_{2\pi/L}^{\tilde\Lambda} \frac{d^2q}{(2\pi)^2} \langle |{\bf u}^T({\bf q})|^2\rangle \nonumber \\&\approx& \frac{T_c}{2\mu_R} \int_{2\pi/L}^{\tilde\Lambda} \frac{d^2q}{(2\pi)^2} \frac{1}{[q^2\{\ln (\Lambda/q)\}^{1/3}]}\nonumber \\&\approx& \frac{3T_c}{8\pi\mu_R}[\ln ( L/a_0)]^{2/3},\label{vari-u}
\end{eqnarray}
in the limit of large $L$. This though rises with $L$ and eventually diverges in the thermodynamic limit,
it does so significantly more slowly than the QLRO, indicating an order {\em stronger} than QLRO. Henceforth, we call it  SQLRO. In (\ref{vari-u}) above, $\tilde\Lambda$ is an upper momentum cut off below which (i.e., $q<\tilde\Lambda$) Eq.~(\ref{lambda-q}) holds; see also Appendix~\ref{vari-2d-re}.

We next calculate the renormalized equal-time correlation function $C_{TT}(|{\bf x-x'}|)$ of ${\bf u}^T({\bf x})$ near $T=T_c$. We get

\begin{widetext}

\begin{eqnarray}
 C_{TT}(|{\bf x-x'}|)&\equiv& \langle \left[ u_i^T({\bf x})-u_i^T({\bf x})\right]^2\rangle \approx \int_0^{\tilde\Lambda} \frac{d^2k}{(2\pi)^2}\frac{T_c}{2\mu_R k^2 (\ln (\Lambda/k))^{2/3}}\left[1-\exp(i{\bf k}\cdot ({\bf x-x'}))\right]\nonumber \\&\approx&\frac{T_c}{4\pi\mu_R}[\ln |{\bf x-x'}|\tilde\Lambda|]^{2/3}, \label{corr-uTT}
\end{eqnarray}
in the limit of large $|{\bf x-x'}|$.
\end{widetext}
See details in Appendix~\ref{corr-2d}. Equations~(\ref{vari-u}) and (\ref{corr-uTT}) are the essence of SQLRO, as already mentioned above; the alert reader will find them in Introduction above. Equations~(\ref{vari-u}) and (\ref{corr-uTT}) define the new universality class close to second order phase transitions of incompressible systems.

This is to be contrasted with the corresponding result for the displacement correlator $C^0_{TT}(|{\bf x-x'}|)$ in the harmonic theory, where the same correlation scales as $\ln( |{\bf x-x'}|\Lambda)$. Thus for large $| {\bf x-x'}|$,
\begin{equation}
 C_{TT}(|{\bf x-x'}|)\ll C^0_{TT}(|{\bf x-x'}|), 
\end{equation}
showing SQLRO as a distinctly stronger order than the well-known QLRO.

 { Let us consider the physics on the other side of, {\em i.e., above} the separatrix (\ref{sep-eq}). The RG trajectories starting with initial values lying above the separatrix, but not far from the origin (so as to stay within the validity regime of our one-loop RG, at least for small $\alpha_{10},\,\beta_{10}$) flow away, since $\alpha_1$ still flows to zero, whereas $\beta_1$ flows away, giving $\beta_1(l)\gg\alpha_1(l)$, which we show {\em aposteriori}.  This follows directly from (\ref{delta-eq}) with the replacement of $\delta_1$ by $-\delta_1$ corresponding to an ``initial condition'' above the separatrix. It then follows
\begin{equation}
 \frac{d}{dl}\ln\Gamma_1(l)=\frac{1}{\Gamma_{1c}}\frac{d}{dl}\left(\frac{\beta_1(l)}{\alpha_1(l)}\right)>0,
\end{equation}
meaning that starting from an initial condition that lies slightly above the separatrix, the RG trajectories will move towards to the origin along the $\alpha_1$-direction but move away from it along the $\beta_1$-direction; see the flow lines in the flow diagram (\ref{sep-2d}). It is useful to find out {\em how} $\beta_1(l)$ grows and $\alpha_1(l)$ decays in the limit of large $l$. (It should be kept in mind though that on the unstable side of the separatrix, the accuracy of our one-loop RG gets progressively poorer with larger RG time $l$, since as $l$ gets larger, $\beta_1(l)$ gets larger as well, eventually making the perturbative approximation untenable.} Retaining the most dominant terms for large $l$, we find from (\ref{flowb1})
\begin{equation}
 \frac{d\beta_1(l)}{dl}=2\beta_1^2,
\end{equation}
giving
\begin{equation}
 \beta_1(l)=\frac{\beta_{10}}{1-2l\beta_{10}}.\label{beta-l2d}
\end{equation}
where $\beta_{10}=\beta_1(l=0)$ appears as a constant of integration. Thus, as $\beta_1(l)$ diverges as $l\rightarrow \ell_\beta\equiv 1/(2\beta_{10})$ from below, i.e., as the system size exceeds an initial condition-dependent finite size $L_\beta=a_0\exp[1/(2\beta_{10})]$, a finite, model parameter-dependent nonuniversal  size, from below. Similarly, retaining the most dominant terms we find from (\ref{flowa1}) for large $l$
\begin{equation}
 \frac{d\alpha_1}{dl}=-\frac{\beta_1^2}{8} = \frac{-1}{8}\frac{\beta_0^2}{(1-2l\beta_0)^2}.
\end{equation}
Thus $\alpha_1(l)$ decreases monotonically as $l$ increases. Solving, we find
\begin{equation}
 \alpha_1(l) = c_\alpha+\frac{1}{32} \frac{1}{ l-1/(2\beta_{10})}.
\end{equation}
Here, $c_\alpha$ is a constant of integration that can be fixed by demanding that at $l = 0$, $\alpha_1(l=0)=\alpha_{10}\equiv T_c g_1(l=0)/\mu(l=0)$, the ``initial'' or unrenormalized value of $\alpha_1$. This gives $c_\alpha = \beta_{10}/16 + \alpha_{10}$, yielding
\begin{equation}
 \alpha_1(l)=\frac{-1}{32}\frac{1}{1/(2\beta_{10}) -l } +\frac{\beta_{10}}{16}+\alpha_{10}.
\end{equation}
Thus, $\alpha_1(l)$ continuously decreases as $l$ increases, and eventually {\em vanishes} as the system size crosses a finite threshold $\tilde\ell$. This scale $\tilde\ell$ may be found by setting $\alpha(\tilde\ell)=0$ giving
\begin{equation}
 \tilde\ell=\frac{8\alpha_{10}}{\beta_{10}(\beta_{10}+16\alpha_{10})}<\ell_\beta,
\end{equation}
which unsurprisingly is finite and {\em nonuniversal}.

We  now study the fate of $\mu(l)$ on this unstable side. In the limit $l\gg 1$, we get from (\ref{mu-flow})
\begin{equation}
 \frac{d\mu}{dl}=-\mu\frac{\beta_1}{2}=-\frac{\mu\beta_{10}}{2(1/(2\beta_{10}) -l)},
\end{equation}
 giving
\begin{equation}
 \mu(l)=\tilde \mu_0 |\frac{1}{2\beta_{10}}-l|^{1/4},
\end{equation}
as $l\rightarrow \ell_\beta$ from below,
where $\tilde \mu_0$ is a constant of integration: $\tilde \mu_0= \mu (l =0) \left(\frac{1}{2\beta_{10}}\right)^{1/4}$.
Thus, $\mu(l)$ depends sensitively on $\beta_{10}$, and as $l\rightarrow \ell_\beta$ from below, $\mu(l)$ vanishes. Of course, we cannot follow $\mu(l)$ all the way to $l \rightarrow 1/(2\beta_{10})$ from below, as precisely there $\beta_1(l)$ diverges. 

It is clear from the discussions above that for the stability of the system, RG ``time'' $l$ should satisfy
\begin{equation}
 l<\ell_c = l_{\beta}\equiv \frac{1}{2\beta_{10}}=\frac{1}{2}\frac{\mu^2(l=0)}{T_c\overline g_{10}^2(l=0)},
\end{equation}
whereas if $l>\ell_c$, the system becomes unstable.
Equivalently, the length scale for stability $L_c$, which is a measure of the positional correlation length, is given by
\begin{equation}
 L_c = L_{\beta} = a_0\exp \left[\frac{1}{2T_c}\frac{\mu^2(l=0)}{\overline g_{10}^2(l=0)}\right],\label{pos-corr-len1}
\end{equation}
such that for system size $L< (>)L_c$, the system is stable and the system remains positionally ordered (unstable without any positional order).
For a fixed $L_c$, this relation may be written in an alternative form:
\begin{equation}
 \mu(l=0)=|\overline g_1(l=0)|\sqrt{ T_c \ln (L_c/a_0)} \times {\cal O}(1).
\end{equation}
Thus, $L_c$ depends very sensitively on $\overline g_{10}$. For small $g_{10}$, $L_c$ becomes very large. Readers who are interested in a perturbation theory argument for divergence of $L_c$ for small $\overline g_{1}(l=0)$ can find one in Appendix~\ref{pos-corr}. The interpretation of the separatrix as the threshold for breakdown of positional order is in fact supported from the fact that on the unstable side of the separatrix, an arbitrarily large system will not be able to sustain any positional order. 

The instability discussed above occurs if
\begin{equation}
 \Gamma_1(l=0)> \Gamma_{1c}=\frac{1}{2}\left[-12 + \sqrt{240}\right],
\end{equation}
or equivalently, the bare or unrenormalized model parameters satisfy
\begin{equation}
 \frac{\overline g_1^2(l=0)}{\mu(l=0)g_1(l=0)}>\Gamma_{1c}.
\end{equation}
In the unstable region of the parameter space, the elastic sheet can remain stable if it is sufficiently small, for in that case, $\mu (l=\ell_c)>0$, stabilizing the positional order.  So long as the system size $L<L_c=a_0\exp(\ell_c)$, the positional order is stable since $\mu(l\lesssim \ell_c)$ remains positive. On the unstable side of the separatrix, starting from small ``initial'' values of $\alpha_1$ and $\beta_1$, anharmonic effects will not be visible until $\beta_1(l=\ell^*)\sim {\cal O}(1)$. For $l<\ell^*$, the system remains stable and shows scaling as given by the Gaussian theory.
Using (\ref{beta-l2d}), we get an estimate about $\ell^*$:
\begin{equation}
 \ell^*=\frac{1-\beta_{10}}{\beta_{10}}\times {\cal O}(1).
\end{equation} For $l\ll \ell^*$, there is no substantial renormalization of $\mu$, and hence, the variance $\langle (\mathbf{u}^T({\bf x}))^2 \rangle$ should scale as $\ln (L/a_0)$, implying conventional QLRO. 

The flow along the separatrix can be calculated. We find on the separatrix
\begin{eqnarray}
 \frac{d\alpha_1}{dl}&=&\alpha_1^2\left(-3 - \frac{\Gamma_{1c}^2}{8}+\frac{\Gamma_{1c}}{2}\right) <0,\\
 \frac{d\beta_1}{dl}&=&2\alpha_1\beta_1 \left(\Gamma_{1c} -3\right) <0,
\end{eqnarray}
implying a flow {\em towards} the origin.

\begin{figure}[htb]
 \includegraphics[width=8cm]{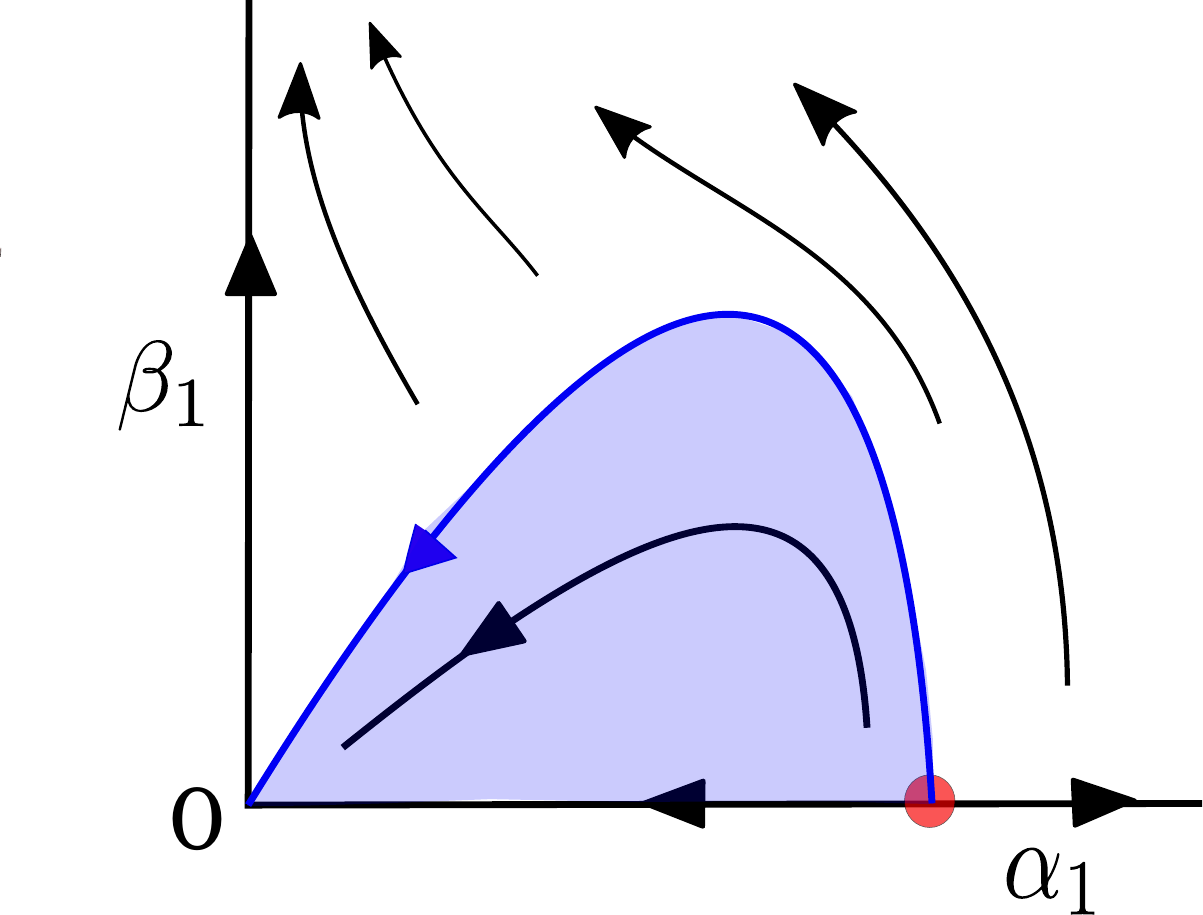}
 \caption{Conjectured ``Occam's razor'' global RG flows in the $\alpha_1-\beta_1$-plane at 2D. Arrows indicate the flow directions. The (light blue) shaded region is the stable region with positional SQLRO. The red circle on the $\alpha_1$-axis is the putative unstable critical point not accessible in our perturbative RG (see text). }\label{occ-1}
\end{figure}

 Let us further analyze the behavior of the system qualitatively on the unstable side of the separatrix (\ref{sep-eq}), where the RG trajectories flow out, eventually going out of the regime of validity of our
perturbation theory. The system will then behave differently than what it does below the separatrix, i.e., in the stable regime. What might this different behavior be? Since the flows in
this regime lead out of the region of validity of our perturbation theory, we cannot follow these flow lines, but can only speculate about it. For this, we are guided by the expectation that for large enough $\alpha_1$ (which can be accessed by, e.g., high enough $T=T_c$), there should be a phase where positional order breaks down giving way to a ``new phase'' that only has short range order. Accordingly, there must be an unstable critical point on the $\alpha_1$-axis, controlling this transition to this putative phase. If we now consider the full RG flows for an elastic sheet near a second order phase transition in the two dimensional parameter space ($\alpha_1,\,\beta_1$)
and connect this putative flow on the $\alpha_1$ axis with our
flows near the origin in the simplest possible way (i.e., one that does not involve introducing any other new fixed points), we are then led to Fig.~\ref{occ-1}. 
This is essentially an ``Occam’s razor''-style argument: Fig.~\ref{occ-1} has the simplest flow topology that naturally reduces to the known flow trajectories for small $\alpha_1,\,\beta_1$ (as shown in Fig.~\ref{sep-2d}).  It at the same time gives the putative global flow allowing for a continuous transition to a phase with breakdown of positional order with breakdown of elasticity that we are tempted to identify with the liquid-like phase with SRO, where obviously the mean square displacement is unbounded.  It for instance suggests that starting very close to this ``conjectured unstable fixed point'' on the $\alpha_1$-axis, but outside of the separatrix (i.e., in the unstable region), we expect the resulting RG trajectories to follow the separatrix for a long ``RG time'', eventually moving away from it to flow away towards unbounded $\beta_1$ for large $l$. In 2D, this melting transition could be a first or second order.  What if the putative transition to this phase is first order in nature? In that case, the conjectured unstable fixed point or the critical point with, e.g., a diverging correlation length and a continuously vanishing order parameter cannot exist.   Still there should be a phase boundary schematically similar to that drawn in Fig.~\ref{occ-1} that should presumably be a first order boundary between a phase characterized by positional order and a new phase with only short range order. In an RG description, the separatrix should still end at a critical point of certain type, where the associated exponents are such that they represent, for instance, a jump in the order parameter (e.g., a vanishing order parameter exponent in the ordered phase). While the true nature of the transition may still be somewhat unclear till the date, we expect the topology of the RG flow lines to remain the same regardless of the precise nature of the transition, and hence, we speculate Fig.~\ref{occ-1} to hold in general.

For nearly incompressible systems, $\tilde \lambda\rightarrow \infty$, and $u_{ii}\approx 0$. However, for compressible systems with vanishing strain in the zero-stress states,
an identical analysis holds for the renormalization and scaling of the longitudinal displacements ${\bf u}^L({\bf x})$.
In summary: We again have $(0,0)$ as the only fixed point in the $\alpha_2-\beta_2$ plane, which is attractive along the $\alpha_2$-axis and repulsive along the $\beta_2$-axis. Similar to the separatrix in the $\alpha_1-\beta_1$-plane, there is a separatrix in the $\alpha_2-\beta_2$-plane, such that for initial conditions lying below the separatrix, the RG flow is towards the origin, where as for initial conditions lying above the separatrix, the flow lines go away from the origin and eventually go out of the validity of the perturbation theory.
The equation of the separatrix in the ($\alpha_2,\beta_2$) plane, that is the direct analog of Eq.~(\ref{sep-2d}) in the $(\alpha_1,\,\beta_1)$, is given by
\begin{equation}
 \Gamma_{2c}=\frac{1}{2}\left[-6\pm \sqrt{60}\right],\label{gamma-val22}
\end{equation}
that is the analog of $\Gamma_{1c}$ in (\ref{gamma-val}). Proceeding as for ${\bf u}^T$, on the stable side of the separatrix, we find that
\begin{equation}
 \frac{d\tilde\lambda}{dl} = 2g_2 = \frac{\tilde\lambda}{3l},
\end{equation}
giving for the scale-dependent, renormalized $\tilde\lambda (q)$ as
\begin{equation}
 \tilde \lambda(q)=\tilde\lambda_R \left[\ln \left(\frac{\Lambda}{q}\right)\right]^{1/3},
\end{equation}
where $\tilde \lambda_R$ is the amplitude of renormalized bulk modulus. This
 in turn gives
\begin{equation}
 \langle (\mathbf{u}^L({\bf x}))^2 \rangle \approx \frac{T_c}{2\pi\tilde\lambda_R} \left[\ln ( L/a_0)\right]^{2/3},\label{ren-uL}
\end{equation}
that is identical with the corresponding result for $\langle (\mathbf{u}^L({\bf x}))^2 \rangle$. Unsurprisingly, renormalized correlator $C_{LL}(|{\bf x-x'}|)\equiv \langle \left[ u_i^L({\bf x})-u_i^L({\bf x'})\right]^2\rangle$ scales as $(T_c/\tilde\lambda_R)[\ln (|{\bf x -x'}|)\tilde\Lambda]^{2/3}$ for large $|{\bf x -x'}|$; here, $\tilde\lambda_R$ is the amplitude of the scale-dependent, renormalized shear modulus. This together with (\ref{ren-uL}) are analogs of (\ref{corr-uTT}) and (\ref{vari-u}). These four results establish the new universality class with SQLRO near $T_c$ for second order phase transitions in compressible elastic media.

A RG flow diagram analogous to (\ref{sep-2d}) may be drawn in the $\alpha_2-\beta_2$ plane; we do not show it here.

For initial conditions 
\begin{equation}
\overline g_2^2(l=0)/(\tilde \lambda (l=0)g_2(l=0))<(>)\Gamma_{2c}\equiv \frac{1}{2}\left[ -6 +\sqrt{60}\right],
\end{equation}
the RG flow lines in the $\alpha_2\,-\,\beta_2$-plane flow to  (away from) the origin, with $\beta_2$ eventually diverging at a finite $l$ and $\alpha_2$ vanishing in the latter case.

It is possible that the initial conditions are such that $\Gamma_1<(>)\Gamma_{1c}$ and $\Gamma_2>(<)\Gamma_{2c}$, since all the phenomenological parameters are free parameters in our theory. In that case $\mu(l)$ will stiffen (soften) and $\tilde\lambda(l)$ will soften (stiffen) due to the order parameter fluctuations near the critical point. Therefore, the longitudinal modes will have enhanced (reduced) fluctuations, whereas the fluctuations of the transverse modes will be suppressed (enhanced) due to the anhamornic effects. Thus, separate measurements of the longitudinal and transverse modes fluctuations should reveal important information about the microscopic parameters of the system. It should however be remembered as the fluctuations of either the longitudinal or transverse modes rise (being on the unstable side of the respective sepatarrix), eventually the anhamornic terms neglected in the strain (on the ground of being RG irrelevant) are going to be important, ultimately leading to disorder and overall loss of positional order for a sufficiently large system size $L$. We do not discuss that further here.

\subsection{Bulk sample: $d>2$}

\begin{figure}[htb]
 \includegraphics[width=7cm]{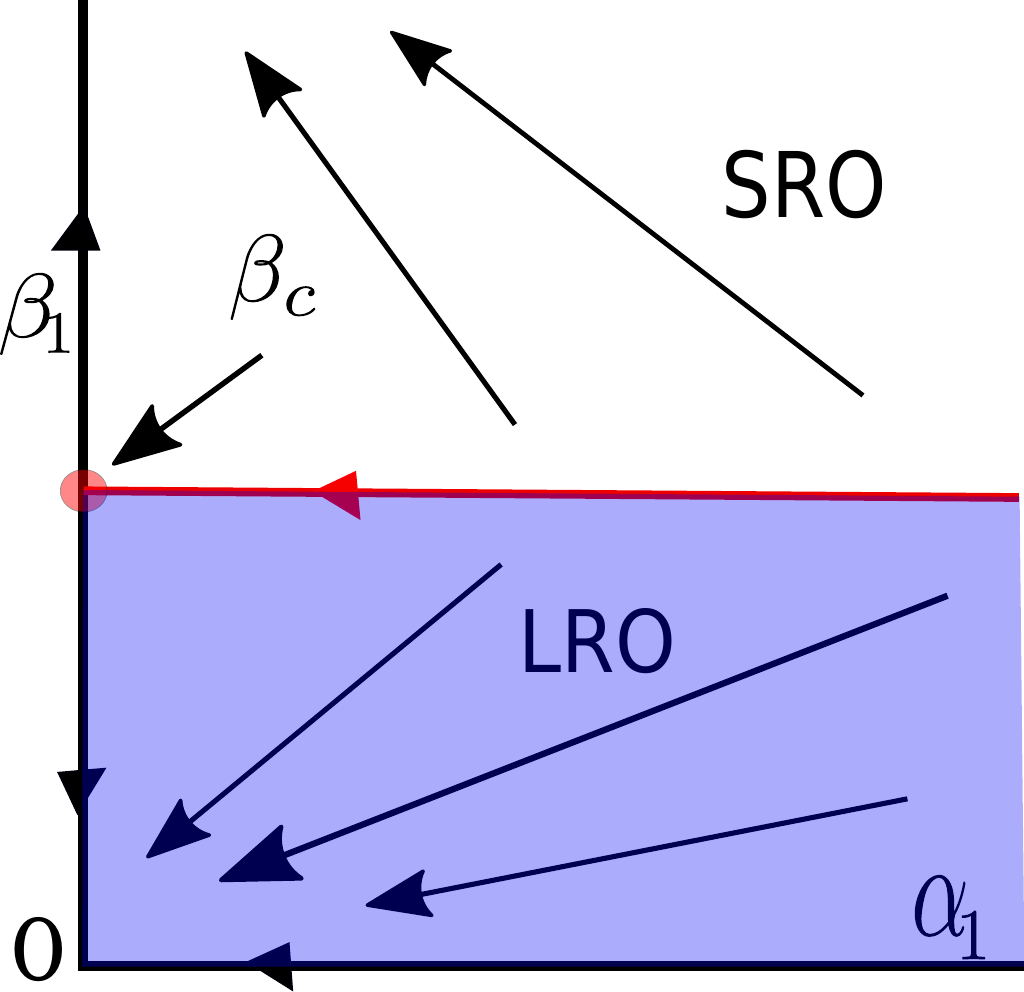}
 \caption{RG flow diagram in the $\alpha_1-\beta_1$ plane in 3D. The blue line is the separatrix given by $\beta=\epsilon/2$. The small circle on the $\beta_1$-axis is the unstable fixed point $(0,\,\beta_c)$. Arrows indicate the flow directions (see text).}\label{rg-3d}
\end{figure}

We now study the universal scaling properties at higher dimensions, i.e., $d=2+\epsilon>2$ near $T=T_c$. We use the flow equations (\ref{flowa1}-\ref{flowb2}). There are now two fixed points for the RG flow equations (\ref{flowa1}) and (\ref{flowb1}). These are $(0,0)$ which is linearly stable and $(0,\,\epsilon/2)$ which is linearly stable along the $\alpha_1$-direction, but unstable along $\beta_1$-direction. There could in principle be a third possibility of a fixed point, in which $\alpha_1\neq 0$, and we solve
\begin{eqnarray}
 2\beta_1 - 6\alpha_1&=&\epsilon,\\
 -3\alpha_1^2 - \frac{\beta_1^2}{8} + \frac{\alpha_1\beta_1}{2} &=& \epsilon \alpha_1
\end{eqnarray}
simultaneously. Eliminating $\beta_1$, we find
\begin{equation}
 84\alpha_1^2 + 36\alpha_1\epsilon + \epsilon^2=0,
\end{equation}
that has no real positive solution for $\alpha_1$. Thus, $\alpha_1=0$ is the only possible physically acceptable fixed point. RG trajectories for all initial values for $\alpha_1$ and for $\beta_1<\epsilon/2$ flow to the stable fixed point (0,\,0), where as RG trajectories for all initial values for $\alpha_1$ together with all initial $\beta_1 > \epsilon/2$ flow away from the origin, until they are out of the validity of our perturbation theory. This signifies instability arising from break down of linear elasticity (see below). Further, $\beta_1=\epsilon/2\equiv\beta_c$ is the separatrix between the stable and unstable phases.  The RG flow diagram is shown in Fig.~\ref{rg-3d}. 


In the stable region of the phase space, linearizing about the stable fixed point (0,0), we get
\begin{eqnarray}
 \frac{d\alpha_1}{dl}&=&-\epsilon \alpha_1,\\
 \frac{d\beta_1}{dl}&=&-\epsilon \beta_1.
\end{eqnarray}
These give
\begin{equation}
 \alpha_1(l)\sim \exp(-\epsilon l),\;\beta_1(l)\sim \exp(-\epsilon l).\label{alpha-beta-3d}
\end{equation}
Thus, $\alpha_1(l)$ and $\beta_1(l)$ vanish exponentially in $l$. 
This gives for $\mu(l)$
\begin{equation}
 \frac{d\mu}{dl}=g_1 - \frac{\overline g_1^2}{2\mu}=\mu\left[\alpha_1-\frac{\beta_1}{2}\right].
\end{equation}
Thus, 
\begin{equation}
 \frac{d\mu}{dl}\rightarrow 0
\end{equation}
as $l\rightarrow \infty$, implying $\mu(l\rightarrow \infty)\rightarrow \mu_\infty$, a constant. Similar argument gives $\tilde\lambda(l\rightarrow \infty)\rightarrow \tilde\lambda_\infty$, another constant. Thus, there are {\em no infinite renormalizations} of $\mu(l)$ and $\tilde\lambda(l)$, unlike their 2D counterparts. Hence, there is {\em no} anomalous elasticity.
This further means that in the thermodynamic limit, the elastic deformation fluctuations are statistically identical to that in the noninteracting theory. This immediately gives $\langle (\mathbf{u}^T({\bf x}))^2 \rangle$  as finite in the thermodynamic limit. This naturally corresponds to long range order.

The separatrix is linearly unstable along the $\beta_1$ direction. This can be seen easily linearizing Eq.~(\ref{flowb1}) by writing $\beta_1(l)=\beta_{1c}+\delta_2(l)$, where $\beta_{1c}=\epsilon/2$. To the linear order in $\delta_2$, we find
\begin{equation}
\frac{d\delta_2}{dl}=4\beta_c\delta_2>0,\label{delta2-eq}
\end{equation}
clearly showing instability along both the directions parallel to the $\beta_1$-axis. On the other hand the flow along the separatrix is {\em towards} the critical point $(0,\,\beta_{1c}$):
\begin{equation}
 \frac{d\alpha_1(l)}{dl}|_{\beta_1(l)=\beta_{1c}}=-\epsilon\alpha_1+\frac{\alpha_1\beta_1}{2} - 3\alpha_1^2 -\frac{\beta_1^2}{8}<0,
\end{equation}
indicating a flow towards to the $\beta_1$-axis. 

The linear instability of the fixed point $(0,\,\epsilon/2)$ together with the RG flow diagram in Fig.~\ref{rg-3d} 
implies that on the unstable side of the separatrix, $\beta_1(l)$ diverges for any $L$, so long as the initial values of $\alpha_1,\,\beta_1$ lie in the unstable side of the separatrix $\beta_1=\epsilon/2$. This further gives that for all such initial conditions renormalized $\mu$ vanishes independent of $L$.


Having known the structure of the RG flow lines near the fixed points for small values of the coupling constants, we can now use a Occam's razor-type argument to speculate on the global RG flow lines for arbitrary coupling constants. 

\begin{figure}[htb]
 \includegraphics[width=8cm]{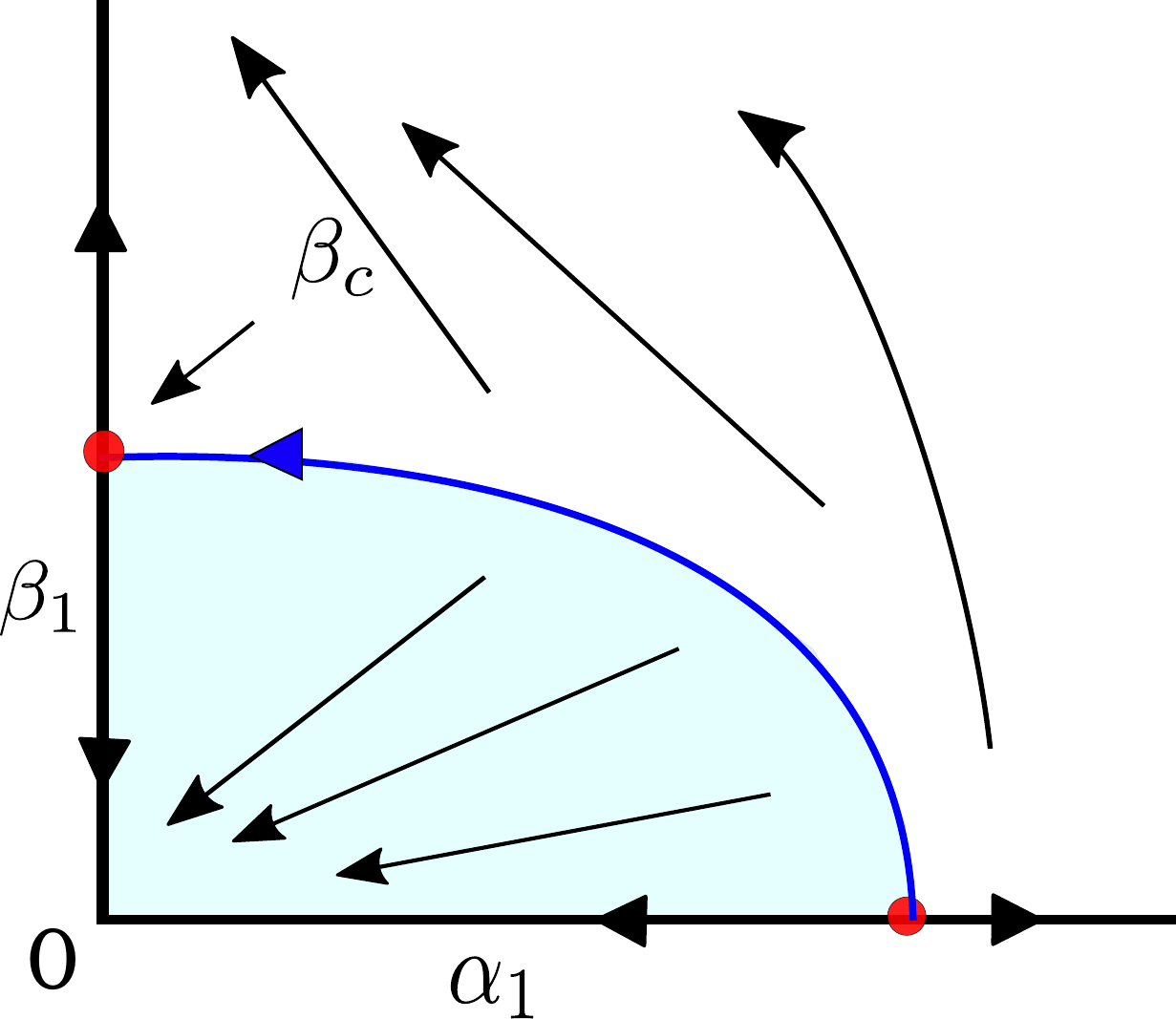}
 \caption{Conjectured ``Occam's razor'' global RG flows in the $\alpha_1-\beta_1$-plane at $d>2$. Arrows indicate the flow directions. The light blue shaded region is the stable region. The small red circle on the $\alpha_1$-axis is the putative unstable critical point not accessible in our perturbative RG. The small red circle on the $\beta_1$-axis is the unstable fixed point (see text).}\label{occ-2}
\end{figure}
It is generally expected, as in 2D, that for large enough $\alpha_1$ (with $\beta_1=0$), which may be achieved at high enough $T_c$, an elastic medium should melt into a liquid-like phase with SRO.
It is also known that the melting transitions of crystals in 3D is indeed {\em first order}. Nonetheless, as argued above for 2D, the topology of the global RG flow lines should be such that the separatrix should terminate on the $\alpha_1$ axis demarcating the phase with long range positional order and the high temperature liquid phase. 

We can now infer the fluctuation properties of ${\bf u}^L({\bf x})$ directly from those of ${\bf u}^T({\bf x})$. Here, $\beta_{2}=\beta_{2c}\equiv \epsilon/8$ is separatrix, that separates the stable phase with positional order for $\beta_2(l)<\beta_{2c}$ and an unstable phase without any positional order for $\beta_{2}(l=0)>\beta_{2c}$. The separatrix is linearly unstable along the $\beta_2$-direction. Again the flow along the separatrix is towards the critical point $(0,\,\beta_{2c})$. The schematic RG flow diagram in the $\alpha_2-\beta_2$ plane is identical to Fig.~\ref{rg-3d}. Again, we can construct an Occam's razor argument global RG flow lines; the corresponding flow diagram looks schematically identical to Fig.~\ref{occ-2}. 

Notice that since both $\mu(l)$ and $\tilde\lambda (l)$ rapidly vanish (essentially as soon as the system size grows beyond a small microscopic size) on the unstable side of the separatrix, the instability is practically independent of the system size. This is in contrast to the situation at 2D. Readers interested in a perturbation theory argument for this should find one in Appendix~\ref{pos-corr}.

\section{First order transitions}\label{first-order}

We have argued above that the nonlinear couplings of $\phi$ with ${\bf u}^T$ or ${\bf u}^L$ are irrelevant (in a RG sense) near the Heisenberg fixed point of the Ising model. This means the coupling with the elastic medium does not affect the second order phase transition of the Ising model and the corresponding universal critical scaling near $T_c$. This however implies that the fluctuation corrected coupling constant $v$ in (\ref{free2}) is {\em always} positive at any length scale. Can effective $v$ be turned negative at any finite scale by large enough fluctuations? We carefully consider this question in this Section.

{ We first consider the case with microscopic Ising symmetry, i.e., the inversion symmetry breaking couplings $\overline{g}_1,\,\overline g_2$ vanish. In this case, in order to have a second order transition, it should be ensured  that under mode elimination, $v_e$, the fluctuation-corrected $v$ at any intermediate scale never  turns negative. This need not be the case always. In fact, this may not hold true for sufficiently strong order parameter-strain couplings. In the anticipation that $v_e$ can actually turn negative, we extend ${\cal F}$ by adding a $v_6\phi^6$-term in it with $v_6>0$ for thermodynamic stability reasons. We consider the inhomogeneous fluctuation corrections to $v$ that originate from $g_1,\overline g_1, g_2,\overline g_2$, that themselves do not depend upon $v$ explicitly. These contributions are {\em finite}, but {\em negative}: Neglecting the homogeneous corrections to $v$ for simplicity,  and  using the expressions of the diagrams in Appendix~\ref{first-order-diag}, (with  $\overline g_1=0=\overline g_2$ for the microscopic inversion symmetric case) we get
\begin{eqnarray}
 \beta_c v_e&\equiv& \beta_c v-2dT_c^2\left(\frac{\beta_c^2 g_1^2}{4\mu^2}+\frac{\beta_c^2 g_2^2}{\tilde\lambda^2}\right)\frac{\Lambda^d}{(2\pi)^d},\label{v-tot1}
\end{eqnarray}
valid for all $d \geq 2$. Now, for $v_e>0$, the $v_6\phi^6$-term is unnecessary. The phase transition of $\phi$ is then unaffected by the order parameter-strain couplings, and remains a continuous transition belonging to the Ising universality class. 
If, however, $v_e<0$, then a $v_6\phi^6$-term must be taken into account for reasons of thermodynamic stability. In that case, $\phi$ now undergoes a first order transition with the order parameter $m\equiv \langle\phi\rangle$ jumping of magnitude $[|v_e|(2v_6)]^{1/2}$. We thus conclude that in ZTE systems with microscopic Ising symmetry, sufficiently strong spin-lattice couplings necessarily turn the second order transition into a first order one. }

{ The general case with nonzero $\overline g_1,\,\overline g_2$ requires more careful scrutiny.}
To proceed further systematically, we generalize the free energy functional ${\cal F}_\phi$ to ${\cal \tilde F}_\phi$ in general $d$-dimensions given by
\begin{equation}
 {\cal \tilde F}_\phi = {\cal F}_\phi + \int d^dx \left[g\phi^3 - h \phi\right],\label{extnd-f}
\end{equation}
as appropriate for an {\em asymmetric} binary system (i.e., without any symmetry under the inversion of $\phi$); parameters $g$ and $h$ can be of any sign. For a symmetric system $g=0=h$, which corresponds to an Ising magnet with no net magnetization. { For a conserved system (relevant in soft matter version of this model) $\int d^dx \phi({\bf x})$ is a conserved quantity (a constant), and hence drops out from (\ref{extnd-f}).} Clearly, free energy $\cal \tilde F_\phi$ in Eq.~(\ref{extnd-f}) in the limit of a rigid lattice  (all strains vanish) generically describes a first order transition akin to the standard liquid-gas transition. This naturally implies the existence of a coexistence curve identical to that for the liquid-gas transition, with an associated finite jump in the order parameter~\cite{chaikin}. Nonetheless, it still admits a second order transition at a critical point belonging to the Ising universality class. This can be shown by expanding ${\cal \tilde F}_\phi$ about $\phi = \phi_0$, with $\phi_0$ is chosen in a way such that the $g \phi^3$ -term in ${\cal \tilde F}_\phi$ above vanishes. The
resulting transformed free energy has the form same as that of the Ising model at a finite external magnetic field $h_0$ (related to $h$ in (\ref{extnd-f}) above and depends in general upon the model parameters) that has generic first order transitions (or
 no transitions at all) if $h_0$ is tuned at any general $T$~\cite{chaikin}; furthermore, second order transitions belonging to the 
Ising universality class is found only if both $T$ and $h_0$ are simultaneously tuned; the corresponding critical point is found in the
$T-h_0$-plane at $r=0$ or $T = T_c$ (in a mean-field description) and $h_0 = 0$~\cite{chaikin}.
The coexistence curve for $\cal \tilde F$ above in fact is similar to that for the  Ising model, except that it is now asymmetric with respect to the order parameter $\langle\phi\rangle$, averaged over the whole system  due to the lack of any symmetry of ${\cal F}_\phi$ under inversion of $\phi$. The order parameter-strain coupling terms do not change this general picture as we now discuss below.

To see the effects of the order parameter - strain couplings, we integrate out the strains from $\cal \tilde F$ perturbatively, a process that produces ${\tilde F}_{\phi e}$, a ``dressed'' ${\cal \tilde F}_\phi$ given by
\begin{equation}
 {\tilde F}_{\phi e}=\int d^dx \left[\frac{r}{2}\phi^2 + \frac{1}{2}({\boldsymbol\nabla}\phi)^2 + v_e\phi^4+ g_e\phi^3 + v_6\phi^6-h_e\phi\right],\label{eff-F}
\end{equation}
where $h_e$ is the ``effective magnetic field'', and $v_e$ and $g_e$ are the ``effective'' coupling constants, produced by integrating over the strains; $v_6>0$ is added for thermodynamic stability (see below). Here, we have ignored any corrections to $r$, obtained by eliminating the strain field, as they represent just a shift in $T_c$, a fact that is present but of little significance to the present discussion. { Again, the last term in the rhs of (\ref{eff-F}) can be dropped in a conserved system.}

We focus on the fluctuation-corrections to $v$ and $g$ that arise solely from the order parameter-strain couplings (this suffices for our purposes here). Consider the two inhomogeneous fluctuation corrections to $v$ that originate from $g_1$ and $\tilde g_1$; which are discussed in Appendix~\ref{first-order-diag} (see Fig.~\ref{inhom-v}). Contributions from these diagrams are {\em finite}. and independent of $v$ itself, i.e., inhomogeneous in. More importantly, these contribute {\em negatively} to $v$. While in the standard wisdom of RG, these {\em finite} corrections do not matter and are to be neglected, there is a possibility that for sufficiently large $g_1$ and $\overline g_1$, effective $v$ actually turns negative. This then immediately destroys the assumed second order transition, and with it all the diverging fluctuation corrections to the model parameters, since fluctuations are bounded in a first order transition.  Neglecting the homogeneous corrections to $v$ and using the expressions of the diagrams in Appendix~\ref{first-order-diag}, we define an effective $v$, that we denote by $v_e$, as follows:
\begin{eqnarray}
  v_e&\equiv&  v-2dT_c\left(\frac{ g_1^2}{4\mu^2}+\frac{ g_2^2}{\tilde\lambda^2}\right)\frac{\Lambda^d}{(2\pi)^d} \nonumber \\&-&2dT_c \left(\frac{\overline  g_1^4}{32 \mu^4} + \frac{\overline g_2^4}{\tilde\lambda^4}\right)\frac{\Lambda^d}{(2\pi)^d}.\label{v-tot}
\end{eqnarray}
Similarly, the inhomogeneous one-loop corrections to $g$ are given in the Feynman graphs in Appendix~\ref{first-order-diag} (see Fig.~\ref{inhom-g}).
 The resulting effective parameter $g_e$ is thus given  (neglecting any homogeneous correction) by

 
\begin{eqnarray}
 g_e&=&g+  g_1\overline g_1\frac{T_c}{2\mu^2}d\frac{\Lambda^d}{(2\pi)^d} +2g_2\overline g_2\frac{T_c}{\tilde\lambda^2}d\frac{\Lambda^d}{(2\pi)^d} \nonumber \\&-& \overline g_1^3\frac{T_c}{2\mu^3}d\frac{\Lambda^d}{(2\pi)^d} -  4\overline g_2^3 \frac{T_c}{\tilde\lambda^3}\frac{\Lambda^d}{(2\pi)^d}.\label{g-tot}
\end{eqnarray}


{Two distinct situations can arise. First consider $g_e\neq 0$ which is the more general case. In this case, there is a generic first order transition, akin to the liquid-gas first order transition with an order parameter jump $m=-g_e/(12 v)$ at a transition temperature $T^*=T_c+9g_e^2/(16 v)$~\cite{chaikin}. On the other hand, $g_e$ can be turned zero by tuning $g-1,\,\overline g_1,\,g_2,\,\overline g_2$. }
Consider $v_e>0$. The $v_6\phi^6$-term in (\ref{eff-F}) can now be ignored. Then (\ref{eff-F}) has the same structure as ${\cal \tilde F}_\phi$ in (\ref{extnd-f}) above; the discussion that follows immediately after (\ref{extnd-f}) applies here as well. By making a suitable shift in $\phi$, the cubic term $g_e \phi^3$ may be
eliminated from ${\cal F}_{\phi e}$, yielding $F_e$, a modified form for ${\cal F}_{\phi e}$, identical to the free energy for the Ising model in the
presence of an external magnetic field $h_\phi$. Then, the order parameter $\phi$ clearly generally undergoes a first order transition below a
transition temperature. A critical point with a second order transition may still be accessed only by suitable tuning of both
$T$ and $h_\phi$: in fact the critical point is located in the $T-h_\phi$- plane at $r=0$ or $T = T_c$ and $h_\phi = 0$, with an
associated universal scaling behavior belonging to the 2D Ising universality class. Since $h_\phi$ in general depends on the order parameter - strain couplings, it is possible to tune it to zero by tuning $g_1,\,\overline g_1,\,g_2,\,\overline g_2$. Further, $T_c$ also receives fluctuation
corrections that depends on the strain - order parameter couplings (not shown here). Thus the critical point is accessed by tuning $g_1,\,\overline g_1,\,g_2,\,\overline g_2$. At the simplest level, the role of $g_e \neq 0$ is only to introduce an asymmetry of
the order parameter $\langle\phi\rangle$, which is reflected in the curvature of the coexistence curve at the
criticality~\cite{chaikin}. Since $g_e$ can also be varied continuously and made positive, negative or zero by tuning the order parameter - strain coupling constants, the curvature at criticality and hence the location of the coexistence curve in
the $\langle\phi\rangle - T$ plane should change continuously with these coupling constants. Experimental measurements of the
coexistence curve for a given system can thus reveal valuable quantitative information about these coupling constants.

\begin{figure}[htb]
 \includegraphics[width=7cm]{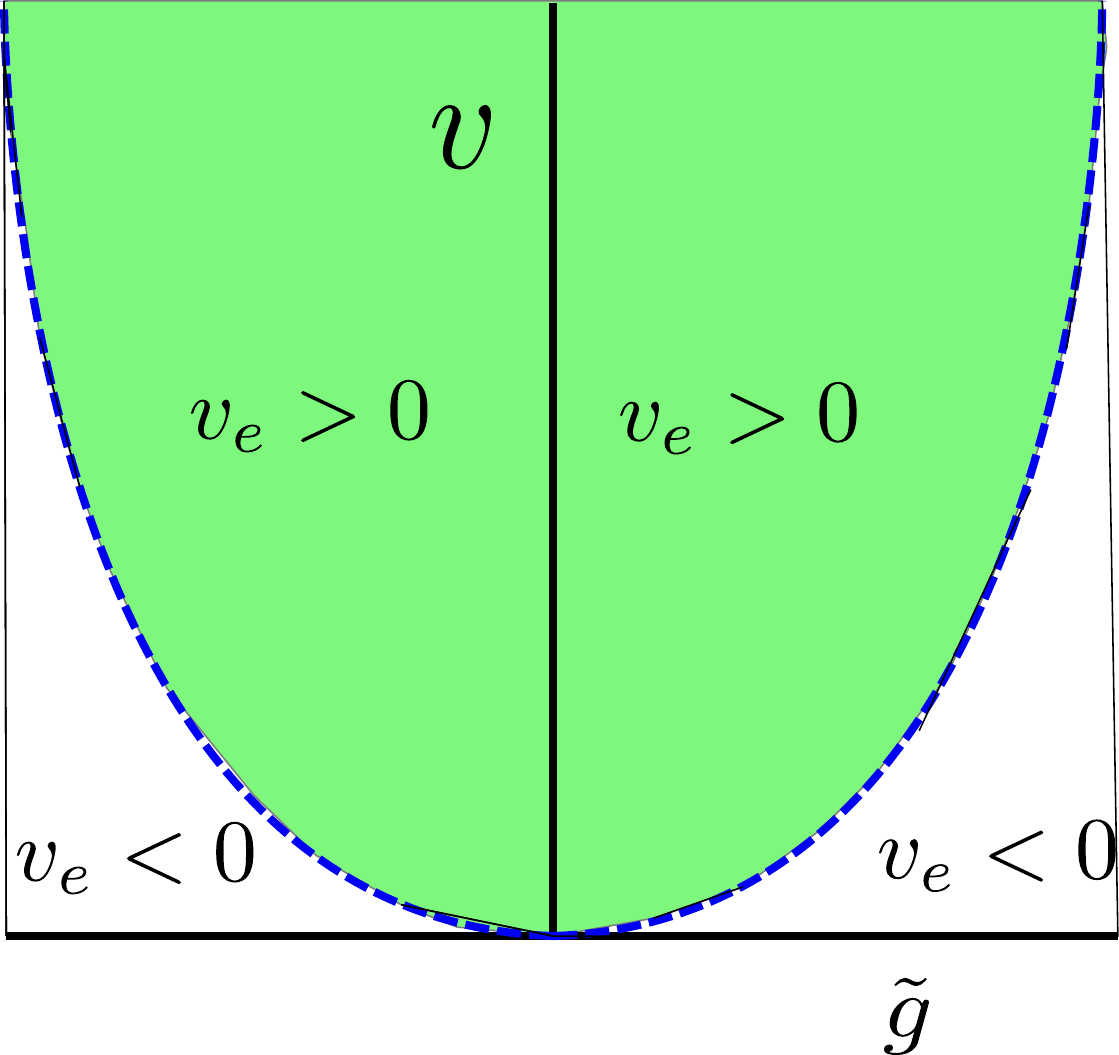}
 \caption{Schematic phase diagram of the order parameter in the $\tilde g-v$ plane, where for simplicity we have set $g_1=g_2=\overline g_1=\overline g_2=\tilde g$. We have assumed $g_e=0$, which rules out a first order transition of the liquid gas kind. Second order phase transition is obtained in the green region with $v_e>0$, whereas the transition is first order in the outside white region with $v_e<0$. The broken blue curved line is given by (see~(\ref{v-tot})) $v_e=0$ (along with $g_e = 0$), corresponding to a line of tricritical points.}\label{order-diag}
\end{figure}

If $v_e<0$, then conditions of thermodynamic stability dictates that the $v_6\phi^6$-term in (\ref{free2}) must be taken into consideration, where $v_6$ is positive definite. We can still set $g_e$ to zero by  tuning $g$ (in a technical language this is akin to adding a ``counter term'' in $\cal F$ to as to keep the net coefficient of a $\phi^3$-term to zero). Now with $v_e<0$, instead of a second order transition, this now allows a {\em first order transition} with an order parameter jump $\phi_c = \pm [|v_e|(2v_6)]^{1/2}$ at the transition temperature $T^*=T_c+ 2|v_e|^2/(3v_6)$~\cite{chaikin}. In fact, there now exists a {\em tricritical point} that is determined by the condition $v_e=0$ (along with $g_e = 0$).  Notice that the dependence of $v_e$ on the selectivity parameters $\overline g_1$ or $\overline g_2$ by a factor of 2 changes their contributions to $v_e$ by a factor of 16, whereas similar changes in $g_1,\,g_2$ change their contributions to $v_e$ by just a factor of 4. Thus carefully prepared samples with different selectivity parameters should enable one to test the possibilities of both first and second order transitions. See Fig.~\ref{order-diag} for a schematic phase diagram.

{ We thus conclude that  even in the presence of Ising-symmetry breaking spin-lattice coupling terms, the transition is generically first order. Nonetheless, a second order Ising transition can be accessed by tuning the model parameters reminiscent of the second order transition in liquid-gas systems. Intriguingly, this second order transition can get converted into a {\em different} first order one for sufficiently strong spin-lattice interactions. Across such first order transitions, the elastic modulii are finite, but still anomalous in the sense discussed below.}



If there is a first order transition, there are no instabilities since all corrections to $\tilde\lambda$ and $\mu$ are finite. Nonetheless, there are corrections which can be measured. The fluctuation corrections are finite and small for small (bare) anharmonic coupling constants $g_1,\,g_2,\,\overline g_1,\,\overline g_2$. With this and neglecting the contributions from the one-loop corrections, effective $\mu$ and effective $\tilde\lambda$ are given by [see free energy (\ref{free2}) above]
\begin{eqnarray}
 \mu_e&=&
 \mu + g_1   \langle \phi^2({\bf x})\rangle+\overline g_1 \langle\phi ({\bf x})\rangle,\\
 \tilde\lambda_e &=& \tilde\lambda  +2g_2 \langle \phi^2({\bf x})\rangle +\overline g_2 \langle\phi ({\bf x})\rangle
\end{eqnarray}
 to the lowest order in $g_1,\,g_2,\,\overline g_1,\,\overline g_2$.
For $T>T^*$, $\langle \phi^2({\bf x}) \rangle$ is negligible, where for $T<T^*$, $\langle \phi^2({\bf x})\rangle$ can be approximated by $m^2$;  where $m=\langle\phi({\bf x})\rangle \neq 0$ is the mean field value of the order parameter below $T^*$. For simplicity, we neglect the one-loop corrections to $g_1, g_2,\overline g_1$ and $\overline g_2$ in this discussion. Thus, 
\begin{eqnarray}
\mu(T<T^*)&=&\mu(T>T^*) + g_1 m^2+\overline g_1 m\nonumber \\&\neq& \mu(T>T^*),\\
 \tilde\lambda(T<T^*)&=&\tilde\lambda(T>T^*) + 2g_2 m^2+2\overline g_2 m\nonumber \\&\neq& \tilde\lambda(T>T^*).
\end{eqnarray}
 Whether or not $\mu (T<T^*)$ or $ \tilde\lambda(T<T^*)$ is larger or smaller than their counterparts at $T>T^*$ depends on the relative values of $g_1$ and $\overline g_1$, or $g_2$ and $\overline g_2$, which are free parameters in our theory, but are actually controlled by the microscopic material properties, and the signs of $\overline g_1, \overline g_2$ and $m$. In fact, it is entirely possible that one among $\mu (T<T^*)$ and $\tilde\lambda (T<T^*)$ larger than its counterpart at $T>T^*$, whereas the other is smaller, since all of $g_1,\,\overline g_1,\,g_2,\,\overline g_2$ can in principle vary freely. Therefore, measurements of the elastic fluctuations should show a sudden jump across $T^*$ and should give valuable information about the material properties. As before, for nearly incompressible systems $\tilde \lambda$ diverges, and we need to be concerned only with the variation of $\mu$ across the first order transition.
It is in fact possible to have instability in the ordered phase of the order parameter, leading to loss of positional order. That is any one among $\mu(T<T^*)$ and $\tilde\lambda(T<T^*)$ or both may be {\em negative}, if $\overline g_1$ and/or $\overline g_2$ are sufficiently large. Notice that this instability is independent of the system size at any dimension. A schematic phase diagram in the $\overline g_2^2 - \mu$ plane may be drawn, which is topologically identical to the phase diagram valid for $T\approx T_c$, valid when there is a second order transition; see the phase diagram in Fig.~\ref{phase-diag3-2d}. In the case of first order transition the instability exists in the entire ordered phase $T<T^*$, whereas for the second order transition case, it is confined to the neighborhood of $T_c$ only. More intriguingly, if the order parameter $m$ is conserved, then below $T^*$, there will be (at least) two macroscopically large domains corresponding to $+m$ and $-m$. This means there is a possibility that the system remains stable in one domain, but gets unstable in the other. At the very least, the effective Lam\'e coefficients will be different in the different domains.

\section{Correspondence between the order of the transition and displacement fluctuations}
\label{corres}

We now elucidate the correspondence between the displacement fluctuations and the order of the associated transitions, as one crosses the transition temperature. We make the following general conclusions about the interrelations between the variances and correlations of the 
displacement fluctuations. 
\begin{enumerate}
\item At 2D
\begin{enumerate}
    \item If there is a second order phase transition (i.e., with no jump in the order parameter across the transition) with the unrenormalized model parameters falling in the stable region of the phase space, then the Lam\'e coefficients increase as the transition temperature $T_c$ is approached from the above. The Lam\'e coefficients decrease as  $T$ is further reduced below $T_c$. If the system size diverges, the Lam\'e coefficients too diverge as $T_c$ is approached. The system shows novel {\em anomalous elasticity} resulting into positional SQLRO in the thermodynamic limit, different from the well-known QLRO at $T\neq T_c$, or in a single-component elastic medium. The displacement correlation function for a large separation is much smaller than what it is away from $T_c$, or in a single component medium. 
 \item On the other hand, if there is a second order phase transition and the unrenormalized model parameters fall in the unstable region of the phase space, then for a system with a finite size $L<L_c$, a critical size, the Lam\'e coefficients decrease as temperature $T$ as the transition temperature $T_c$ is approached from the above. The Lam\'e coefficients increase as $T$ is further reduced below $T_c$. Close to $T_c$, the Lam\'e coefficients vanish as the system size $L$ approaches $L_c$ from below; in fact, for $L>L_c$, the system gets unstable with the attendant loss of any positional order.
 \item If there is a first order transition at the transition temperature $T^*$ with a finite jump in the order parameter, then there is a finite jump in the Lam\'e coefficients directly related to the jump in the order parameter. The displacement correlator shows conventional QLRO, with its amplitude showing a jump across the transition temperature.
 \item Independent of the order of the phase transition, the values of the Lam\'e coefficients above the transition temperature can be lower or higher than the corresponding values below the transition. This is controlled by the model parameters.
 \end{enumerate}
 \item At higher dimensions $d>2$
 \begin{enumerate}
  \item If there is a second order phase transition with the unrenormalized model parameters falling in the stable region of the phase space, the Lam\'e coefficients do not diverge as the critical point is approached. The system shows conventional LRO. When the unrenormalized model parameters fall in the unstable region of the phase space for any value of the system size $L$. This corresponds to positional short range order only.
  \item If there is a first order transition, the Lam\`e coefficients show a jump across the transition temperature, concomitant with a jump in the displacement correlation function that shows conventional LRO below and above the transition temperature.
 \end{enumerate}

\end{enumerate}

The above correspondences  are pictorially shown in the schematic diagram in Fig.~\ref{schematic-lame} and
Fig.~\ref{schematic-lame1}.

\begin{widetext}

\begin{figure}[htb]
 \includegraphics[width=5cm]{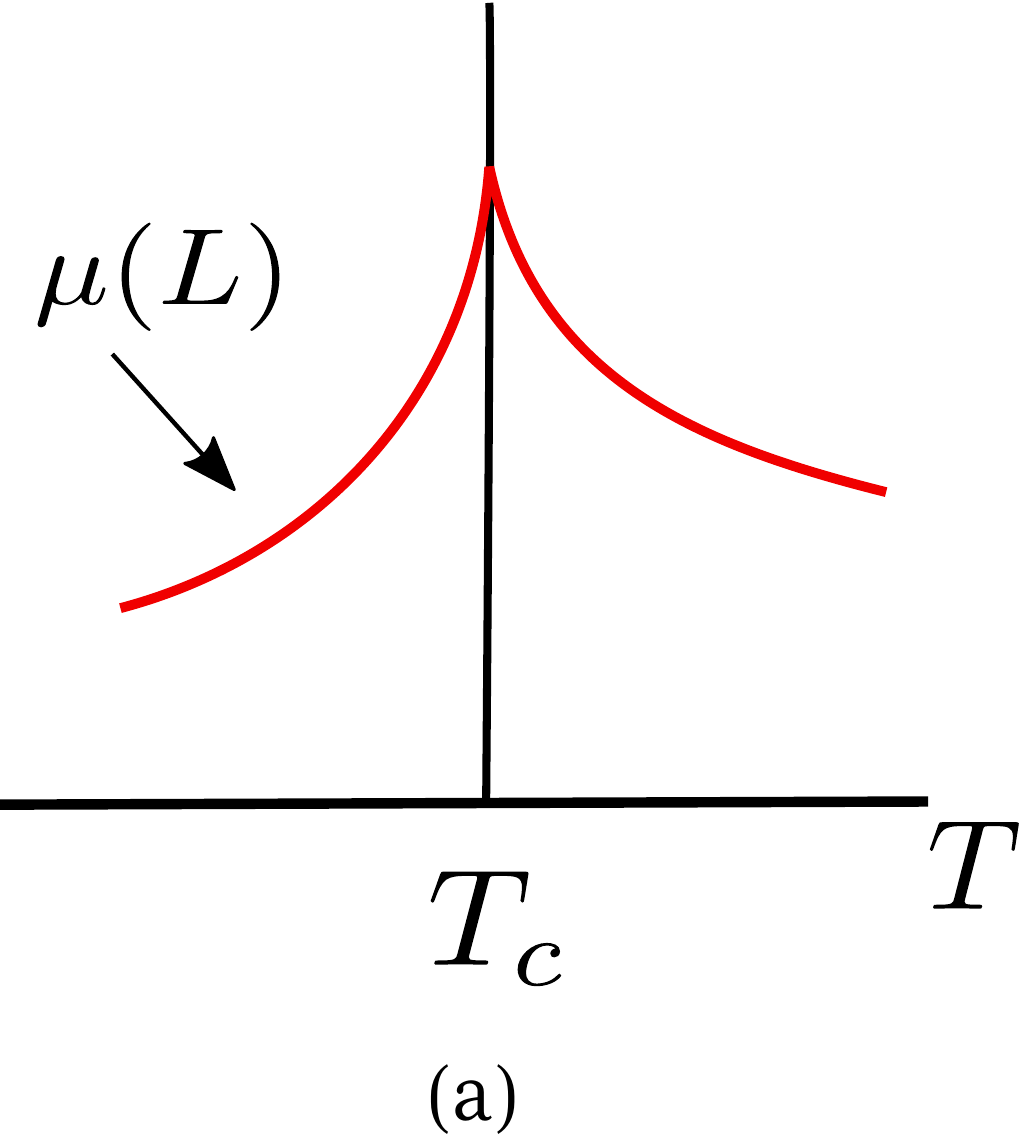}\hfill
 \includegraphics[width=5cm]{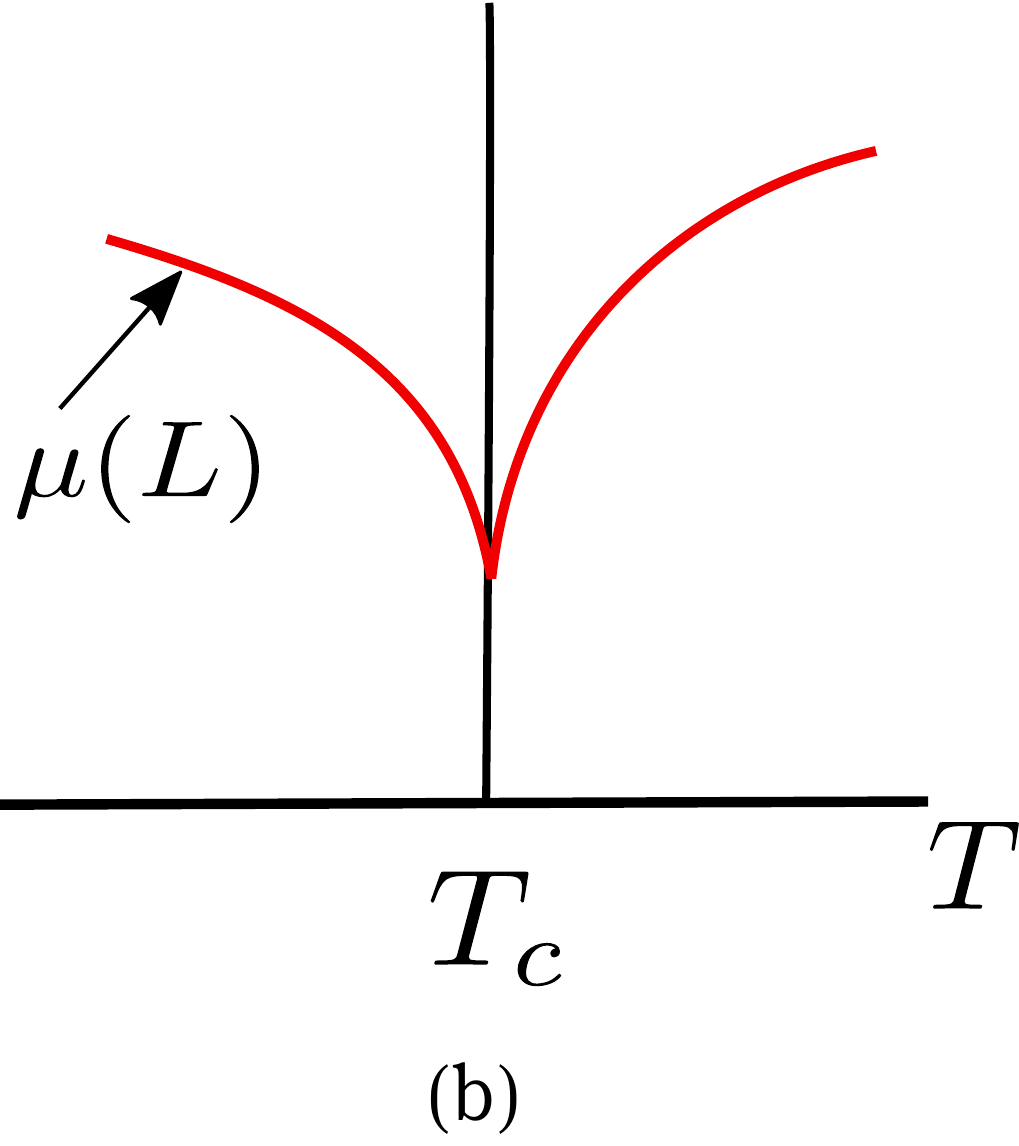}\hfill
 \includegraphics[width=5cm]{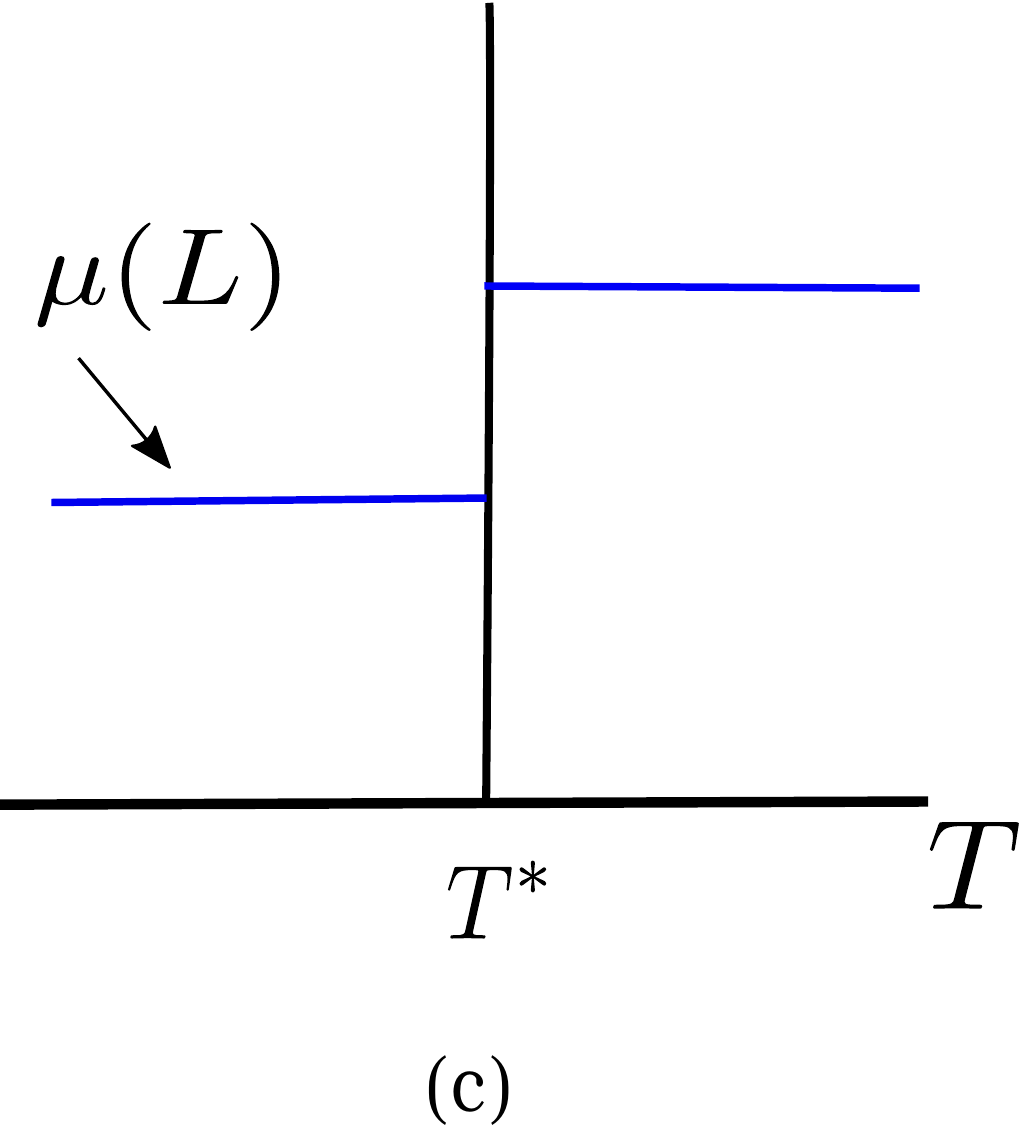}
 \caption{Schematic variation of $\mu$ in a 2D system of finite size $L$ across (a) a second order transition at $T_c$ when the unrenormalized parameters fall on the stable side of the separatrix, (b) a second order transition at $T_c$ when the unrenormalized parameters fall on the unstable side of the separatrix and $L<L_c$, the instability threshold, (c) a first order transition at $T^*>T_c$, with $\mu$ having no significant $L$-dependence. At 3D, (c) holds regardless of the order of transitions. In all these cases, $\mu(L)$ is {\em smaller} in the ordered low-$T$ phase than its value in the disordered high $T$-phase, which is controlled by the bare (unrenormalized) model parameters. Note the smooth variations (shown schematically) of the Lam\'e coefficients across second order transitions, as opposed to their discontinuity across first order transitions.}
 \label{schematic-lame}
\end{figure}

\begin{figure}[htb]
 \includegraphics[width=5cm]{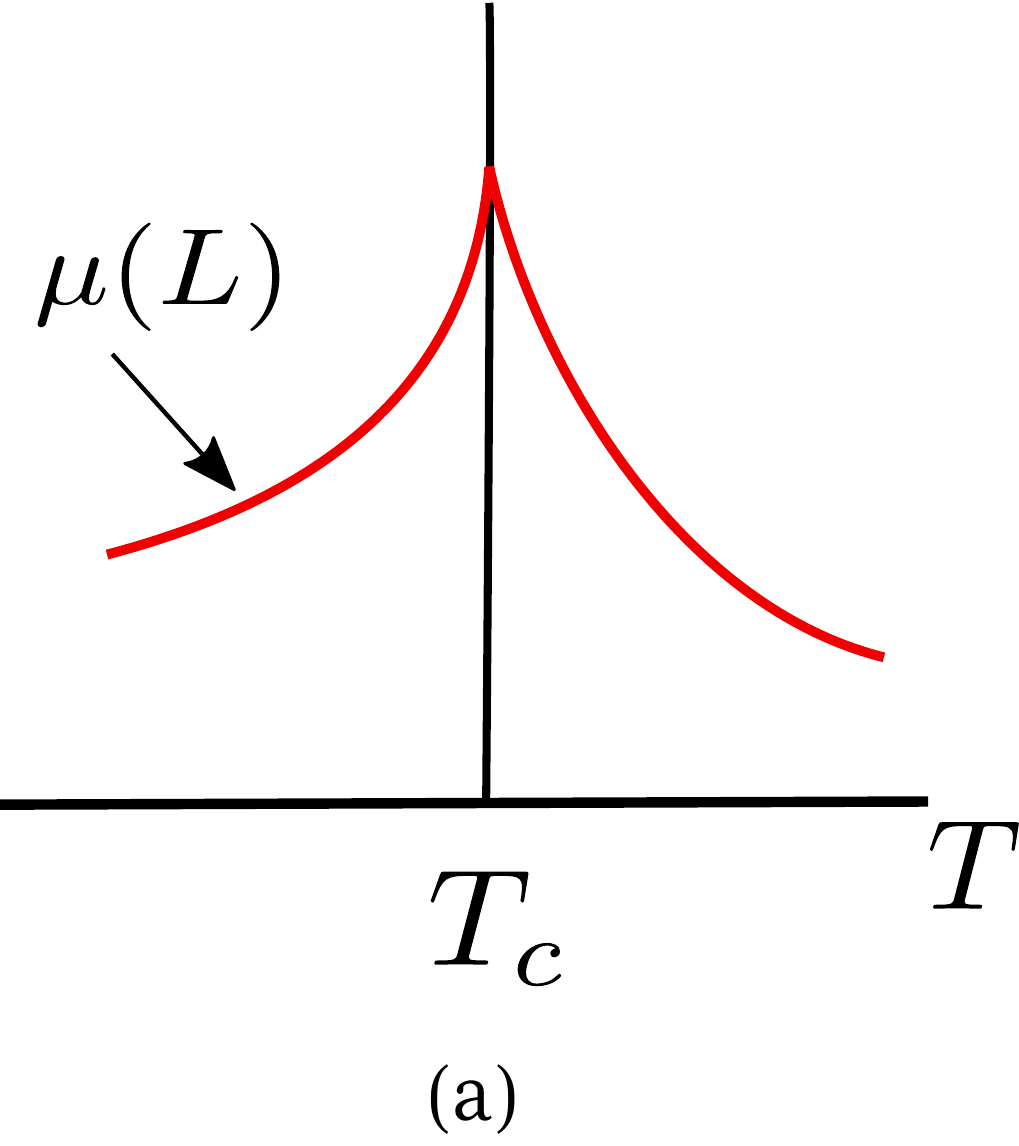}\hfill
 \includegraphics[width=5cm]{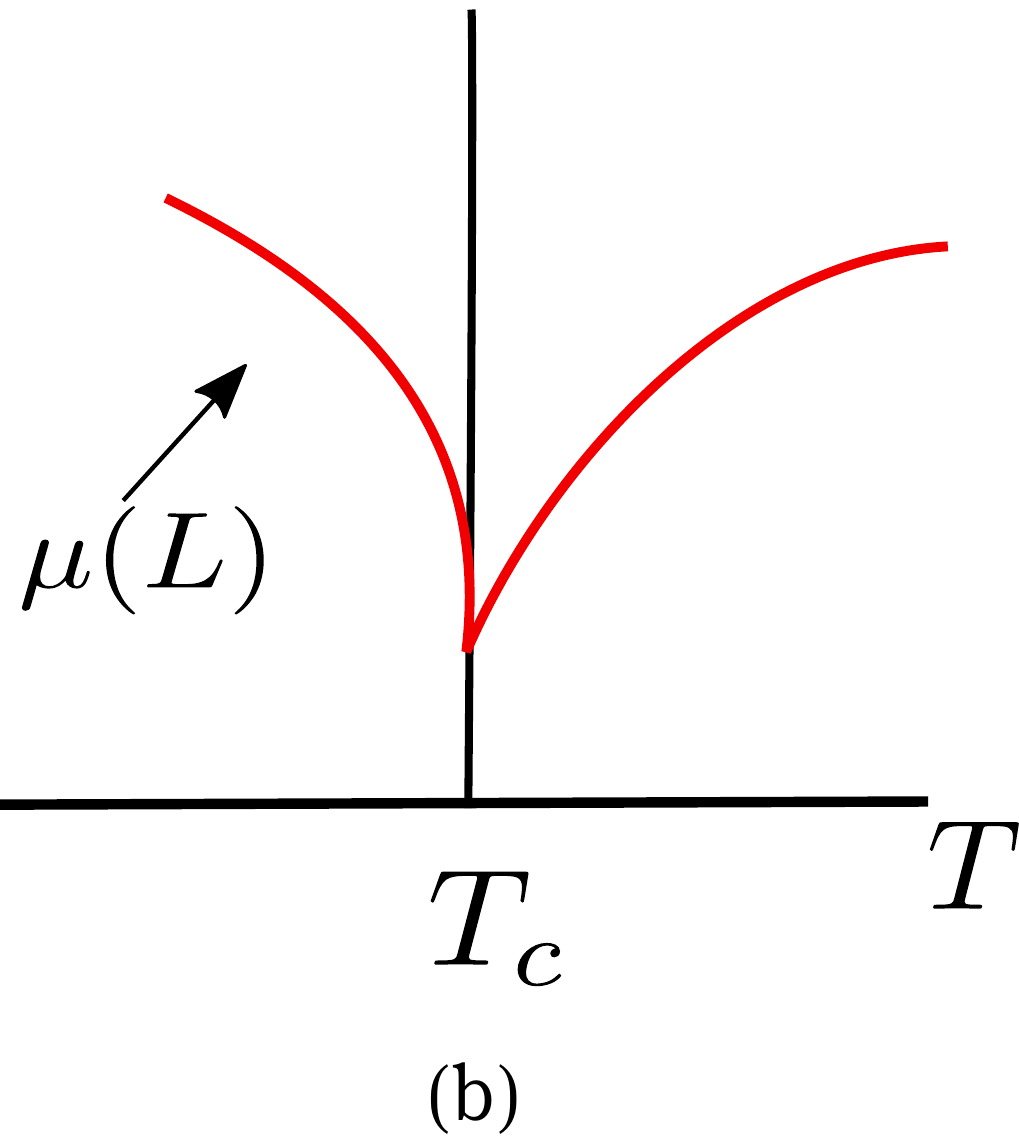}\hfill
 \includegraphics[width=5cm]{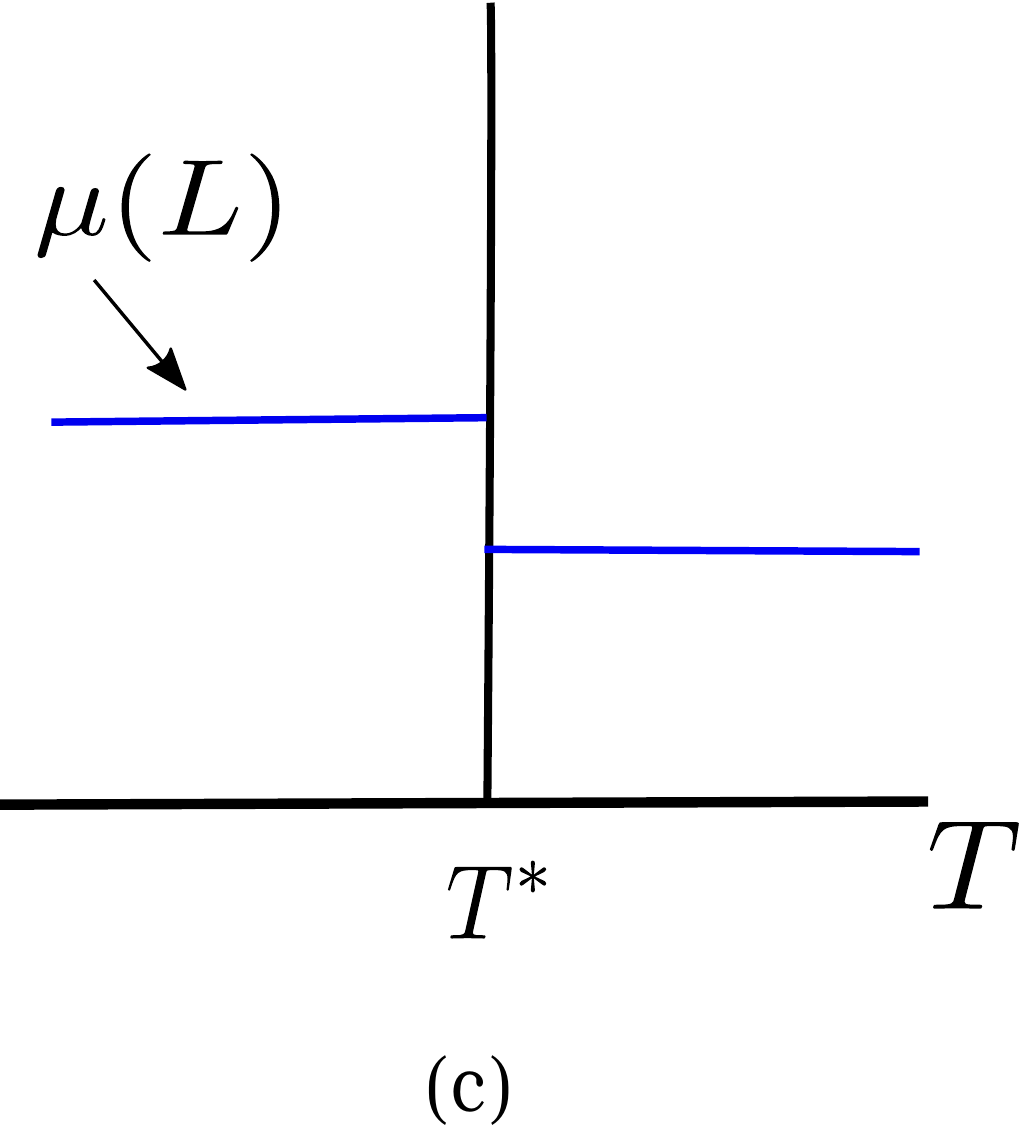}
 \caption{Schematic variation of $\mu$ in a 2D system of finite size $L$ across (a) a second order transition at $T_c$, when the unrenormalized parameters fall on the stable side of the separatrix, (b) a second order transition at $T_c$ when the unrenormalized parameters fall on the unstable side of the separatrix and $L<L_c$, the instability threshold, (c) a first order transition at $T^*>T_c$, with $\mu$ having no significant $L$-dependence. At 3D, (c) holds regardless of the order of transitions. In all these cases, $\mu(L)$ is {\em larger} in the ordered low-$T$ phase than its value in the disordered high $T$-phase, which is controlled by the bare (unrenormalized) model parameters. Note the smooth variations (shown schematically) of the Lam\'e coefficients across second order transitions, as opposed to their discontinuity across first order transitions.}
 \label{schematic-lame1}
\end{figure}

\end{widetext}

\section{Summary}\label{summ}
We have here developed a continuum theory of Ising transitions in a deformable isotropic zero thermal expansion elastic medium, e.g., a gel, and investigate the existence of anomalous elasticity. We consider an Ising-type scalar order parameter to describe the phase transition.  Our theory includes  anharmonic couplings between local in-plane lattice dilations or strains with the order parameter, such that $dT_c/dV=0$, or $\langle u_{ij}\rangle =0$.  Further, these couplings contain two distinct anhamornic contributions, one of which respects the Ising symmetry of the order parameter in a rigid lattice, the other explicitly breaking it. The latter effectively implies selective coupling of the local strain with the local states of the order parameter, and makes the system inversion (i.e., ${\cal Z}_2$) asymmetric. The breaking of the ${\cal Z}_2$ symmetry in the present study is entirely due to its coupling with the local strain, and hence vanishes in the rigid limit of the model.  These anhamornic terms are irrelevant in the RG sense when $dT_c/dV\neq 0$, and were not considered in Ref.~\cite{berg-halp}. In contrast to the present study, the absence of these anhamornic terms even at at $dT_c/dV=0$ led Ref.~\cite{berg-halp} to conclude that spin and lattice degrees of freedom decouple in this limit.

We study the system both at 2D and $d >2$ close to the phase transition temperature of the order parameter. At 2D, we find anomalous elasticity: When there is a second order transition with the selectivity couplings being  sufficiently weak, the in-plane displacement fluctuations are significantly suppressed in comparison with its behavior  away from the phase transition; the phase transition itself remains second order belonging to the 2D Ising universality class. The elasticity is anomalous and the mean-square in-plane displacement scales as $[\ln\,(L/a_0)]^{2/3}$, a significantly weaker dependence on the system size $L$ than the traditional $\ln (L/a_0)$ behavior observed in a 2D elastic medium. Similarly, the two-point correlation function of the differences in the local displacements at two points separated by a distance $r$ scales as $[\ln (r/a_0)]^{2/3}$ for large $r$, a much weaker $r$-dependence than the well-known $\ln r$ dependence observed in QLRO. Thus our result can be thought as a novel positional {SQLRO} that forms a new universality class. As the selectivity parameters grow in magnitude, the system gets unstable beyond a threshold value of the parameters as the renormalized elastic modulii vanish when the system size $L$ exceeds a finite value. This implies a positional short range order or SRO, reminiscent of a liquid. Thus the selectivity parameters can introduce a novel SQLRO to SRO transition. 
These results are summarized in Fig.~\ref{sep-2d}. Melting of 2D crystals are believed to be defect-mediated. It would be interesting to study how melting proceeds near the critical point, when the positional order is not QLRO, but SQLRO. Sufficiently strong strain-order parameter couplings can turn the phase transition to a first order one. In that case, there is no SQLRO; conventional QLRO is observed at all temperatures. However, there are jumps in the elastic modulii as the system passes through the transition temperature. For sufficiently strong selectivity parameters, the system can get distabilized as well.

At dimensions $d>2$, for sufficiently weak selectivity couplings, the variance of the local elastic displacements is independent of the size of the system, corresponding to positional long range order (LRO), not different  from an ordinary 3D elastic medium, e.g., a 3D crystal. However, as the selectivity increases, the system gets unstable beyond a threshold value of the selectivity parameters, with only positional short range  order reminiscent of a liquid. Thus a transition between LRO and SRO can be induced by turning the selectivity parameter. As in 2D, the phase transition can be turned to first order by tuning the selectivity parameters, across which the elastic modulii display finite jumps.

On the whole, thus, the selectivity parameters can be tuned to distabilize the positional order and also to turn the second order phase transition to a first order one. Assuming the selectivity parameters to be continuously varying control parameters, we can note that such variations lead to {\em re-entrant structural phase transitions} of the sample; this could be easily seen if one moves along the $\overline g_1$-axis in the phase diagrams (\ref{phase-diag1-2d}-\ref{phase-diag3-2d}). 


 The free energy $\cal F$  in Eq.~(\ref{free2}) for ZTE elastic media is constructed in such  a way that $\langle u_{ij}\rangle =0$ identically in the absence of externally applied stresses, ensuring vanishing thermal expansion. 
If we relax this condition, then additional terms of the form $\int d^dx \hat g_A A(\phi) u_{ii}$ can be added to ${\cal F}_{u\phi}$ in (\ref{f-int}) above, where $A(\phi)$ is a generic polynomial function of $\phi$, which would lead to thermal expansion $\propto g_A\langle A(\phi)\rangle$ (which vanishes automatically in the incompressible limit).  Such a term with $A(\phi)=A_0\phi + A_1 \phi^2+...$, being more relevant than the existing order parameter - elastic deformation anhamornic terms, can distablize the RG fixed points discussed here; see also Ref.~\cite{berg-halp}. These terms would then describe materials with finite thermal expansions. However, in nearly compressible systems, $u_{ii}$ is small, and there should be a sufficiently large scales over which the physics described here could be observed.

The stiffening of the system with weak selectivity at 2D holds close to the critical point only. Away from the critical point, all the fluctuation corrections are finite. These finite corrections to $\mu$ and $\tilde \lambda$ can however turn these elastic modulii negative, if the selectivity parameters are large enough. However, these instabilities no longer depend upon the system size. For weak selectivity, $\mu$ and $\tilde \lambda$ remain positive, but finite, leading to the standard QLRO. Thus phase diagrams in Fig.~\ref{phase-diag1-2d} and Fig.~\ref{phase-diag3-2d} still hold with the caveat that the positional order now refers to just QLRO. This picture holds in 3D as well.

 In general, thermal expansion could be controlled or significantly restricted by various means, e.g., by inclusion of additives or impurities with negative thermal expansion coefficient~\cite{thermal}, such that results predicted by theory could be observed over a large range of length-scales. Our theory is generic, and applicable to Ising transitions in any isotropic elastic medium, independent of its microscopic details. This theory can be tested in numerical simulations of appropriately constructed spin-lattice models near phase transitions in models with ZTE or the condition $dT_c/dV=0$, and also by performing controlled experiment on ZTE materials having phase transitions within the temperature range of ZTE behavior. Recent progress in the synthesis of two-component ZTE materials~\cite{two-comp} are promising developments in this direction. We expect future technological breakthrough will make it possible to design specific ZTE materials where our theory can be tested. Possible stiffening of ZTE materials near second order transitions may make such materials highly valuable in making precision engineering equipment.  We look forward to future research in these directions.

Our work can be extended in several ways. First of all, We have assumed an isotropic elastic medium. It would be interesting to study how anisotropy would affect our results. Then, to keep the theory simple, we have just considered a continuous medium coupled with Ising spins. More realistic situations, in particular, biologically relevant ones, may involve several lipids, requiring multiple order parameters. This can potentially give rise to multicritical points, or even simultaneous occurrence of first and second order transitions. The nature of any anomalous elasticity, and the displacement fluctuations in such  systems remain open questions. It will be interesting to study the dynamics of the fluctuations and the spatio-temporal scaling of the time-dependent correlation functions of the displacements near the second order transitions. Furthermore, extending these ideas to ``active'' or nonequilibrium systems should be important, with possible strong relevance to biological systems, e.g., phase separations in biological cells, or flocking phenomena in a correlated background due to a second order phase transition. Further work should be undertaken in this context.

\section{Acknowledgment} 

We thank T. Das for comments and suggestions. S.M. thanks 
the SERB, DST (India) for partial financial support through the TARE scheme [file no.: TAR/2021/000170]. AB thanks the SERB, DST (India) for
partial financial support through the MATRICS scheme
[file no.: MTR/2020/000406].

\appendix

\section{Glossary}\label{gloss}

In this glossary, we list, and give rough definitions for,  the symbols used in this paper, in the order in which they appear in the text. We also refer to the equations that precisely define them, or where they appear first in the text. 

\begin{itemize}
 \item $\tilde \cal F$: Total free energy of the system [Eq.~(\ref{free-tot1})].
 \item ${\cal F}_\phi$: Free energy for the Ising degrees of freedom [Eq.(\ref{f-phi})].
 \item $r=T-T_c$: Difference between the temperature $T$ (not to be confused with superscript $T$ in $\bf u^T$), and the mean-field critical temperature $T_c$ of the Ising degrees of freedom [Eq.(\ref{f-phi})].
 \item $v>0$: anharmonic coupling constant that couples $\phi$'s with itself [Eq.(\ref{f-phi})].
 \item ${\cal F}_u$: Free energy for the local displacements $u_i$ [Eq.~(\ref{f-u})].
 \item $\mu,\,\lambda$: Bare Lam\'{e} coefficients  for the in-plane elasticity of the system [Eq.~(\ref{f-u})].
 \item $u_{ij}=\frac{1}{2}\left(\nabla_i u_j + \nabla_j u_i\right)$ is the local strain [Eq.~(\ref{f-u})]. 
 \item ${\cal F}_{u\phi}$: free energy of interactions between the order parameter $\phi$ and displacement $u_i$ [Eq.~(\ref{f-int})]. 
 \item $\overline g_{10},\,\overline g_2$: anhamornic coupling constants which couple order parameter linearly with the local displacement, and {\em explicitly break} the inversion symmetry of $\phi$ in ${\cal F}_\phi$ [Eq.~(\ref{f-int})]. 
 \item $g_{10},\,g_2$: anhamornic coupling constants which couple order parameter quadratically with the local displacements, and {\em maintain} the inversion symmetry of $\phi$ in ${\cal F}_\phi$ [Eq.~(\ref{f-int})]. 
 \item ${\bf u}^L({\bf q}),\,{\bf u}^T({\bf q})$: Longitudinal and transverse components of the displacement ${\bf u}({\bf q})$, written in the Fourier space [Eq.~(\ref{ul-ut})].
 \item $L$: linear system size (not to be confused with superscript $L$ in ${\bf u}^L$).
 \item $a_0$: short distance cutoff [Eq.~(\ref{uT-sys})].
 \item $\Lambda=2\pi/a_0$: upper wavevector cutoff [Eq.~(\ref{uT-sys3})].
  \item $\tilde \lambda = \lambda + 2\mu$: effective elastic modulus [Eq.~(\ref{free2})].
 \item $C_v$: specific heat at constant volume [Eq.~(\ref{sp-heat})].
 \item $\epsilon=d-2$: small parameter in the RG calculation [Eq.~(\ref{g1-flow})].
 \item $\alpha_1=\frac{T_cg_{1}S_d}{(2\pi)^d\mu}\Lambda^\epsilon,\,\beta_1=\frac{T_c\overline g_1^2S_d}{(2\pi)^d\mu^2}\Lambda^\epsilon$: effective coupling constants [Eq.~(\ref{eff-coup1})].
 \item $\alpha_2=\frac{T_cg_{2}S_d}{(2\pi)^d\tilde\lambda}\Lambda^\epsilon,\,\beta_2=\frac{T_c\overline g_2^2S_d}{(2\pi)^d\tilde\lambda^2}\Lambda^\epsilon$: effective coupling constants [Eq.~(\ref{eff-coup2})].
 \item $\Gamma_{1c}$: slope of the separatrix in the $\alpha_1-\beta_1$-plane in 2D [Eq.~(\ref{gamma-val})].
 \item $\delta_1$: (small) deviation from the separatrix in the $\alpha_1-\beta_1$-plane in 2D [Eq.~(\ref{delta-eq})].
 \item $\tilde\lambda(q)$: renormalized wavevector-dependent elastic modulus [Eq.~(\ref{lambda-q})].
 \item $\xi_{NL}$: the length scale at which the anharmonic effects become important [Eq.~(\ref{nonlin-length})].
 \item $L_c$: position correlation length, or the length at which $\mu(L_c)\approx 0$ [Eq.~(\ref{pos-corr-len1})]. 
 \item $\Gamma_{2c}$: slope of the separatrix in the $\alpha_2-\beta_2$-plane in 2D [Eq.~(\ref{gamma-val22})].
 \item $\delta_2$: (small) deviation from the separatrix in the $\alpha_2-\beta_2$-plane in 3D [Eq.~(\ref{delta2-eq})].
 \item ${\cal \tilde F}_\phi$: Free energy of the Ising degrees of freedom that includes a $\phi^3$ and linear $\phi$ terms [Eq.~(\ref{extnd-f})].
 \item $g$: coefficient of the $\phi^3$-term in ${\cal\tilde F}_\phi$ [Eq.~(\ref{extnd-f})].
 \item $h$: thermodynamic conjugate to $\phi$ - ``magnetic field''; coefficient of the linear $\phi$-term in ${\cal\tilde F}_\phi$ [Eq.~(\ref{extnd-f})].
\end{itemize}

\section{Parameter estimates}\label{param-esti}

We begin by noting that both the elastic modulii have the dimensions of energy/length$^d\sim k_BT/{\rm length}^d$ in $d$ dimensions. Taking this length to be the small scale $\sim a_0$, which is the mesh size of a cross linked polymer network, or the lattice spacing for a crystal, we get
\begin{equation}
 \mu,\lambda \sim k_BT/a_0^d.
\end{equation}
We take $a_0\sim 60 nm$ for a spectrin network~\cite{takeuchi}. Typical shear modulus of a 2D spectrin network are $\mu\sim 10^{-5} J/m^2$~\cite{boal}; for an incompressible medium $\tilde\lambda\gg \mu$. 

We can now find out the dimensions of $g_{1,2}$ and $\overline g_{1,2}$. We start from the fact that
\begin{equation}
 \int d^dx g_{1,2}\phi^2(\nabla_i u_j)^2 \sim k_BT\sim \int d^dx \overline g_{1,2}\phi(\nabla_i u_j)^2,
\end{equation}
giving
\begin{equation}
 [g_{1,2}\phi^2]\sim k_BT/a_0^d\sim [\overline g_{1,2}\phi];
\end{equation}
where $[...]$ implies ``in a dimensional sense''. Now for a two-component system, if we assume $\phi$ to be the concentration or number density difference between the two components, then $[\phi]\sim 1/a^d$. In this case, the dimensions of $g_{1,2}$ differ from those of $\overline g_{1,2}$. However, if $\phi$ is a magnetic (Ising) spin, then $\phi$ may be chosen dimensionless. In this case, $g_{1,2}$ and $\overline g_{1,2}$ have the same dimensions. 


\section{Free energy}\label{free-en}

We first derive the free energy $\cal F$ in (\ref{free2}). We split
\begin{equation}
 u_i({\bf q}) = P_{ij}({\bf q})u_j({\bf q}) + Q_{ij}({\bf q})u_j({\bf q})= u^T_i ({\bf q}) + u^L_i ({\bf q}).
\end{equation}
Next we note that
\begin{eqnarray}
 &&\int d^dx u_i^T({\bf x})\,u_i^L({\bf x})= \int \frac{d^dq}{(2\pi)^d} u_i({\bf -q})^T\,u_i^{L}({\bf q})\nonumber \\ &=& \int \frac{d^dq}{(2\pi)^d} P_{ij}({\bf q})Q_{im}({\bf q}) u_j^T({\bf -q})u_n^T({\bf q}) =0,
\end{eqnarray}
where we have used $P_{ij}({\bf q})Q_{ij}({\bf q})=0$. Furthermore,
\begin{eqnarray}
 &&\int d^dx [u_{ii}({\bf x})]^2 = \int d^dx (\partial_i u_j^L({\bf x}))(\partial_j u_i^L({\bf x}))\nonumber \\&=&\int d^dx (\partial_i u_j^L({\bf x}))(\partial_i u_j^L({\bf x}))\nonumber \\&=&\int d^dx (\partial_i u_j^L({\bf x}))^2.
\end{eqnarray}

Now use that
\begin{eqnarray}
 &&\int d^dx \left(\nabla_i u_j\right)^2 =\int \frac{d^dq}{(2\pi)^d} q^2 {\bf u}({\bf q})\cdot {\bf u}({\bf -q})\nonumber \\ &=& \int \frac{d^dq}{(2\pi)^d} q^2 \left[ {\bf u}^L({\bf q})\cdot {\bf u}^L({\bf -q}) + {\bf u}^T({\bf q})\cdot {\bf u}^T({\bf -q})\right]\nonumber \\
 &=& \int d^dx \left[\left(\nabla_i u^L_j\right)^2 + \left(\nabla_i u_j^T\right)^2\right].
\end{eqnarray}
Substituting this decomposition, we get (\ref{free2}), and also (\ref{free-gauss}) at the harmonic order. 


\section{RG calculation}\label{rg-calc}

\subsection{Upper critical dimensions}\label{upp-dim}

To determine the upper critical dimensions of the various anhamornic terms,
we rescale space and obtain the corresponding scaling of the model parameters. We rescale space and the fields as follows:
\begin{eqnarray}
 &&{\bf x}'=\frac{\bf x}{b},\\
 &&u_i({\bf x})=\zeta_u u({\bf x}')=\zeta_u u_i({\bf x}/b),\\
 &&\phi({\bf x})=\zeta_\phi\phi({\bf x}')=\zeta_\phi \phi({\bf x}/b).
\end{eqnarray}
These rescaling factors may be calculated by demanding that under na\"ive rescaling, bare $\mu,\,\tilde\lambda$ do not scale, and the coefficient of the term $\int d^d x ({\boldsymbol\nabla}\phi)^2$ remains unity under rescaling. This gives
\begin{equation}
 \zeta_u=\zeta_\phi = b^{1-d/2}.\label{rescaling-fac}
\end{equation}
We can use (\ref{rescaling-fac}) to obtain how the anhamornic coupling constants change under na\"ive rescaling. We find
\begin{equation}
 u'=b^{4-d}u,\,g_a'=b^{2-d}g_a,\,\overline g_a'=b^{2-d}\overline g_a,
\end{equation}
$a=1,2$. Thus, critical dimension of $u$ is 4, and the critical dimension of $g_1,g_2,\overline g_1$ and $\overline g_2$ is 2.

\subsection{Feynman diagrams}\label{feyn}

The one-loop integrals for $\mu,\,\tilde\lambda,\,\,g_1,\,\overline g_1,\,g_2$ and $\overline g_2$ are all evaluated at $T=T_c$.


 We first consider the one-loop Feynman graphs in Fig.~\ref{mu-diag} that renormalize $\mu$.
 
\begin{figure}[htb]
 \includegraphics[width=6cm]{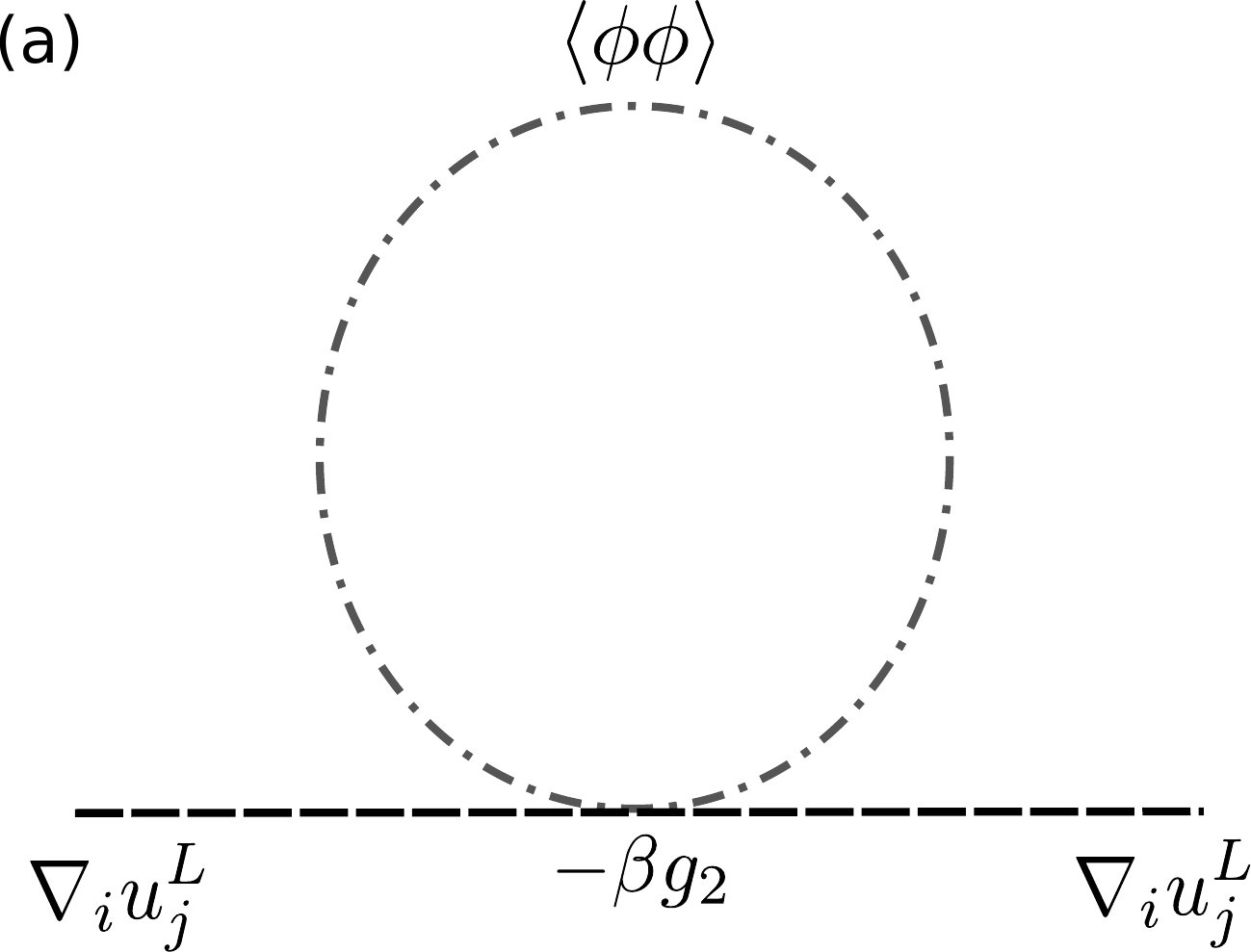}\vspace{.5cm}
  \includegraphics[width=7cm]{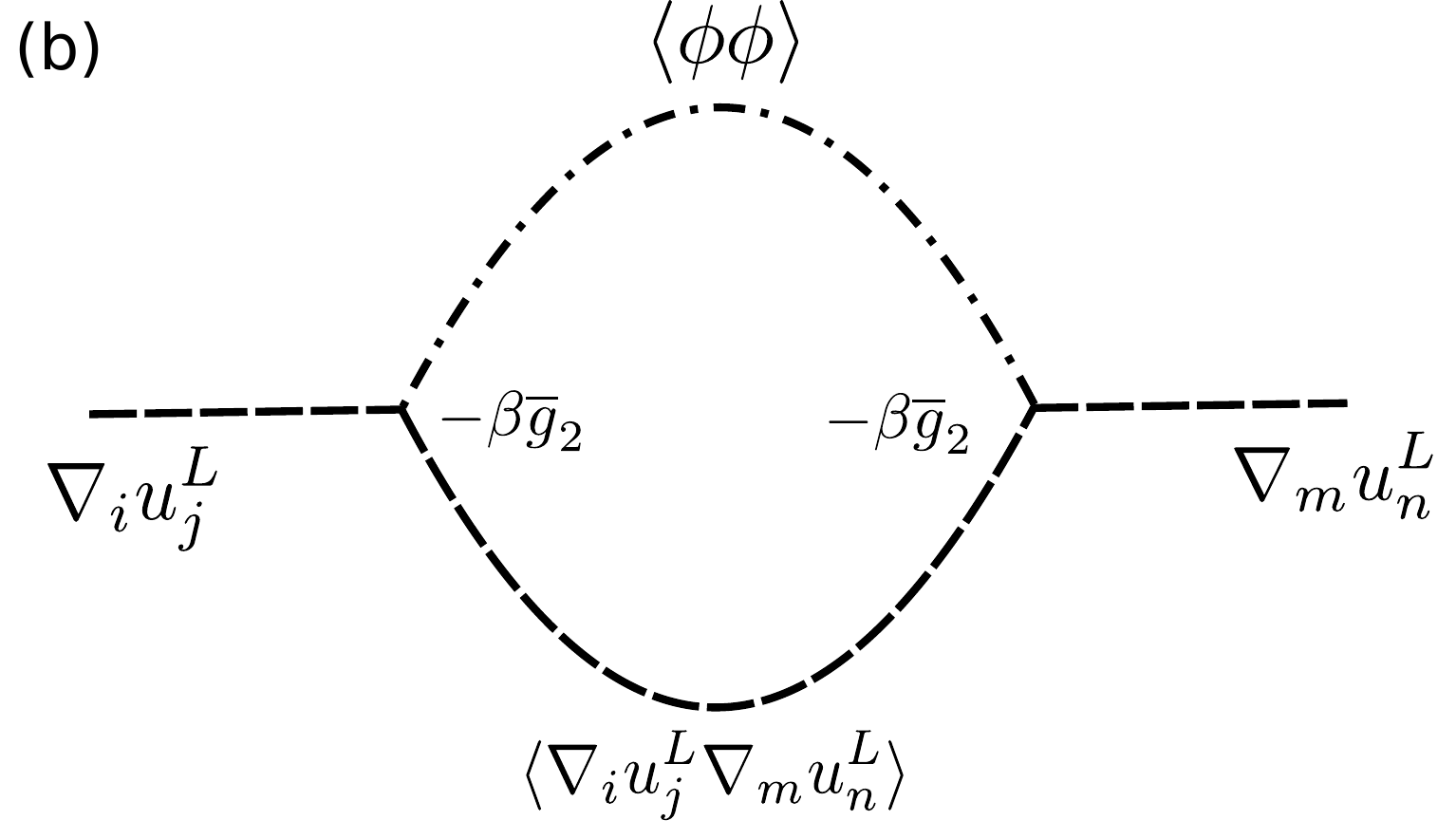}
 \caption{One-loop diagrams that contribute to the fluctuation corrections of $\tilde\lambda$. Diagram (a) comes from the nonlinear coupling $g_2$, where as diagram (b) comes from $\overline g_2$ (see text).}\label{lam-diag}
\end{figure}

Diagram~\ref{mu-diag}(a) has the value 
\begin{equation}
 -g_1 \int_{\Lambda/b}^\Lambda \frac{d^dq}{(2\pi)^d}\langle |\phi({\bf q})|^2\rangle= -g_1\langle\phi^{>2}({\bf x})\rangle=-g_1\frac{T_c}{2\mu}\delta l
\end{equation}
in all dimensions. 
Similarly, the diagram (\ref{mu-diag}(b)) contributes
\begin{equation}
 \overline g_1^2\frac{T_c}{2\mu}\int_{\Lambda/b}^\Lambda \frac{d^dq}{(2\pi)^d}\langle |\phi({\bf q})|^2\rangle = \overline g_1^2\frac{T_c^2}{2\mu}\delta l,
\end{equation}
in all dimensions.
Similarly, evaluating the diagrams for $\mu$ in Fig.~\ref{mu-diag} above, we obtain by combining both the corrections
\begin{equation}
 \mu^<=\mu + g_1 - \frac{T_c\overline g_1^2}{2\mu}.
\end{equation}
The one-loop diagrams which correct $\tilde\lambda$ are given in Fig.~\ref{lam-diag}.
 
 Proceeding as before, we obtain
 \begin{equation}
  \tilde\lambda^< = \tilde\lambda + 2g_2 - \frac{2T_c\overline g_2^2}{\tilde\lambda}.
 \end{equation}

 We now consider the one-loop fluctuation corrections to $g_1$ and $\overline g_1$. We obtain
 \begin{eqnarray}
  g_1^<&=&g_1 - \left(\frac{4 T_c g_1^2}{d\mu^2}+\frac{\overline g_1^4}{4d\mu^3}\right) \langle\phi^2({\bf x})^>\rangle,\\
  \overline g_1^< &=& \overline g_1 +\left( \frac{\overline g_1^3}{d\mu^2}-\frac{4g_1\overline g_1}{\mu}\right)\langle\phi^2({\bf x})^>\rangle.
 \end{eqnarray}
Using the expression for $\langle\phi^2({\bf x})\rangle$, $g_1^<,\,\overline g_1^<$ can be calculated.

\begin{figure}[htb]
 \includegraphics[width=7cm]{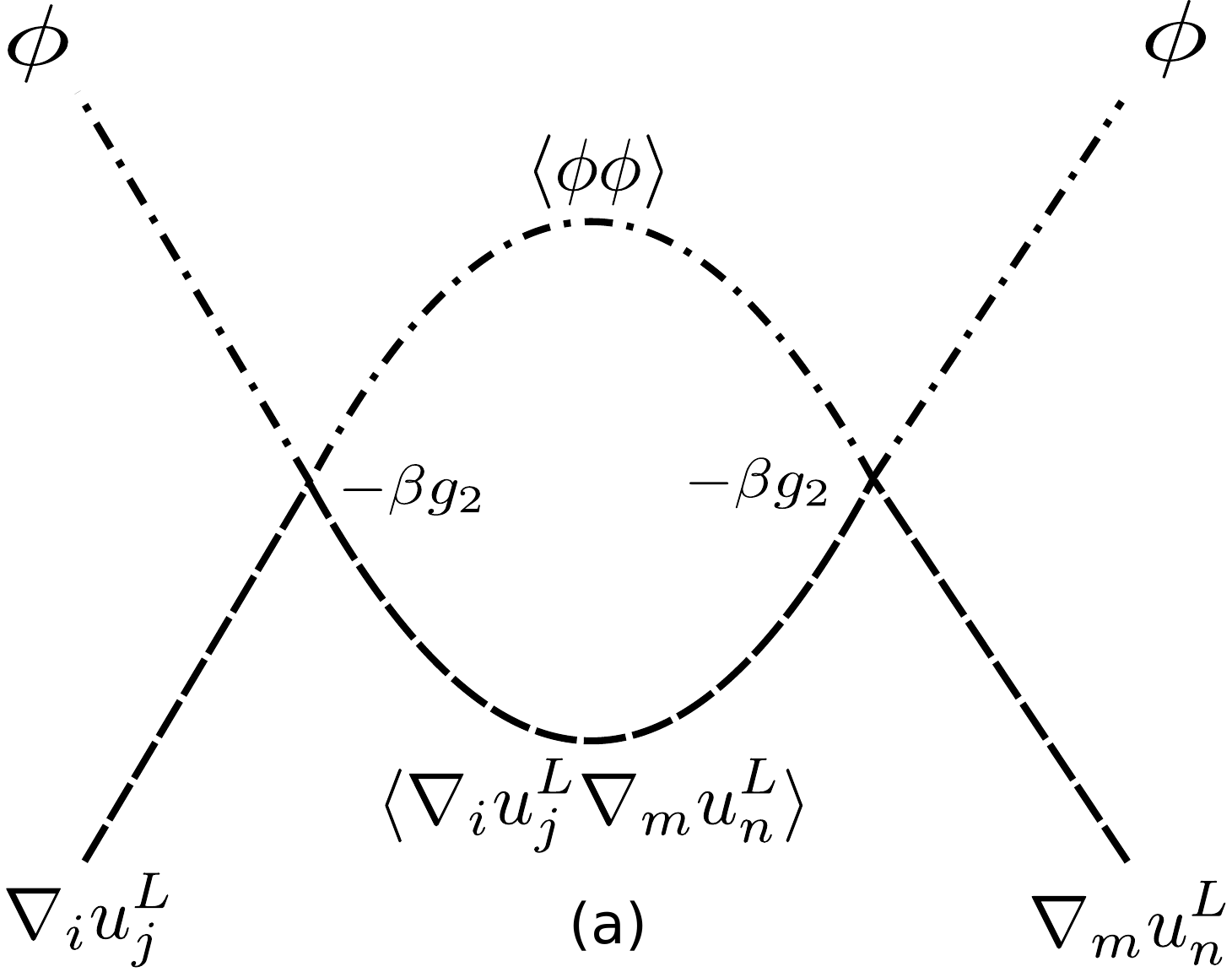}\\
 \includegraphics[width=7cm]{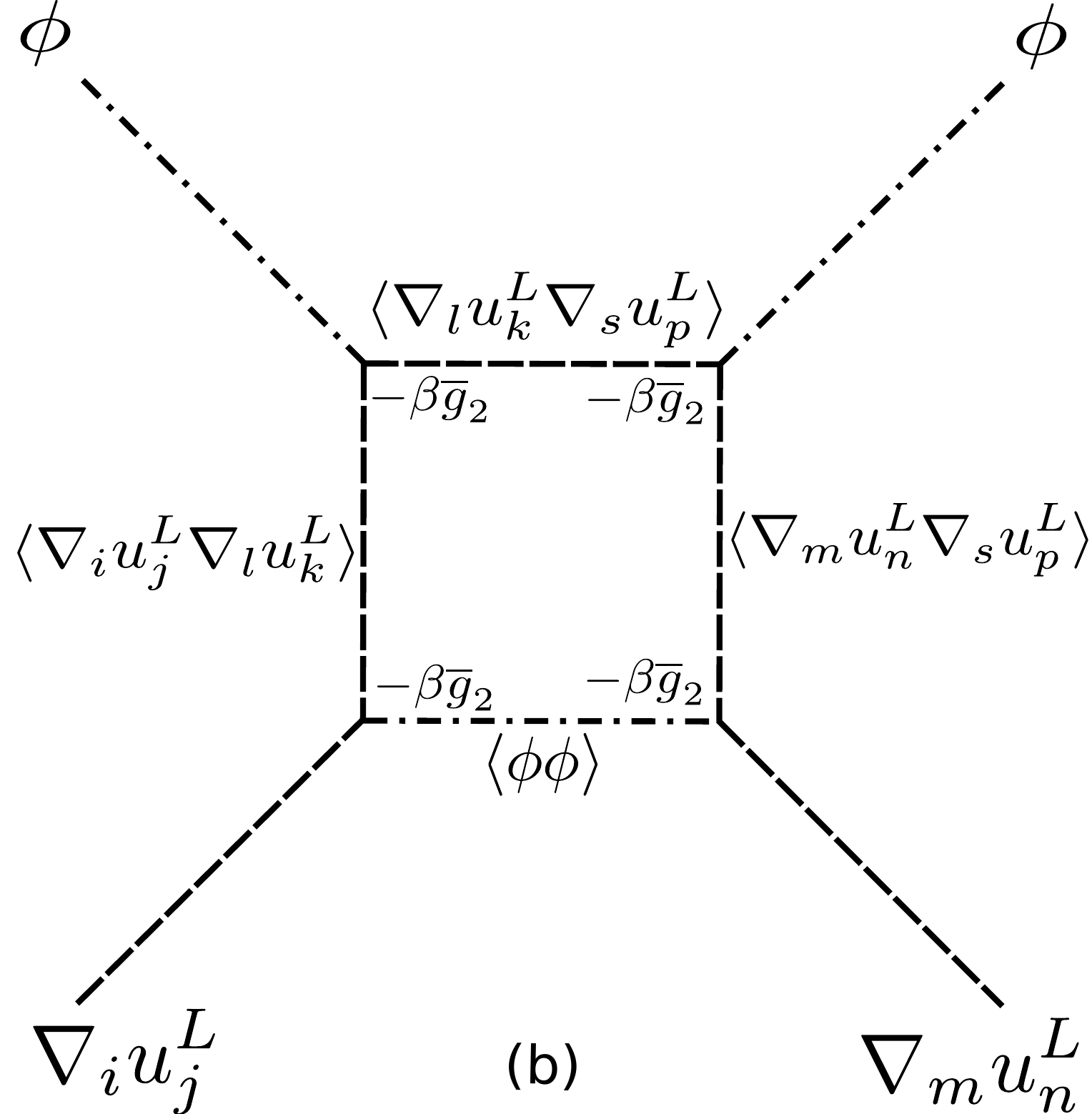}
 \caption{One-loop diagrams that contribute to the fluctuation corrections of $ g_2$. Diagram (a) depends only on $g_2$, whereas diagram (b) depends only on $\overline g_2$.}
 \label{vertex-diag-11}
\end{figure}

\begin{figure}[htb]
 \includegraphics[width=7cm]{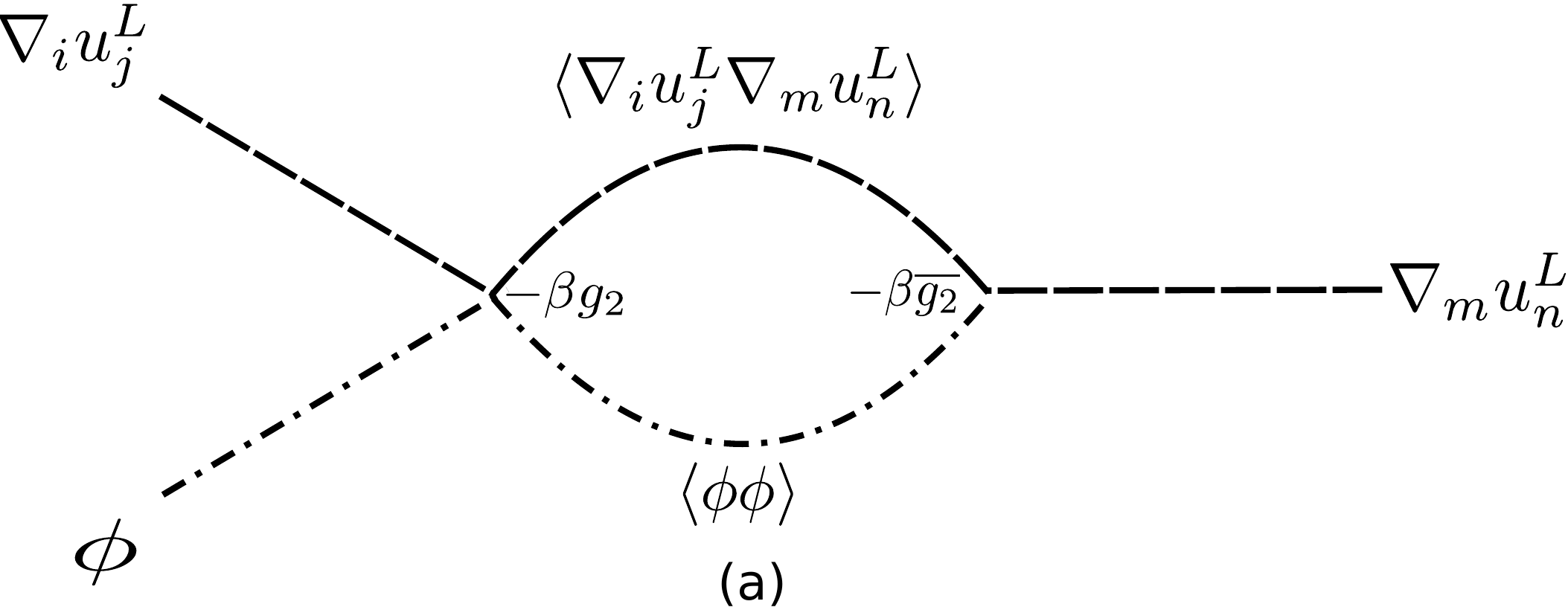}\\
 \includegraphics[width=7cm]{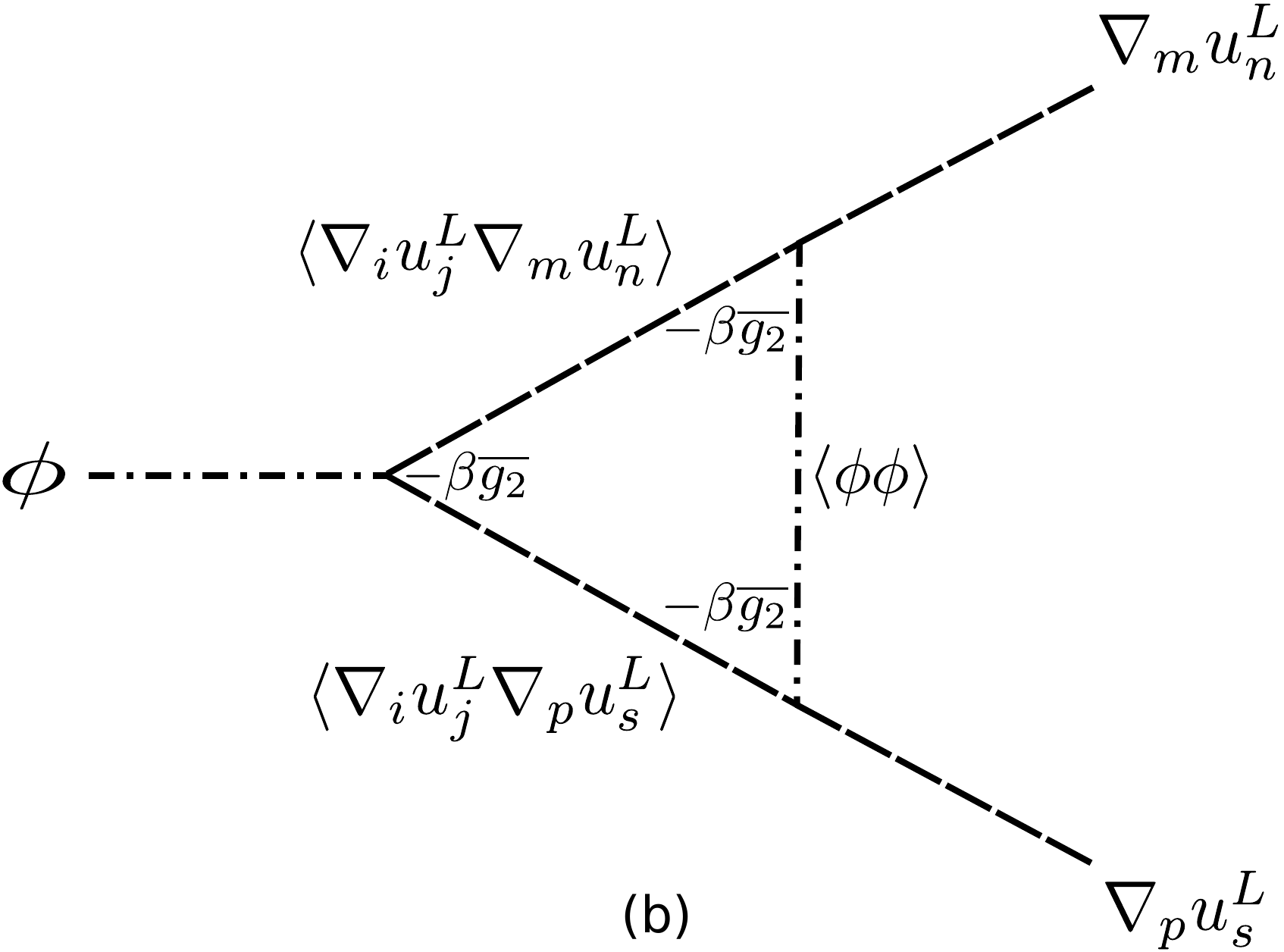}
 \caption{One-loop diagrams that contribute to the fluctuation corrections of $\overline g_2$. Diagram (a) depends both $g_2$ and $\overline g_2$, whereas diagram (b) depends only on $\overline g_2$.}\label{vertex-diag-22}
\end{figure}

The one-loop diagrams that correct $g_2$ and $\overline g_2$ are shown in Fig.~\ref{vertex-diag-11} and Fig.~\ref{vertex-diag-22}.

Having shown that all the one-loop corrections are proportional to $\langle \phi^{>2}({\bf x})\rangle$, we are now obliged to discuss the evaluation of $\langle \phi^{>2}({\bf x})\rangle$.

As we have argued above, at 2D $\langle \phi^2({\bf x})\rangle \sim T_c\ln |(T-T_c)/T_c|$ as $T\rightarrow T_c$. Using $\xi\sim \left[(T-T_c)/T_c\right]^{-\nu}$, we get $\langle \phi^2({\bf x})\rangle \sim T_c \ln\xi/a_0\, \times {\cal O}(1)$. This gives
\begin{equation}
 \langle\phi^{>2} ({\bf x})\rangle \sim T_c \ln b \times {\cal O}(1).
\end{equation}
Thus the contribution from diagram (\ref{mu-diag}(a)) reads
\begin{equation}
  -g_1 T_c\ln b 
 \approx -g_1 T_c\delta l\label{diag-lam1}
\end{equation}
in 2D,
where $b=\exp(\delta\ell)\approx 1+\delta l$ for small $\delta l$.
On the other hand, at $d>2$, proceeding similarly,
\begin{equation}
 \langle\phi^{>2} ({\bf x})\rangle \sim T_c \frac{\Lambda^{-\alpha+1}}{-\alpha+1}[1-b^{-\alpha+1}]\approx T_c \delta l.
\end{equation}
Thus the contribution from the diagram (\ref{mu-diag}(a)) at $d>$2 again reads
\begin{equation}
  -g_1 T_c\delta l. \label{diag-lam2}
\end{equation}
In each of (\ref{diag-lam1}) and (\ref{diag-lam2}), we have absorbed ${\cal O}(1)$ constants, that arises in the evaluation of the diagrams, into the definitions of $g_1,\,g_2,\,\overline g_1$ and $\overline g_2$ without any loss of generality. 



\subsection{Variances $\langle (\nabla_i u_j^T)^2\rangle$ and $\langle (\nabla_i u_j^L)^2\rangle$ in 2D}~\label{vari-2d-re}

We now recalculate $\langle (\nabla_i u_j^T)^2\rangle$ and $\langle (\nabla_i u_j^L)^2\rangle$ by using the forms of the renormalized propagators for $u_i^T({\bf q})$ and $u_i^L({\bf q})$ that is valid up to an upper wavevector limit $\Lambda$.
We note that the form of $\langle |{\bf u}^T({\bf q})|^2\rangle$ valid up to an upper wavevector limit $\Lambda$ should read
\begin{eqnarray}
 \langle |{\bf u}^T({\bf q)}|^2\rangle&\approx&  \frac{T_c}{4\pi}[\mu_R \ln (\Lambda/q)]^{2/3} q^2]^{-1},\,q< 1/\xi_{NL},\nonumber \\
 \langle |{\bf u}^T({\bf q)}|^2\rangle&\approx&  \frac{T_c}{4\pi\mu q^2},\,1/\xi_{NL}<q< \Lambda.
 \label{full-uT-corr}
\end{eqnarray}
Inverse Fourier transform of (\ref{full-uT-corr}) gives
\begin{eqnarray}
 &&\langle (\mathbf{u}^T({\bf x}))^2 \rangle = \int_{2\pi/L}^{\Lambda} \frac{d^2q}{(2\pi)^2} \langle |{\bf u}^T({\bf q})|^2\rangle \nonumber \\&=& \frac{T_c}{2\mu_R}\left[\int_{2\pi/L}^{C'} + \int_{C'}^\Lambda\right]\frac{d^2q}{(2\pi)^2}\langle |{\bf u}^T({\bf q})|^2\rangle\nonumber \\&=&\frac{T_c}{2\mu_R} \int_{2\pi/L}^{C'} \frac{d^2q}{(2\pi)^2} \frac{1}{[q^2\{\ln (\Lambda/q)\}^{1/3}]} \nonumber \\&+& \int_{C'}^\Lambda\frac{d^2q}{(2\pi)^2}\langle |{\bf u}^T({\bf q})|^2\rangle,\label{vari-u1}
\end{eqnarray}
where $C'<1/\xi_{NL},\,C'\sim {\cal O}(1)$ is such that in the range $2\pi/L <q< C'$,  $\langle |{\bf u}^T({\bf q})|^2\rangle$ is well approximated by $(T_c/(2\mu_R))1/[q^2\{\ln|\Lambda/q|\}^{1/3}]$. This integral scales with $L$ as $[\ln (C'L)]^{2/3}$. The remaining integral is independent of $L$. We thus conclude that
\begin{equation}
 \langle (\mathbf{u}^T({\bf x}))^2 \rangle = \frac{T_c}{4\pi\mu_0}[\ln (C'L)]^{2/3} + {\cal O}(1)\approx \frac{T_c}{4\pi\mu_0}[\ln (C'L)]^{2/3},
\end{equation}
for large $L$, same as what we have obtained above.

\subsection{Correlation functions in 2D}\label{corr-2d}

We are interested in calculating the correlation functions of $u_i({\bf x})$, defined as
\begin{equation}
 C^{ a}_{uu}( r)\equiv \langle [u^{ a}_i({\bf x})-u^{ a}_i({\bf x'})]^2\rangle,
\end{equation}
where $ a=L$ or $T$, corresponding to $u^L_i$ or $u^T_i$, in 2D near $T=T_c$; $\tilde r=|{\bf x-x}'|$. We first revisit the correlator in the Fourier space in the harmonic theory, in which at $T_c$
\begin{equation}
 C^{ a}_{uu0}(k)\equiv\langle u^{ a}_i({\bf k})u^{ a}_i({\bf -k})\rangle = \frac{T_c}{\tilde a k^2},\label{har-k}
\end{equation}
where $ \tilde a=2\mu,\,\tilde\lambda$ for $a=T,\,L$, respectively, for the transverse and longitudinal components of ${\bf u}({\bf x})$; a subscript ``0'' refers to $C_{uu}(\tilde r)$ being evaluated in the Gaussian theory, i.e., after setting all the anhamornic couplings to zero. Note that (\ref{har-k}) holds at all $T$. Inverse Fourier transform of (\ref{har-k}) gives the correlation function $C^a_{uu}(\tilde r)$ in the real space in the harmonic theory:
\begin{eqnarray}
 &&C^a_{uu0}(\tilde r)=2\int_0^\Lambda \frac{d^2k}{(2\pi)^2} \left[1-\exp i{\bf k}\cdot ({\bf x-x}')\right]\frac{T_c}{\tilde a k^2}\nonumber \\ &&= \frac{2T_c}{(2\pi)^2} \int_0^\Lambda \frac{dk}{\tilde a k}\int_0^{2\pi}\left(1-\exp [ik \tilde r\cos\theta]\right)\nonumber \\ &&= \frac{2T_c}{(2\pi)^2} \int_0^1 \frac{dq}{\tilde a k} \int_0^{2\pi}\left(1-\exp [iq y\cos\theta]\right) \equiv I_0(y),\nonumber\\
\end{eqnarray}
where $q=k/\Lambda$ and $y=\Lambda \tilde r$. Then,
\begin{eqnarray}
 \frac{dI_0}{dy}&=&\frac{-2iT_c}{\tilde a (2\pi)^2} \int_0^1 dq \int_0^{2\pi} \cos\theta \exp[iqy\cos\theta]\,d\theta\nonumber \\&=& \frac{-2iT_c}{\tilde a y(2\pi)^2}\int_0^y du \int_0^{2\pi} d\theta\,\cos\theta \exp[iqy\cos\theta]\nonumber \\
 &=& \frac{-2iT_c}{\tilde a y(2\pi)^2}\int_0^{2\pi} d\theta\,\cos\theta\int_0^{\infty} du \exp[iqy\cos\theta]\nonumber \\
 &=& \frac{-2iT_c}{\tilde a y(2\pi)^2}\int_0^{2\pi} d\theta\,\cos\theta \left[\pi\delta(\cos\theta) + i {\cal P} \left(\frac{1}{\cos\theta}\right)\right],\nonumber \\
\end{eqnarray}
in the limit $\tilde r\rightarrow\infty$.
Now, $\delta(\cos\theta)$ is even under $\theta \rightarrow \theta+\pi$, but $\cos\theta$ is odd under the same. Hence, the contribution from the $\delta(\cos\theta)$ part of the integrand vanishes. In contrast, ${\cal P} \left(\frac{1}{\cos\theta}\right)$ is odd, and hence the corresponding contribution survives. Therefore,
\begin{eqnarray}
 \frac{dI_0}{dy}&=&\frac{2T_c}{\tilde a (2\pi)^2} \int_0^{2\pi} d\theta\,\cos\theta\,{\cal P} \left(\frac{1}{\cos\theta}\right)\nonumber \\&=& \frac{T_c}{\tilde a\pi y}.
\end{eqnarray}
Therefore, 
\begin{equation}
C^a_{uu0}(\tilde r)=\frac{T}{\pi\tilde a}\ln (\Lambda\,\tilde r), 
\end{equation}
for $\tilde r\rightarrow \infty$, giving QLRO. As expected, this is valid at all temperature $T$.

We now calculate $C^a_{uu}(\tilde r)$ in the anhamornic, renormalized theory at $T=T_c$. We start from 
\begin{equation}
  \langle u^a_i({\bf k})u^a_i({\bf -k})\rangle \approx \frac{T_c}{a_R k^2|\ln (\Lambda/k)|^{1/3}},\label{anhar-k}
\end{equation}
where $a_R=2\mu_R,\,\tilde\lambda_R$.
Expression (\ref{anhar-k}) is no longer valid over the wavevector range from 0 to $\Lambda$, rather it is valid between 0 and $\tilde\Lambda\ll \Lambda$.

The renormalized correlation function in the real space is then given as
\begin{equation}
  C^a_{uu}(r)\approx\int_0^\Lambda \frac{d^2k}{(2\pi)^2}\left[1- \exp i{\bf k}\cdot ({\bf x-x}')\right]\frac{2T_c}{ a_R k^2 [\ln (\Lambda/k)]^{1/3}}.\label{anhar-r}
\end{equation}
Integrating over the angular variable, we get
\begin{eqnarray}
 C^a_{uu}(r)&\approx&\int_0^{\tilde\Lambda}\frac{dq\,2T_c}{a_R q|\ln(q/\Lambda)|^{1/3}}\left[\frac{1}{2\pi} \int_0^{2\pi} d\theta (1-e^{iqr\cos\theta}\right]\nonumber \\
 &=& \int_0^{\tilde\Lambda}\frac{dq\,2T_c}{a_Rq|\ln(q/\Lambda)|^{1/3}}\left[1-J_0(qr)\right]\nonumber \\
 &=& \int_0^{\tilde\Lambda r}\frac{du\,2T_c[1-J_0(u)]}{a_R u |\ln (\frac{u}{x\Lambda})|^{1/3}},
 \end{eqnarray}
where $J_0(u)$ is the Bessel function of order zero. Then

\begin{widetext}
 
 \begin{equation}
  C^a_{uu}(r)=\int_0^1\frac{du\,2T_c[1-J_0(u)]}{a_Ru[\ln u + \ln (1/y)]^{1/3}}+\int_1^{\tilde\Lambda r}\frac{du\,2T_c}{a_Ru[\ln u + \ln (1/(\Lambda r))]^{1/3}} - \int_1^{\tilde\Lambda r} \frac{du\,2T_cJ_0(u)}{a_Ru[\ln u + \ln (1/(\Lambda r))]^{1/3}}. \label{inter}
 \end{equation}
 Since $u_{max}=\tilde\Lambda r \ll \Lambda r$, the second contribution on the right may be evaluated by setting $u=\exp(z)$. This gives 
 \begin{equation}
  \int_1^{\tilde\Lambda\, r}\frac{du}{a_Ru[\ln (\Lambda\,r)]^{1/3}} \approx \frac{2}{3} [\ln (\Lambda r)]^{2/3} + \text{const.}
 \end{equation}
 
 \end{widetext}
 We thus find $C^a_{uu}(r)\approx \frac{T_c}{a_R}|\ln(\Lambda\,r)|^{2/3}$ in the limit of large $r$, with the remaining contributions on the right hand side of (\ref{inter}) being finite or subleading for large $r$. By using the above procedure, we recover the scaling of the harmonic theory or QLRO. We find that $C^a_{uu}(r)(r)$ eventually does diverge in the thermodynamic limit, but does so much slower than the corresponding result with QLRO: 
 \begin{equation}
 C^a_{uu}(r)/C^a_{0uu}(r)\rightarrow 0
 \end{equation}
 for large $r$. Naturally, we call this positional order stronger than QLRO (SQLRO).

\section{Positional correlation length in the perturbation theory} \label{pos-corr}

We can obtain the position correlation length from the fluctuation corrected $\mu$ or $\tilde\lambda$ from the one-loop bare perturbation theory. Let us first focus on 2D. The calculation is essentially same as the RG calculations in Appendix \ref{rg-calc} above, except that we now extend the integrals over wavevectors down to an infra-red cut-off $q_{min}\equiv 2\pi/L$, where $L$ is the length-scale at which we intend to calculate the effective values $\mu_e$ and $\tilde\lambda_e$, respectively, of $\mu$ and $\tilde\lambda$. Evaluating the leading order (i.e., one-loop) perturbative corrections to $\mu$ coming from wavevectors $q>2\pi/L$, we obtain at 2D
\begin{equation}
 \mu_e (L) = \mu + \left(g_{1}-\frac{\overline g_{1}^2}{2\mu}\right)\frac{T_c}{2\pi}\ln (L/a_0),\label{eff-mu}
\end{equation}
where as usual $g_{1},\,\overline g_{1}$ refer to the bare or unrenormalised parameters used in the Free energy $\cal F$ in (\ref{free2}). A similar perturbative expansion for $\tilde\lambda_e$ may be written.
Evidently, for a sufficiently large $L=L_c$, $\mu_e$ can be made zero. We get the following relation for $\mu_e (L=L_c)=0$:
\begin{equation}
 \mu = \left(\frac{\overline g_{1}^2}{2\mu} - g_{1}\right) \frac{T_c}{2\pi}\ln (L_c/a_0).\label{phase-2d1}
\end{equation}
\begin{equation}
 L_c=a_0\exp\left[\frac{\mu}{\left(\frac{\overline g_{1}^2}{2\mu} - g_{1}\right)} \frac{2\pi}{T_c}\right].\label{phase-2d}
\end{equation}
Thus, as $g_{1}\rightarrow \frac{\overline g_{1}^2}{2\mu}$ from below, $L_c\rightarrow \infty$. for $L<L_c$, the system remains stable with positional order. This critical length $L_c$, being the linear size of the system such that $\mu_e(L_c)=0$, can be identified with a persistence length or positional correlation length $\xi$: As $L$ exceeds $\xi$, positional correlations are lost. 
When $L_c$ is plotted as a function of $\overline g_1^2$ for a given $\mu$, the phase diagram in Fig.~\ref{phase-diag2-2d}  in the $\overline g_1^2-L$ plane is obtained. 

We are interested in the parameter regimes where $\mu_e,\,\tilde\lambda_e>0$ (our perturbation theory becomes meaningless outside this regime). Clearly, if 
\begin{equation}
 g_{1}-\frac{\overline g_{1}^2}{2\mu}>0,
\end{equation}
we have $\mu_e>\mu$, i.e., stiffening of the shear modulus. This corresponds to the stable side of the separatrix (\ref{sep-2d}) in the RG calculations. On the other hand, if
\begin{equation}
 g_{1}-\frac{\overline g_{1}^2}{2\mu}<0,
\end{equation}
$\mu_e<\mu$ necessarily. Thus, 
\begin{equation}
  \frac{\overline g_{1}^2}{\mu g_1}=2 \label{border1}
\end{equation}
is the borderline of stability. Note that this essentially same as (\ref{border}) above; the slight difference in the rhs of the two is attributed to the quantitative difference between the renormalized perturbation theory and ordinary perturbation theory. Nonetheless, (\ref{border1}) produces the same phase diagrams as (\ref{phase-diag1-2d}) and (\ref{phase-diag3-2d}).

 Similar analysis for $\tilde\lambda_e$ yields an exactly analogous expression for a critical size for the longitudinal modes. See phase diagrams in Fig.~\ref{phase-diag2-2d} and Fig.~\ref{phase-diag3-2d}.

We now consider the three-dimensional case. An equation that is a direct analogue of (\ref{eff-mu}) can be written as below. We note that at 3D, the leading order corrections, i.e., the relevant integrals, can be performed all the way down to wavevector $q=0$ without encountering any divergence from the lower limits of the integrals. This yields 
\begin{equation}
 \mu_e = \mu + \left(g_{1}-\frac{\overline g_{1}^2}{2\mu}\right)\frac{T_c\Lambda}{2\pi}.\label{3deff-mu}
\end{equation}
Similar to the 2D case, if $g_{1}-\frac{\overline g_{1}^2}{2\mu}>0$, $\mu_e>\mu$; else $\mu_e<\mu$. In that latter case, $\mu_e=0$ sets the threshold for instability and breakdown of the positional order. We find
\begin{equation}
 \mu=\left(\frac{\overline g_{1}^2}{2\mu_0}-g_{1}\right) \frac{T_c\Lambda}{2\pi}.\label{3d-inst}
\end{equation}
Unlike its 2D counterpart, (\ref{3d-inst}) is independent of $L$, i.e., the instability sets in at all scales simultaneously; see Fig.~\ref{phase-diag3-2d}. In fact, for all positive (negative) $g_{1}-\frac{\overline g_{1}^2}{2\mu}$, the positional correlation length $L_c$ diverges (vanishes). 

\section{First order transition}\label{first-order-diag}

\begin{widetext}

Consider the following diagrams that contribute to the one-loop corrections to $v$.

\begin{figure}[htb]
 \includegraphics[width=7cm]{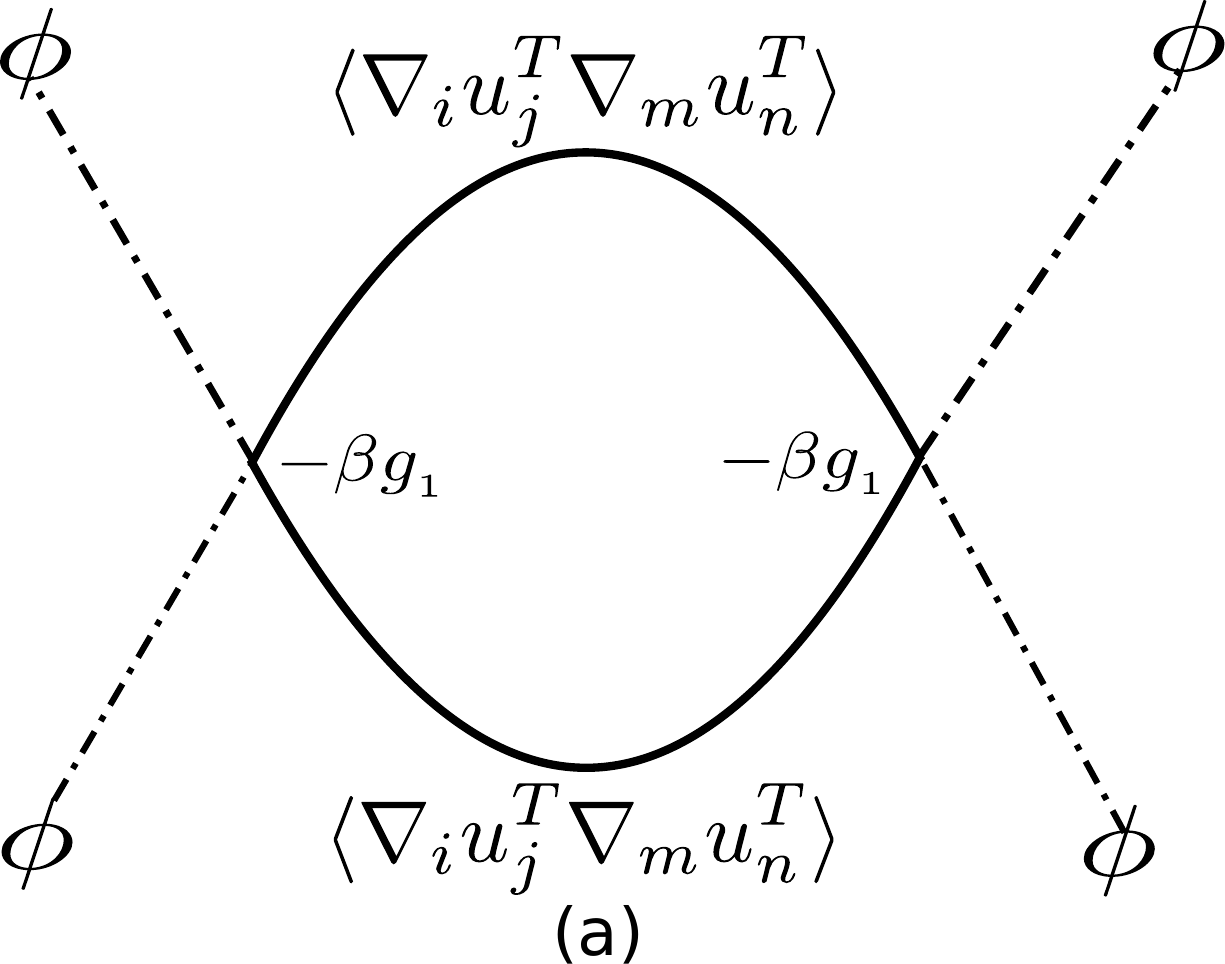}\hfill
 \includegraphics[width=7cm]{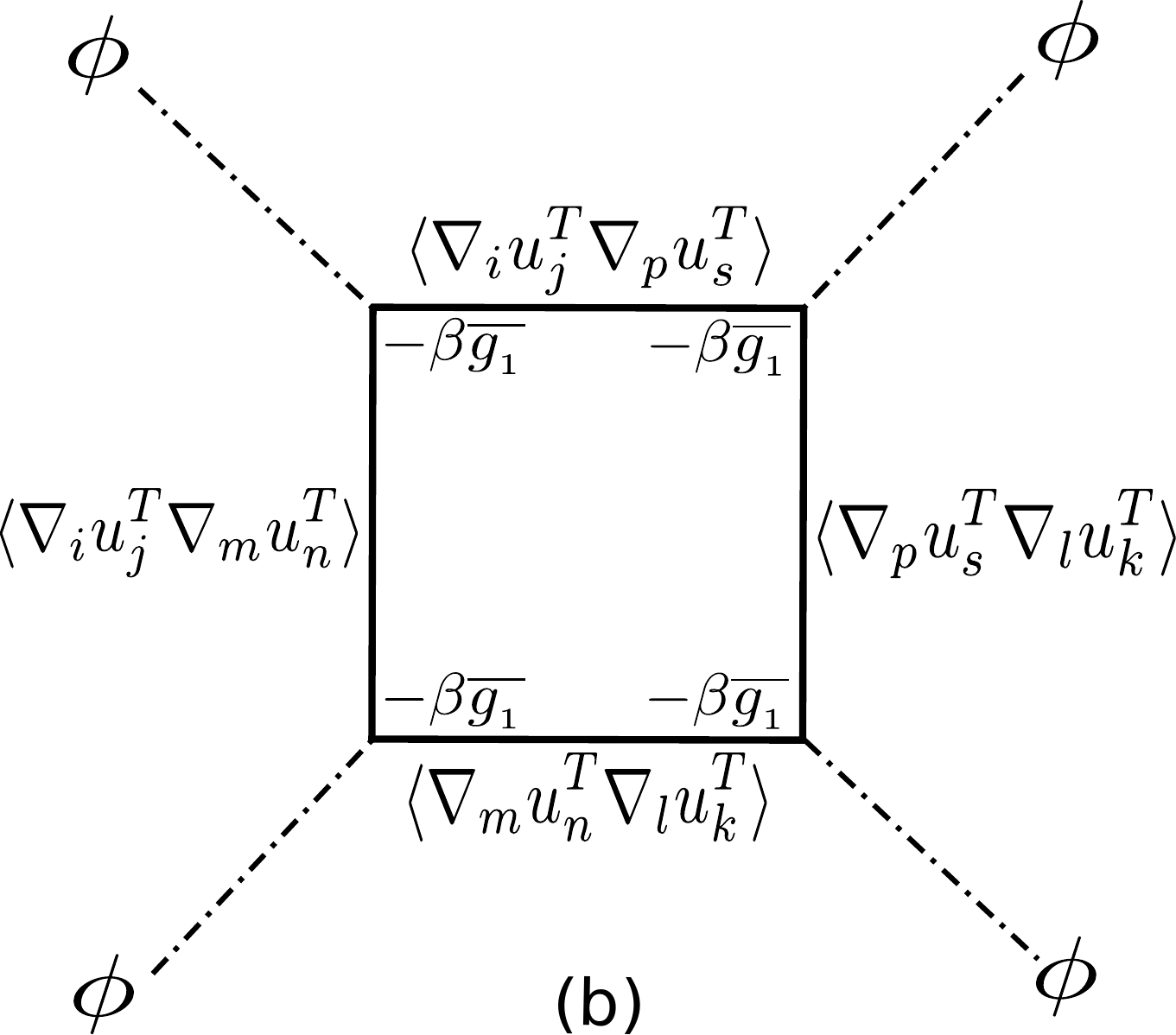}\\
  \includegraphics[width=7cm]{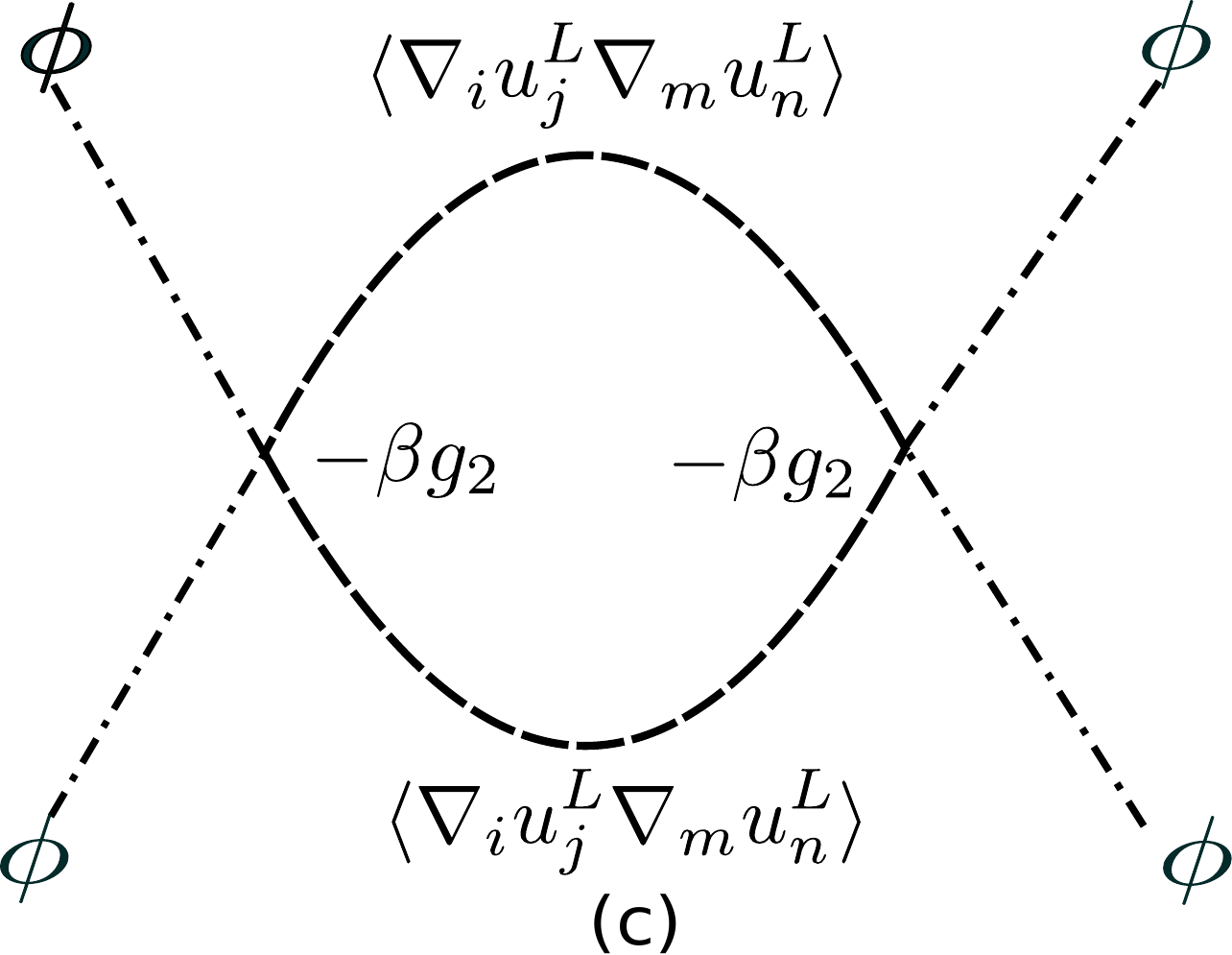}\hfill
 \includegraphics[width=7cm]{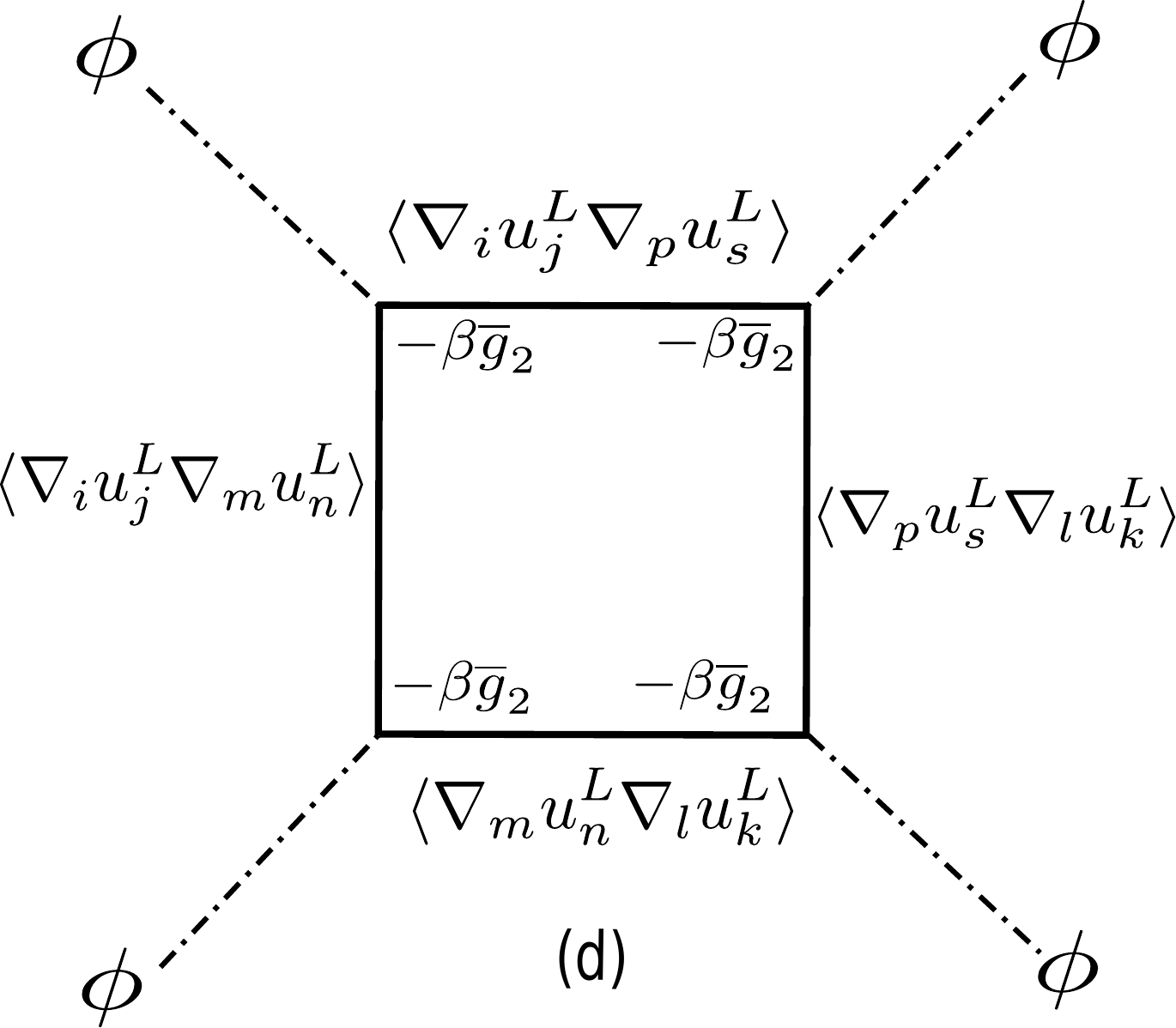}
 \caption{One-loop inhomogeneous Feynman graphs that correct $v$.}\label{inhom-v}
\end{figure}

\begin{figure}[htb]
 \includegraphics[width=7cm]{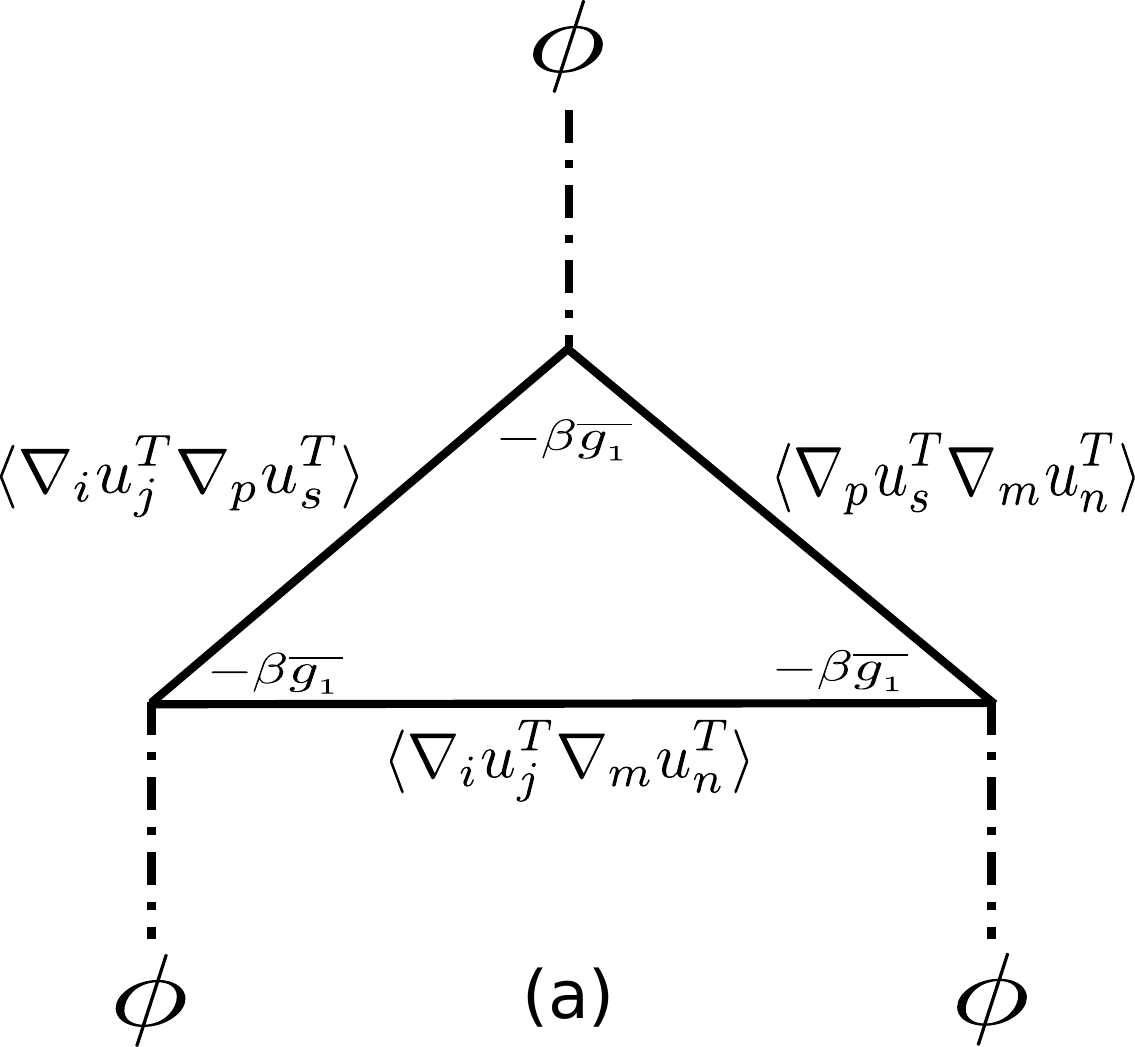}\hfill
 \includegraphics[width=9cm]{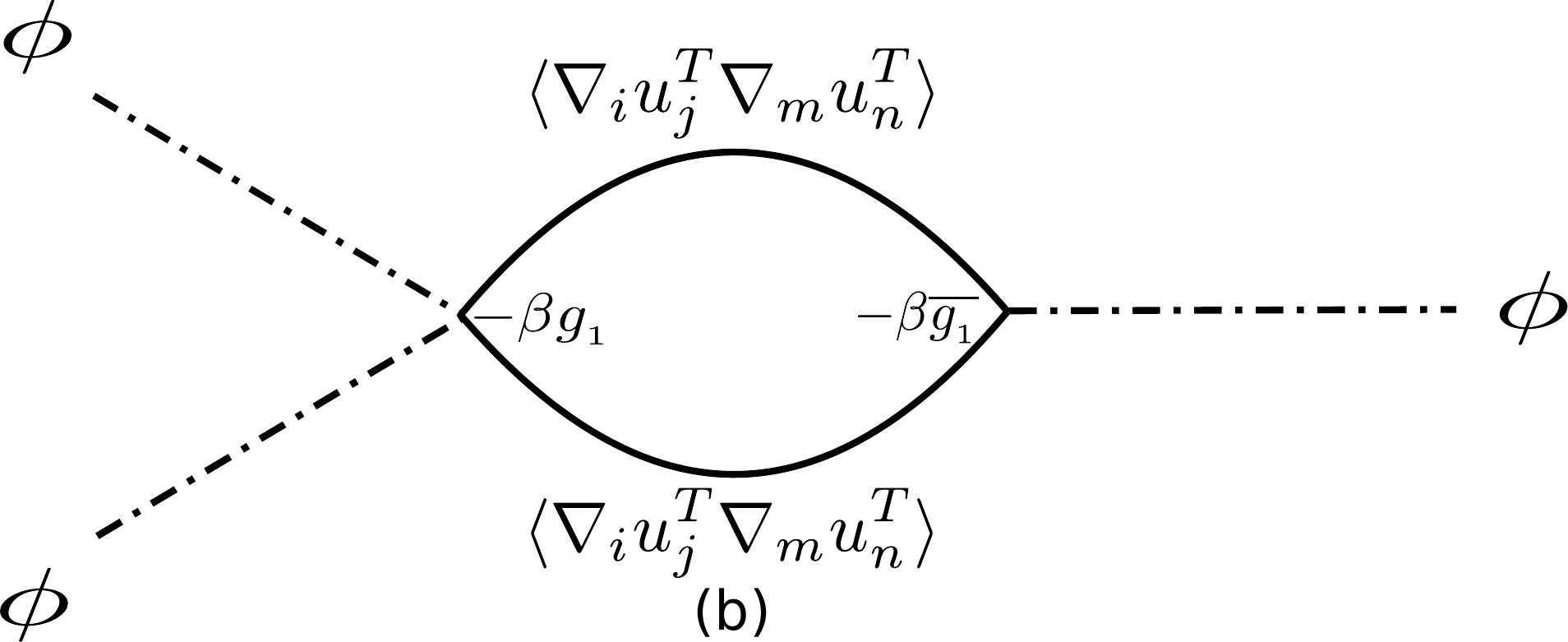}\\
 \includegraphics[width=7cm]{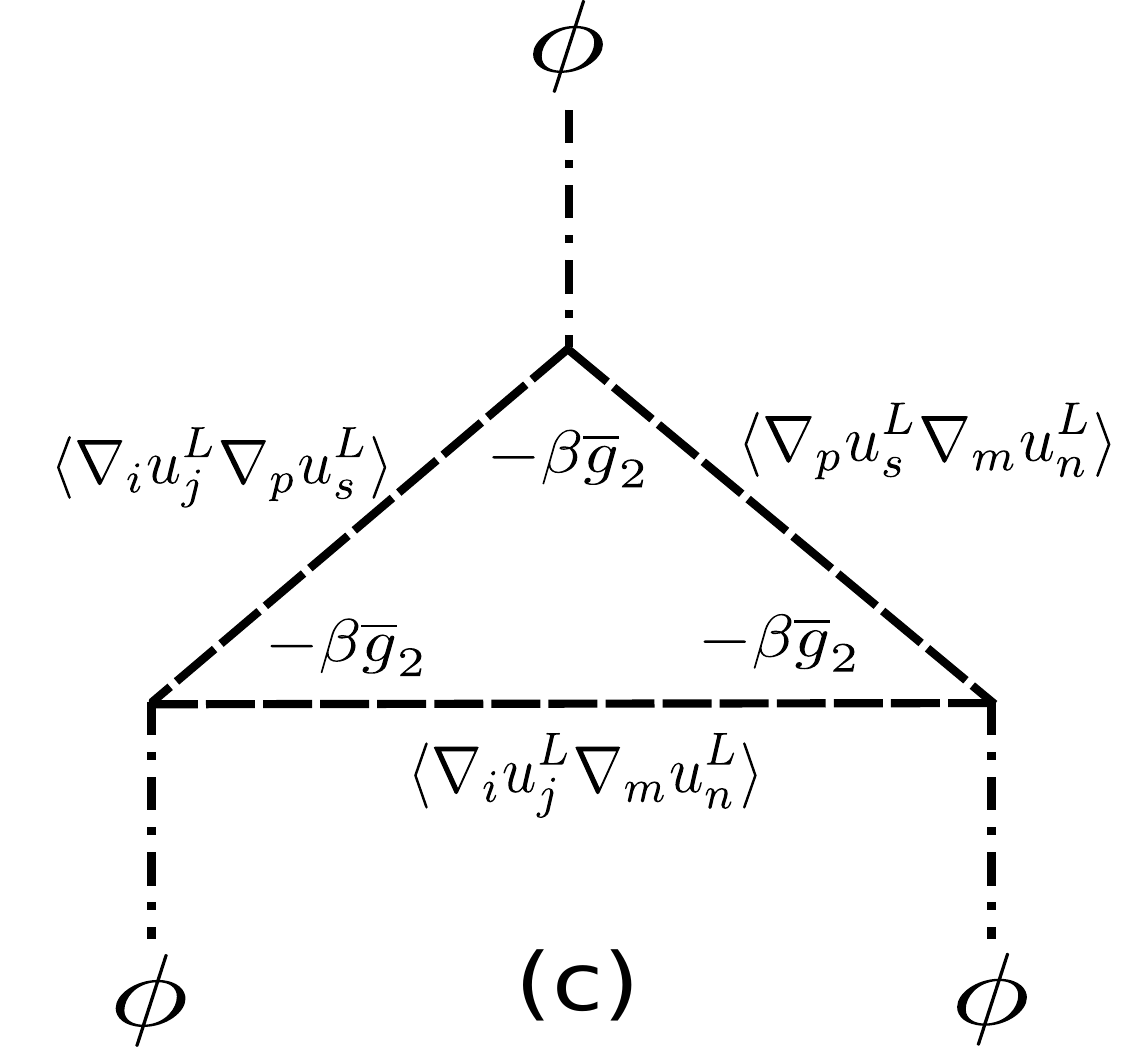}\hfill
 \includegraphics[width=9cm]{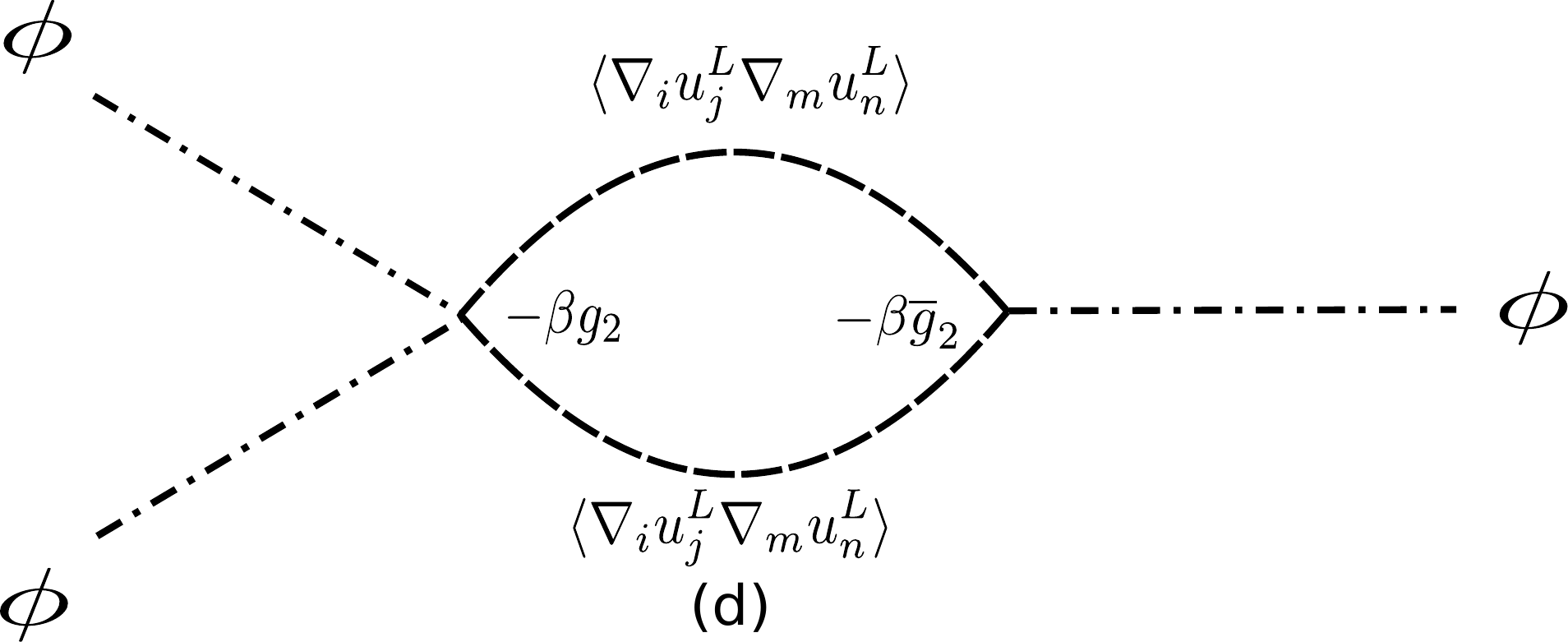}
 \caption{One-loop inhomogeneous Feynman graphs that correct $g$.}\label{inhom-g}
\end{figure}

\end{widetext}

These diagrams are {\em finite}. For instance, diagram \ref{inhom-v}(a) is given by
\begin{eqnarray}
  &&\beta_c^2 g_1^2 \int\frac{d^dq}{(2\pi)^d} \frac{T_c^2 q^4 \delta_{jm}\delta_{jm}}{4\mu^2 q^4} \nonumber \\&=& 2d g_1^2 \frac{1}{4\mu^2}\int\frac{d^dq}{(2\pi)^d} 
 =d  g_1^2 \frac{1}{2\mu^2} \frac{\Lambda^d}{(2\pi)^d}.
\end{eqnarray}
Similarly, diagram \ref{inhom-v}(b) is given by
\begin{eqnarray}
 &&2\overline \beta_c^4 g_1^4 \int\frac{d^dq}{(2\pi)^d} \frac{T_c^4 q^8 \delta_{js}\delta_{sk} \delta_{kn}\delta_{nj}}{16\mu^4 q^8}\nonumber \\&=&d \overline g_1^4 \frac{1}{8\mu^4} \frac{\Lambda^d}{(2\pi)^d}.
\end{eqnarray}
Diagram \ref{inhom-v}(c) is given by
\begin{eqnarray}
 &&\beta_c^2g_2^2 \int\frac{d^dq}{(2\pi)^d} \frac{T_c^2 q^4 \delta_{jm}\delta_{jm}}{\tilde \lambda^2 q^4} \nonumber \\&=& 2d g_2^2 \frac{1}{\tilde\lambda^2}\int\frac{d^dq}{(2\pi)^d} 
 =2d  g_2^2 \frac{1}{\tilde\lambda^2} \frac{\Lambda^d}{(2\pi)^d}.
 \end{eqnarray}
Similarly, diagram \ref{inhom-v}(d) is given by
\begin{eqnarray}
 &&2\beta_c^4 \overline g_2^4 \int\frac{d^dq}{(2\pi)^d} \frac{T_c^4 q^8 \delta_{js}\delta_{sk} \delta_{kn}\delta_{nj}}{\tilde\lambda^4 q^8}\nonumber \\&=&2d\overline g_2^4 \frac{1}{\tilde\lambda^4} \frac{\Lambda^d}{(2\pi)^d}.
\end{eqnarray}
Neglecting the homogeneous fluctuation corrections, we obtain Eq.~(\ref{v-tot}) above for $v_e$.


Consider the one-loop Feynman graphs in Fig.~\ref{inhom-g} that corrects $g$. These are all finite.
Diagram \ref{inhom-g}(a) is
\begin{equation}
 \beta_c^3 4 \overline g_1^3 \int\frac{d^dq}{(2\pi)^d} \frac{T_c^3 q^6\delta_{js}\delta_{sn}\delta_{nj}}{16\mu^4 q^6}.
\end{equation}

Diagram \ref{inhom-g}(b) is
\begin{equation}
 \beta_c^2 2g_1\overline g_1 \int\frac{d^dq}{(2\pi)^d}\frac{T_c^2 q^4\delta_{jm}\delta_{jm}}{4\mu^2 q^4}.
\end{equation}

Diagram \ref{inhom-g}(a) is
\begin{equation}
 \beta_c^4 4 \overline g_2^3 \int\frac{d^dq}{(2\pi)^d} \frac{T_c^3 q^6\delta_{js}\delta_{sn}\delta_{nj}}{\tilde\lambda^4 q^6}.
\end{equation}

Diagram \ref{inhom-g}(c) is
\begin{equation}
 \beta_c^2 2g_2\overline g_2 \int\frac{d^dq}{(2\pi)^d}\frac{T_c^2 q^4\delta_{jm}\delta_{jm}}{\tilde\lambda^2 q^4}.
\end{equation}
Neglecting the homogeneous fluctuation corrections, we obtain Eq.~(\ref{g-tot}) above for $g_e$. Here, $\beta_c=1/T_c$.


\begin{thebibliography}{99}
\bibitem{iso} Although single crystals are not isotropic, polycrystalline materials are macroscopically isotropic, which is due to the random orientations of the constituent crystallites. Similarly, gels are also isotropic at large enough scales.
\bibitem{conju} It is clear that if we replace $(u_{ij})^2$ by its average (which is non-zero), a linear term in $\phi$ is generated, with a coefficient that is the ``external magnetic field''. Accessing the critical point would then require adding a suitable counter term in $\cal F$; see also later.
\bibitem{dTc} Equation (\ref{eff-tc}) may be written schematically as $\hat r \equiv T-T_c^* = r+ 2 (\delta V/V)^2\times {\cal O}(1)$. Very close to the critical point $\hat r\approx 0$, giving $dT_c/dV\sim \delta V$ that approaches zero in the limit of small $\delta V$.
\end{thebibliography}
\end{document}